\documentclass[prd,nofootinbib,twocolumn]{revtex4}

\usepackage{amssymb}
\usepackage{amsmath}
\usepackage{graphicx,subfigure}
\usepackage{longtable}
\usepackage{verbatim}
\usepackage{amsfonts}
\usepackage{hyperref}
\usepackage{cancel}
\usepackage{color}

\DeclareMathAlphabet\mathbfcal{OMS}{cmsy}{b}{n} 

\newcommand{\be}{\begin{equation}}
\newcommand{\ee}{\end{equation}}
\newcommand{\bea}{\begin{eqnarray}}
\newcommand{\eea}{\end{eqnarray}}
\newcommand{\beq}{\begin{equation}}
\newcommand{\eeq}{\end{equation}}
\def\beqa{\begin{eqnarray}}
  \def\eeqa{\end{eqnarray}}
\newcommand{\bv}{\left(\begin{array}{c}}
\newcommand{\ev}{\end{array}\right)}
\def\lsim{\mathrel{\rlap{\lower4pt\hbox{\hskip1pt$\sim$}}
    \raise1pt\hbox{$<$}}}         
\def\gsim{\mathrel{\rlap{\lower4pt\hbox{\hskip1pt$\sim$}}
    \raise1pt\hbox{$>$}}}         
    
\newcommand{\bra}[1]{\left\langle{#1}\right\vert}
\newcommand{\ket}[1]{\left\vert{#1}\right\rangle}
\newcommand{\nn}{\nonumber}
\newcommand{\ov}{\overline}
\newcommand{\eq}[1]{Eq.~(\ref{#1})}
\newcommand{\eqsand}[2]{Eqs.~(\ref{#1}) and (\ref{#2})}
\newcommand{\eqsto}[2]{Eqs.~(\ref{#1}--\ref{#2})}
\newcommand{\Kbar}{\,\overline{\!K}}
\newcommand{\KorKbar}{\raisebox{5.4pt}{$\scriptscriptstyle(\hspace*{9.9pt})$}
  \hspace*{-13.4
pt}\Kbar{}^0\,}

\newcommand*{\vcenteredhbox}[1]{\begingroup
\setbox0=\hbox{#1}\parbox{\wd0}{\box0}\endgroup}

\begin{document}

\begin{flushright}
TTP15-015 
\end{flushright}

\vspace*{-30mm}

\title{\boldmath 
Topological amplitudes in $D$ decays to two pseudoscalars:\\
a global analysis with linear $SU(3)_F$ breaking
}

\author{Sarah M\"uller$^{\,a,b}$}
\email{sarah.mueller2@kit.edu} 
\author{Ulrich Nierste$^{\,b}$}
\email{ulrich.nierste@kit.edu}
\author{Stefan Schacht$^{\,b}$}
\email{stefan.schacht@kit.edu}
\affiliation{
$^{\,a}$ Institut f\"ur Experimentelle Kernphysik, Karlsruher
  Institut f\"ur Technologie, D-76021 Karlsruhe, Germany\\
$^{\,b}$ Institut f\"ur Theoretische Teilchenphysik, Karlsruher
  Institut f\"ur Technologie, D-76128 Karlsruhe, Germany}

\vspace*{1cm}

\begin{abstract}
   We study decays of $D^0$, $D^+$, and $D_s^+$ mesons into two
    pseudoscalar mesons by expressing the decay amplitudes in terms of
    topological amplitudes. Including consistently SU(3)$_F$ breaking to
    linear order, we show how the topological-amplitude decomposition
    can be mapped onto the standard expansion using reduced amplitudes
    characterized by SU(3) representations. The tree and annihilation
    amplitudes can be calculated in factorization up to corrections
    which are quadratic in the color-counting parameter $1/N_c$.  We
  find new sum rules connecting $D^+\rightarrow K_SK^+$,
  $D_s^+\rightarrow K_S\pi^+$ and $D^+\rightarrow K^+\pi^0$, which test
    the quality of the $1/N_c$ expansion.  Subsequently, we determine the
    topological amplitudes in a global fit to the data, taking the
    statistical correlations among the various measurements into
    account.  We carry out likelihood ratio tests in order to
  quantify the role of specific topological contributions. While the
  SU(3)$_F$ limit is excluded with a significance of more than five
    standard deviations, a good fit (with $\Delta \chi^2 <1$) can be
    obtained with less than $28\%$ of SU(3)$_F$ breaking in the
    decay amplitudes.  The magnitude of the penguin amplitude
    $P_{\mathrm{break}}$, which probes the Glashow Iliopoulos Maiani
  (GIM) mechanism, is consistent with zero; the hypothesis
    $P_{\mathrm{break}}=0$ is rejected with a significance of just
    $0.7\sigma$. We obtain the Standard-Model correlation
    between $\mathcal{B}(D^0\rightarrow K_L \pi^0)$ and
    $\mathcal{B}(D^0\rightarrow K_S\pi^0)$, which probes doubly
    Cabibbo-suppressed amplitudes, and find that
    $\mathcal{B}(D^0\rightarrow K_L \pi^0)<\mathcal{B}(D^0\rightarrow
    K_S\pi^0)$ holds with a significance of more than~$4\sigma$. 
   We finally predict 
   $\mathcal{B}(D_s^+\rightarrow K_L K^+)= 0.012^{+0.006}_{-0.002}$ at
   $3\sigma$ CL.  
\end{abstract}

\maketitle

\section{Introduction}\label{sec:intro}

{While there is a plethora of experimental information on hadronic charm
  decays, no theoretical method for dynamical, QCD-based predictions for
  the corresponding decay amplitudes is known. The best theoretical
  approach uses the approximate SU(3)$_F$ symmetry of the QCD {Lagrangian}
  to relate the amplitudes of different decay modes to each other. If
  one assumes this symmetry to be exact, one can express the amplitudes
  of all measured decay modes in terms of a smaller number of
  parameters, which are the reduced amplitudes characterized by
  SU(3)$_F$ quantum numbers. Then one can predict the less precisely
  measured branching fraction on the basis of exact SU(3)$_F$ or assess
  the validity of this assumption from the overall quality of the fit
  \cite{Altarelli:1974sc, Kingsley:1975fe, Einhorn:1975fw,
    Voloshin:1975yx, Cabibbo:1977zv, Quigg:1979ic, Wang:1979dx,
    Eilam:1979mn, Wang:1980ac, Zeppenfeld:1980ex, Chau:1982da,
    Golden:1989qx, Gronau:1994rj, Bhattacharya:2009ps}.  SU(3)$_F$ is
  broken, because the masses $m_{u,d,s}$ of the three lightest quarks
  are not equal. Comparing the differences among these masses with a
  typical hadronic scale one estimates SU(3)$_F$ breaking to be around
  30\%. In practice the quality of SU(3)$_F$ symmetry can be much better
  (e.g.\ in heavy-hadron spectroscopy) or much worse (e.g.\ in
  heavy-quark fragmentation) and should be critically assessed for each
  system to which it is applied. Linear (i.e.\ first-order) SU(3)$_F$
  breaking can be rigorously included into the parameterization of the
  amplitudes, at the expense of a larger number of reduced
  amplitudes. In the case of $D\to PP^\prime$, where $D=D^0$, $D^+$ or
  $D_s^+$ and $P$, $P^\prime$ represent pseudoscalar mesons, such
  studies have been performed in Refs.~\cite{Chau:1986du, Chau:1987tk,
    Chau:1991gx, Savage:1991wu, Gronau:1995hm, Grinstein:1996us,
    Cheng:2010ry, Pirtskhalava:2011va, Hiller:2012xm, Franco:2012ck,
    Cheng:2012wr, Feldmann:2012js, Brod:2012ud, Bhattacharya:2012ah}.
  (Remarkably, one can find relations between amplitudes which even hold
  to first order in SU(3)$_F$ breaking \cite{Grossman:2012ry}.) Since
  there are fewer $D\to PP^\prime$ branching fractions than {real
    parameters}, there is a multi-dimensional space of solutions (all
  giving a perfect $\chi^2$) for the latter. Many of these solutions
  involve reduced SU(3)$_F$-breaking amplitudes whose sizes are indeed
  of order 30\% or less than the SU(3)$_F$-leading ones, giving evidence
  (but no proof) that the SU(3)$_F$ expansion works. The redundancy
  associated with the multi-dimensional space of solutions poses a
  challenge for the numerical method to find the best-fit solutions
  because of the many flat directions in the space of reduced
  amplitudes.}

{An alternative way to parameterize decay amplitudes {involves}
  topological amplitudes which are characterized by the flavor flow in
  the decays \cite{Fakirov:1977ta, Cabibbo:1977zv, Eilam:1979mn, Bernreuther:1980bw, Wang:1980ac, Zeppenfeld:1980ex,Chau:1982da, Chau:1986du, Chau:1987tk, Chau:1991gx, Gronau:1994rj, Gronau:1995hm, Bhattacharya:2009ps, Bhattacharya:2012ah, Li:2012cfa}. The
  building blocks of this approach are shown in
  Tab.~\ref{tab:su3limit-diagrams} and
  Fig.~\ref{fig:su3limit-penguin-annihilation}. The topological
  amplitudes permit an easy and intuitive implementation of SU(3)$_F$
  relations. They further have the merit that they categorize the decays
  by dynamical criteria (i.e.\ whether the valence quark takes part in
  the weak interaction and which meson picks it up) and
  permit the combination of SU(3)$_F$ methods with other calculational
  methods. In this paper we take a first step in this direction and
  apply the $1/N_c$ expansion (first applied to $D$ decays in
  Ref.~\cite{Buras:1985xv}) to the tree ($T$) and annihilation ($A$) amplitudes of
  Tab.~\ref{tab:su3limit-diagrams}. ($N_c=3$ is the number of colors.)
  $T$ and $A$ each factorize into the product of a form factor and a
  decay constant up to corrections of order $1/N_c^2$. We further
  include linear SU(3)$_F$ breaking in the topological-amplitude
  decomposition, similar to the study of $B$ decays in
  Ref.~\cite{Gronau:1995hm}. For fixed values of $T$ and $A$ (obtained by
  adding a chosen $1/N_c^2$ deviation to the factorized expressions) the
  number of fitted {complex topological amplitudes is reduced from 17 to nine}, 
  so that the problem of flat directions is substantially alleviated.}

{The purpose of this paper is a systematic determination of the
  topological amplitudes including linear SU(3)$_F$ breaking from a
  global fit to 16 $D\to PP^\prime$ branching fractions and the measured
  strong-phase difference $\delta_{K^+\pi^-}$. For each topological
  amplitude we quantify the amount of SU(3)$_F$ breaking with
  statistical likelihood-ratio tests using the statistical package
  \textit{my}Fitter \cite{Wiebusch:2012en}. {The latter is especially convenient in 
  order to include nonlinear constraints in a  
  frequentist analysis using the SLSQP algorithm implemented in \texttt{SciPy} \cite{Kraft:1988,SciPy:2001}.}
  As a novel feature our statistical analysis fully includes the statistical correlations
  between the different experimental inputs.  The {ranges} of
  the topological amplitudes found by us are an important input for the
  prediction of CP asymmetries. However, the latter also involve
  quantities which cannot be extracted from branching fractions
  (SU(3)$_F$ triplet amplitudes), so that additional input is needed for
  this purpose. This is one reason why we do not include measurements of
  CP asymmetries in our fit input. The other reason is their sensitivity
  to new physics, whose quantification should be separated from the
  determination of hadronic parameters as much as possible. In this
  paper we also do not consider decays into final states with $\eta$ or
  $\eta^\prime$, which involve additional parameters.}

{The paper is organized as follows:} In Sec.~\ref{sec:topocharm} we
present the parameterization of $D$ decay amplitudes using topological
amplitudes.  We discuss the inclusion of linear SU(3)$_F$ breaking and
the appearing parametric redundancies in the diagrammatic language.
In Sec.~\ref{sec:theoryinput} we 
{combine the method with} $1/N_c$
counting {and define our} measures of SU(3)$_F$ breaking.
In Sec.~\ref{sec:fits} we present the result of our fit. Finally, we
conclude.  

\section{Diagrammatic parameterization of charm decays}
\label{sec:topocharm}

\subsection{Notation}\label{sec:notation}

{We choose the following conventions for the meson states:}
\begin{align}
\ket{K^+}       &=  \ket{u\bar{s}}\,, &
\ket{K^0}       &=  \ket{d\bar{s}}\,, \label{eq:states-1}\\
\ket{K^-}       &= -\ket{s\bar{u}}\,, &
\ket{\bar{K}^0} &=  \ket{s\bar{d}}\,, \label{eq:states-2}\\
\ket{\pi^+}     &= \ket{u\bar{d}}\,,  &
\ket{\pi^0}     &= \frac{1}{\sqrt{2}} \left(\ket{d\bar{d}} - \ket{u\bar{u}}\right)\,, \label{eq:states-3}\\
\ket{\pi^-}     &= -\ket{d\bar{u}}\,, &     
\ket{D^0}	&= -\ket{c\bar{u}}\,, \label{eq:states-4}\\ 
\ket{D^+}	&= \ket{c\bar{d}}\,, &
\ket{D_s^+}	&= \ket{c\bar{s}}\, . \label{eq:states-5}
\end{align} 
{Here the ``$=$'' sign means that the flavor quantum numbers of the
  meson state on the left-hand side {equal} those of the quark-antiquark
  state on the right-hand side. Tab.~\ref{tab:su3limit-diagrams}
  shows the  topological (flavor-flow) amplitudes. 
The cross denotes the $W$-boson exchange encoded in the 
$\Delta C=1$ {Hamiltonian}.} 
We write the Cabibbo-favored (CF), singly Cabibbo-suppressed (SCS), and
doubly Cabibbo suppressed (DCS) decays as
\begin{align}
  \mathcal{A}^{\mathrm{CF}}(d)  &\equiv  V_{cs}^* V_{ud} \mathcal{A}(d) \equiv V_{cs}^* V_{ud} \sum_i c^d_i
  \mathcal{T}_i\,, 
   \label{eq:amps-cf}  \\
  \mathcal{A}^{\mathrm{SCS}}(d) &\equiv  \lambda_{sd} \mathcal{A}(d) \equiv \lambda_{sd} \sum_i c^d_i
  \mathcal{T}_i\,, 
   \label{eq:amps-scs} \\
  \mathcal{A}^{\mathrm{DCS}}(d) &\equiv  V_{cd}^* V_{us} \mathcal{A}(d) \equiv V_{cd}^* V_{us} \sum_i c^d_i
  \mathcal{T}_i\, \label{eq:amps-dcs}.
\end{align}
Here, we defined 
\begin{align}
\lambda_{sd} := (\lambda_s-\lambda_d)/2 := (V_{cs}^* V_{us} - V_{cd}^* V_{ud})/2\,,   
\end{align}
where $\lambda_{sd} \simeq \lambda_s \simeq -\lambda_d$. 
$\mathcal{T}_i$ is a {topological} amplitude (see
Tabs.~\ref{tab:su3limit-diagrams},\,\ref{tab:su3breaking-diagrams}) and
$c_i^{{d}}$ is the corresponding coefficient from
Tab.~\ref{tab:topoparametrization} {and $d=D\to PP^\prime$ labels
  the decay mode}.  {There is a CKM-suppressed 
part $\propto V_{cb}^*V_{ub}$ in SCS amplitudes which can be safely
neglected in all  branching ratios.}

{In the limit of unbroken SU(3)$_F$ symmetry only the} tree ($T$),
annihilation ($A$), color-suppressed ($C$), and exchange ($E$)
{amplitudes are needed to parameterize all $D\to P P^\prime$ decays.
  While the penguin amplitude $P_{s,d,b}$ (labeled with the quark
  flavor running in the loop) is also non-vanishing in  unbroken
  SU(3)$_F$, it only appears in the combination 
\begin{align}
P_{\mathrm{break}} &\equiv P_s - P_d ,
\label{eq:pingubreak}
\end{align}
where we have adopted the notation of Ref.~\cite{Brod:2012ud}.
$T$,$A$,$C$, and $E$ are commonly fitted together with the penguin
amplitude $P_{\mathrm{break}}$, which vanishes in the SU(3)$_F$ limit}
\cite{Zeppenfeld:1980ex,Wang:1980ac,Chau:1982da, Gronau:1994rj,
  Gronau:1995hm, Buras:1998ra, Bhattacharya:2012ah}.  The normalization
of the amplitudes is such that
\begin{align}
\mathcal{B}(D\rightarrow P_1 P_2) &= 
    \vert \mathcal{A}^X(D\rightarrow P_1 P_2)\vert^2 \times \mathcal{P}(D,P_1,P_2)\,, \\
\mathcal{P}(D,P_1,P_2) &\equiv\tau_D \times \frac{1}{ 16 \pi m_D^3} \times \nn\\ 
&\hspace{-1.5cm}\sqrt{ (m_D^2 - (m_{P_1} - m_{P_2})^2 ) ( m_D^2 - (m_{P_1} + m_{P_2})^2) } \,. \label{eq:phasespace}
\end{align}
with $X=\mathrm{CF},\mathrm{SCS},\mathrm{DCS}$.  
In the following, we will only make use of the notation $\mathcal{A}(d)$ without superscript, see \eqsto{eq:amps-cf}{eq:amps-dcs}.

\begin{table}[tp]
\begin{center}
\begin{tabular}{c|c}
\hline \hline
Name & Diagrams \\\hline
$T$ &  \vcenteredhbox{\includegraphics[width=.35\linewidth,clip=true]{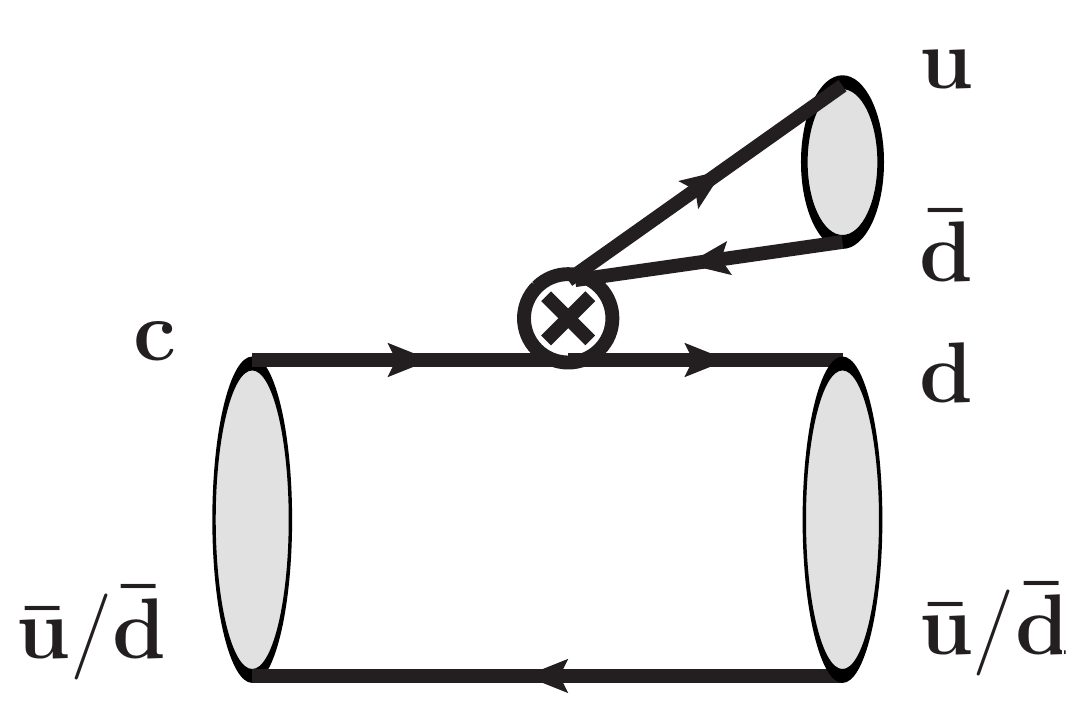}} \\\hline
$A$ &  \vcenteredhbox{\includegraphics[width=.35\linewidth,clip=true]{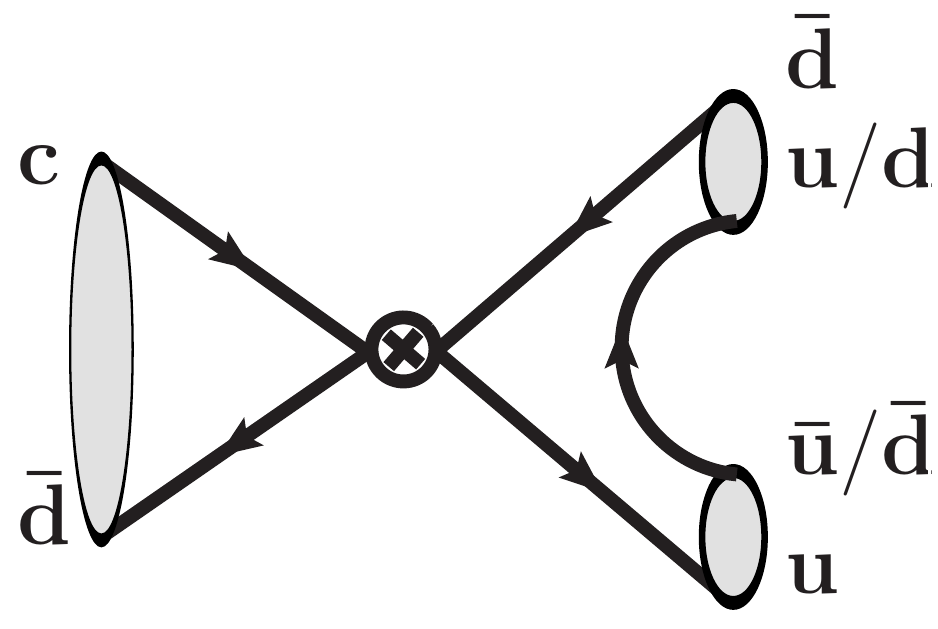}} \\\hline
$C$ &  \vcenteredhbox{\includegraphics[width=.4\linewidth]{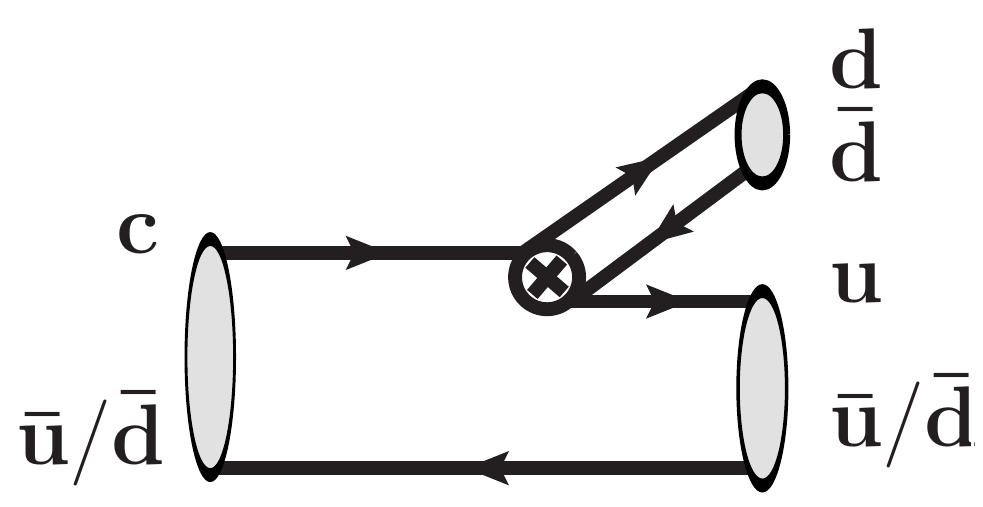}} \\\hline
$E$ &  \vcenteredhbox{\includegraphics[width=.4\linewidth]{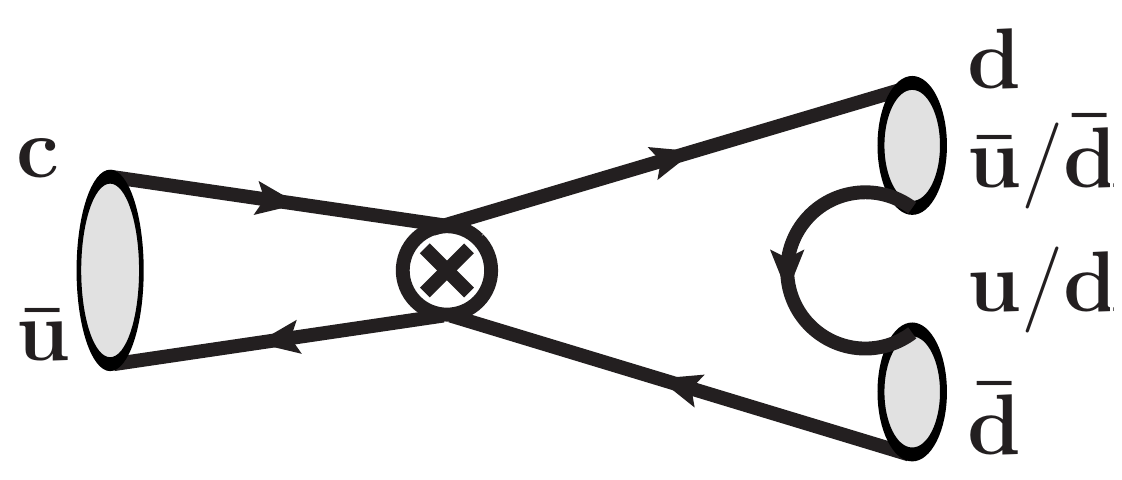}} \\\hline
$P_d$ &  \vcenteredhbox{\includegraphics[width=.4\linewidth]{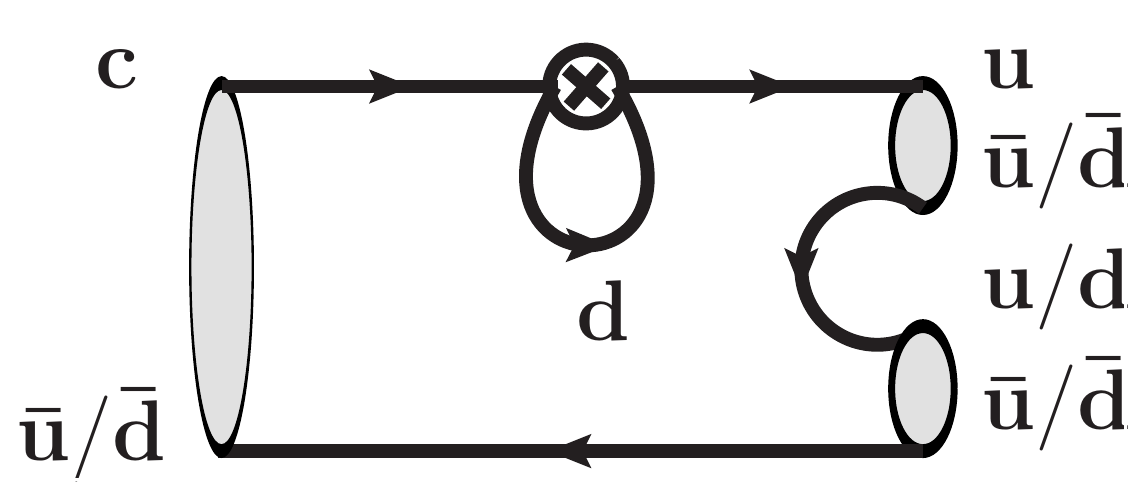}} \\\hline\hline
\end{tabular}
\caption{SU(3)$_F$-limit {topological amplitudes.} 
\label{tab:su3limit-diagrams}}
\end{center}
\end{table}

\begin{table*}[tp!]
\begin{center}
\begin{tabular}{c|c|c}
\hline \hline
Name & $s-d$ difference of topologies & denoted by Feynman rule \\\hline
$T^{(1)}_{1}$ &  \vcenteredhbox{\includegraphics[width=.17\linewidth]{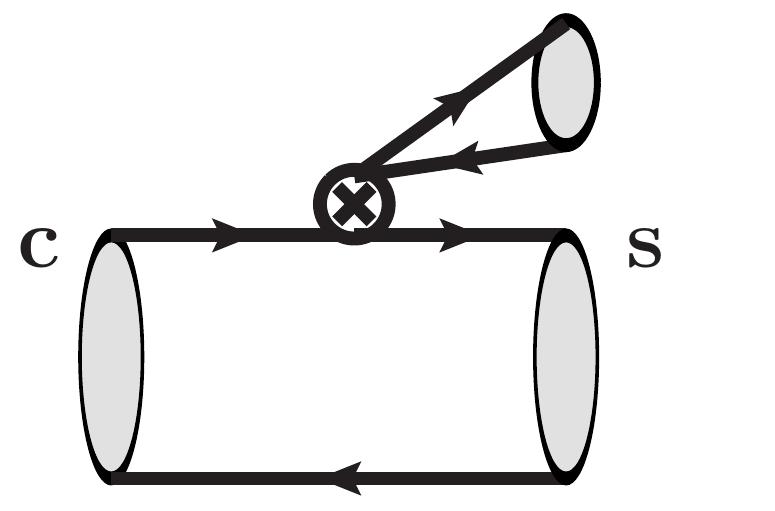}} $-$ 
		 \vcenteredhbox{\includegraphics[width=.17\linewidth]{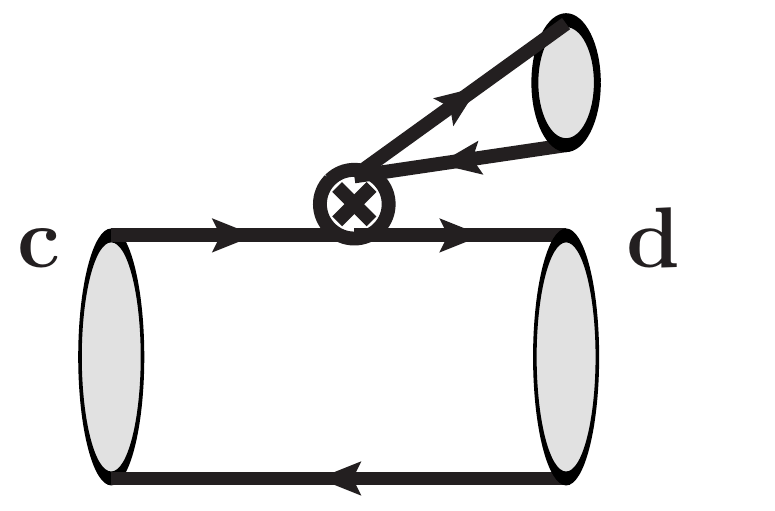}} & 
		 \vcenteredhbox{\includegraphics[width=.14\linewidth]{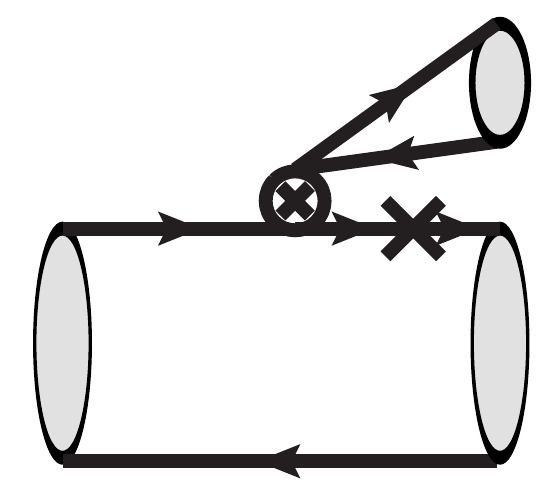}} \\\hline
$T^{(1)}_{2}$ &  \vcenteredhbox{\includegraphics[width=.17\linewidth]{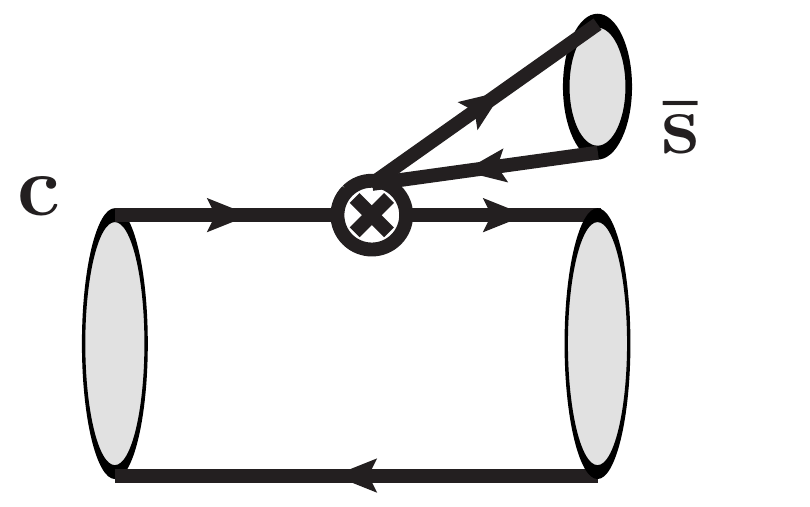}} $-$ 
		 \vcenteredhbox{\includegraphics[width=.17\linewidth]{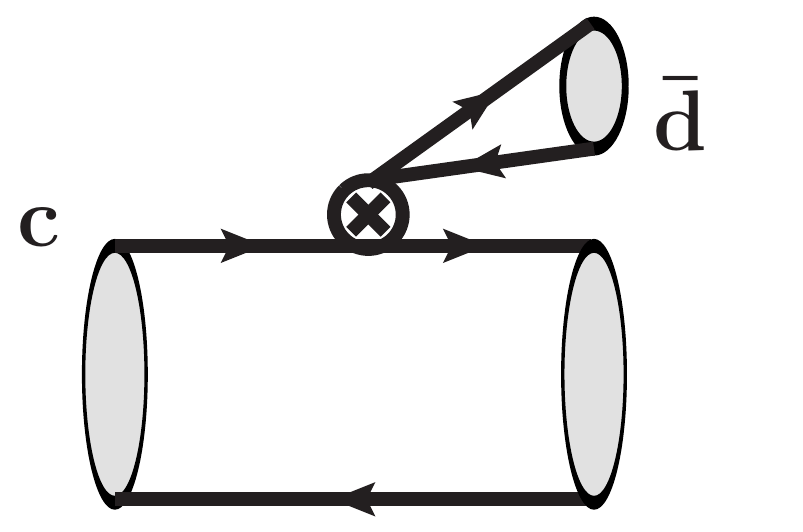}}  & 
		 \vcenteredhbox{\includegraphics[width=.14\linewidth]{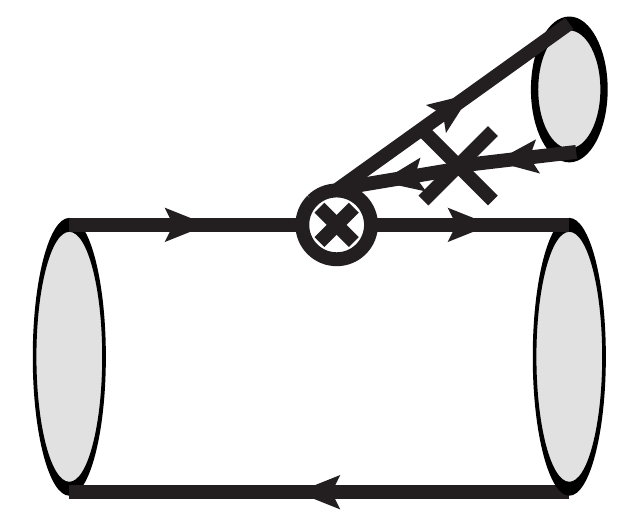}} \\\hline
$T^{(1)}_{3}$ &  \vcenteredhbox{\includegraphics[width=.17\linewidth]{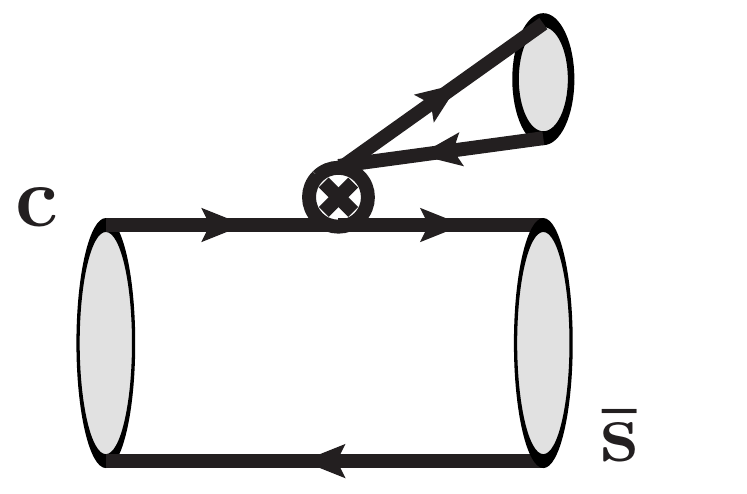}} $-$ 
		 \vcenteredhbox{\includegraphics[width=.17\linewidth]{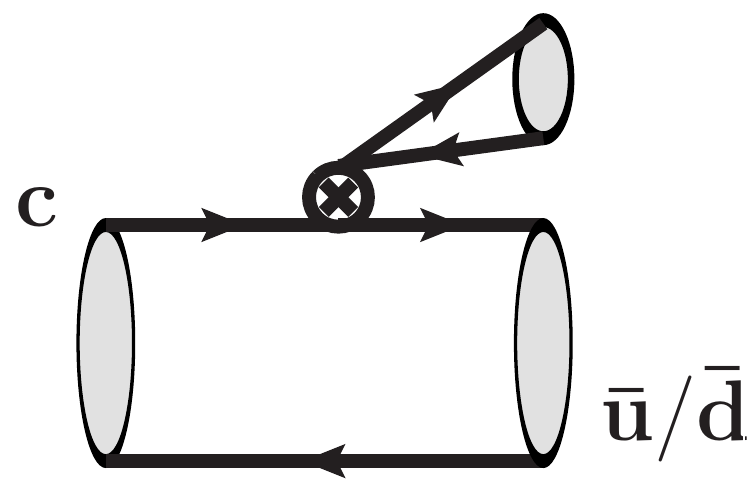}}  & 
		 \vcenteredhbox{\includegraphics[width=.14\linewidth]{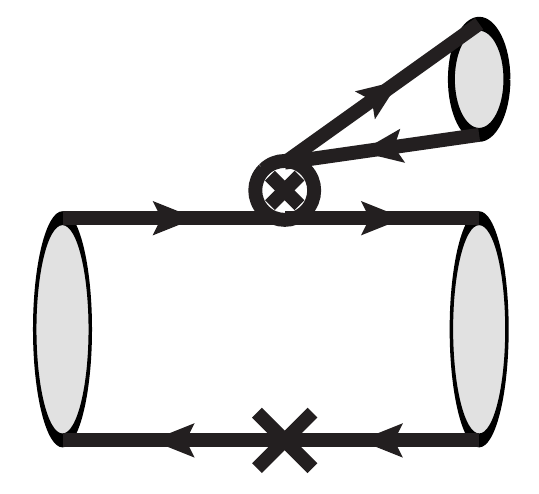}} \\\hline
$A^{(1)}_{1}$ &  \vcenteredhbox{\includegraphics[width=.17\linewidth]{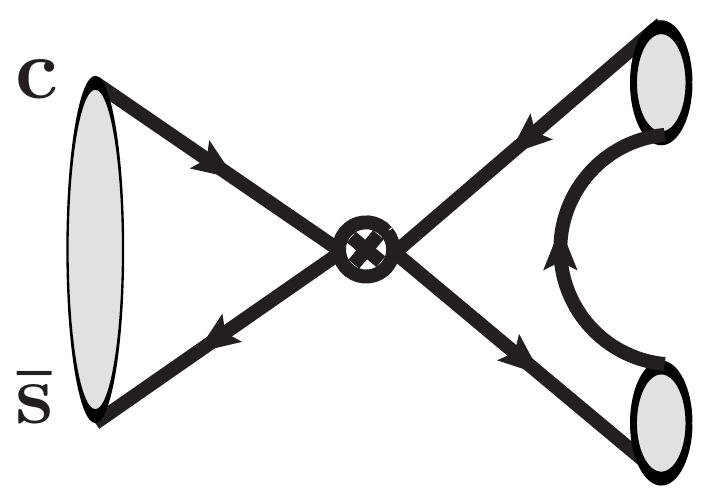}} $-$ 
		 \vcenteredhbox{\includegraphics[width=.17\linewidth]{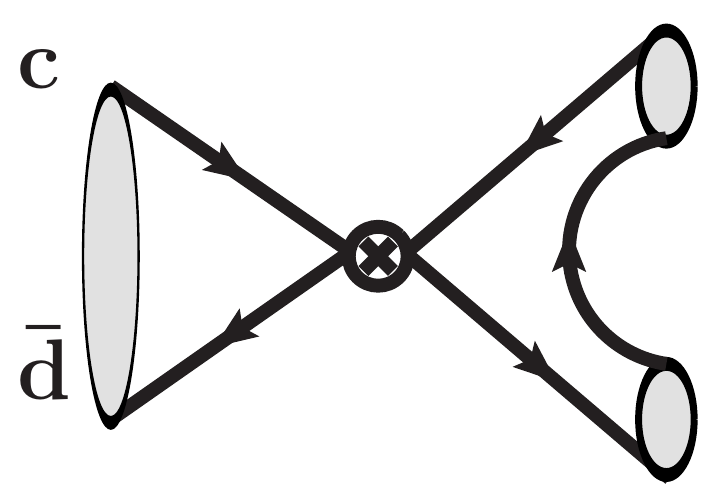}}  & 
		 \vcenteredhbox{\includegraphics[width=.14\linewidth]{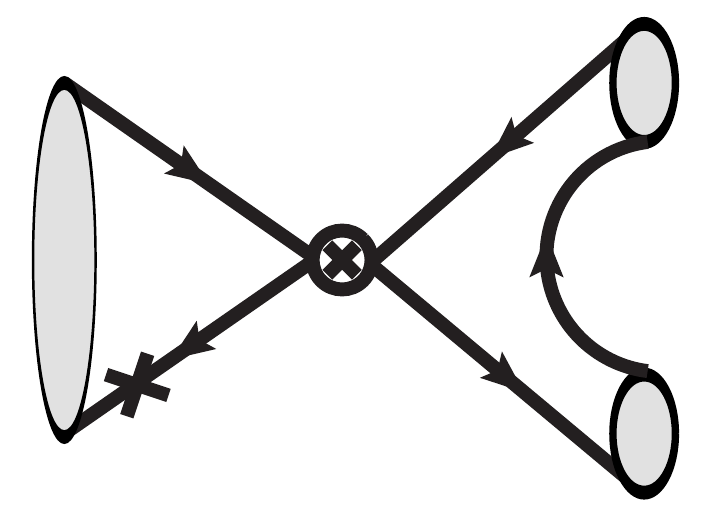}} \\\hline
$A^{(1)}_{2}$ &  \vcenteredhbox{\includegraphics[width=.17\linewidth]{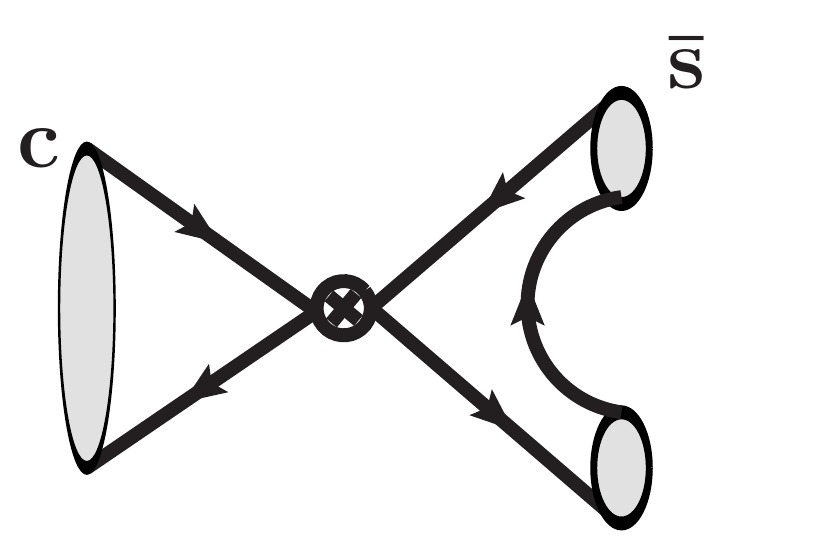}} $-$ 
		 \vcenteredhbox{\includegraphics[width=.17\linewidth]{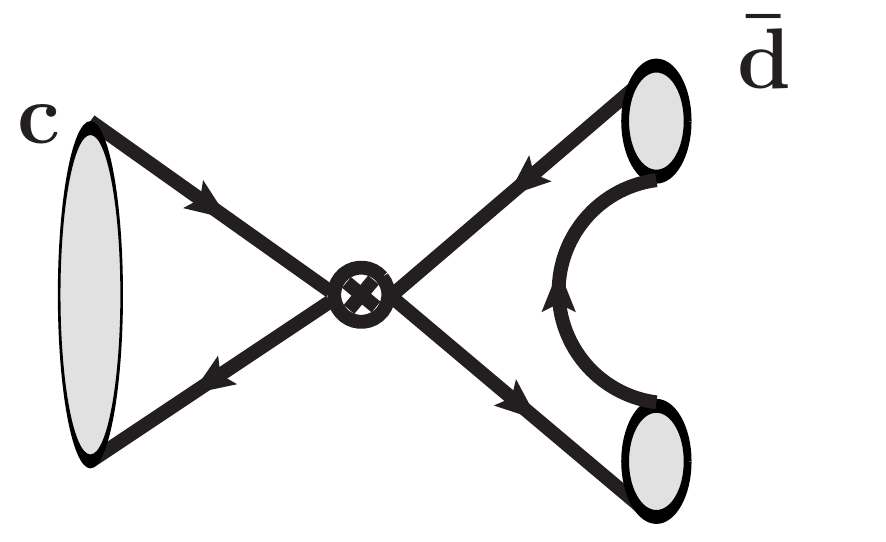}}  & 
		 \vcenteredhbox{\includegraphics[width=.14\linewidth]{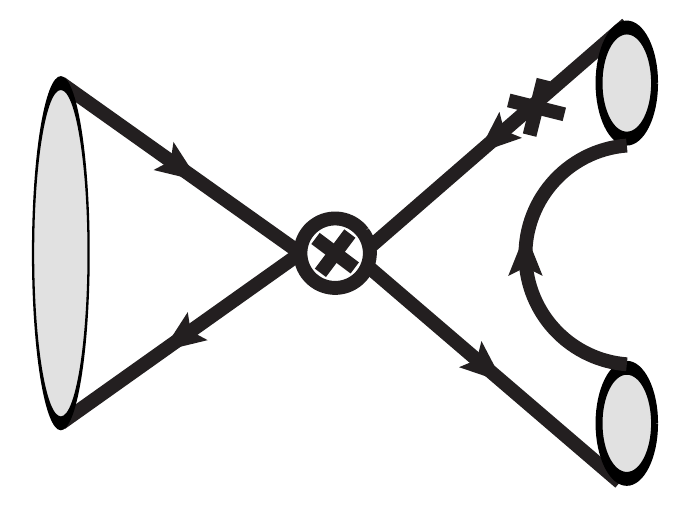}} \\\hline
$A^{(1)}_{3}$ &  \vcenteredhbox{\includegraphics[width=.17\linewidth]{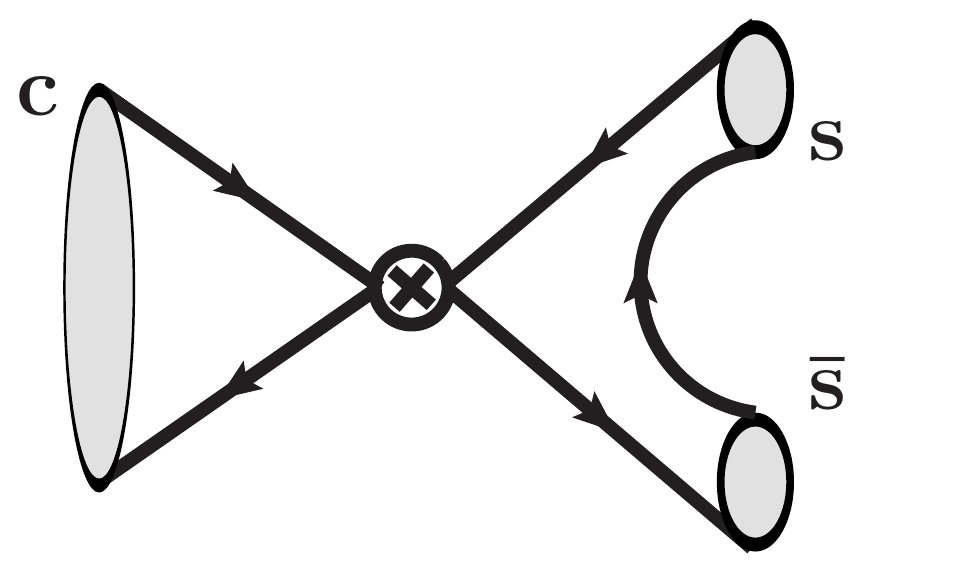}} $-$ 
		 \vcenteredhbox{\includegraphics[width=.17\linewidth]{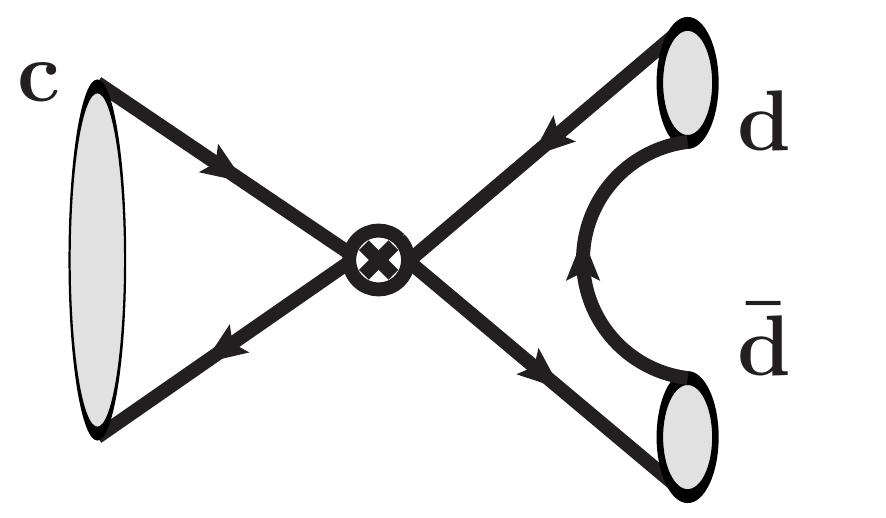}}  & 
		 \vcenteredhbox{\includegraphics[width=.14\linewidth]{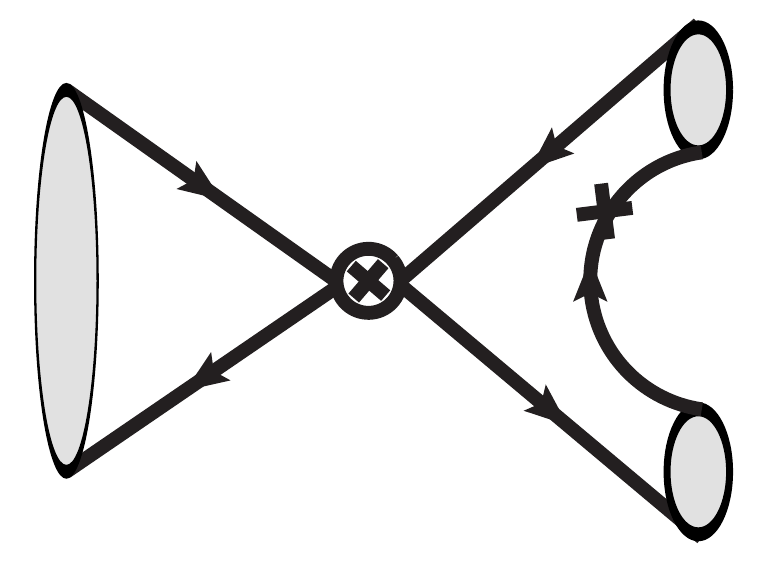}} \\\hline
$C^{(1)}_{1}$ &  \vcenteredhbox{\includegraphics[width=.19\linewidth]{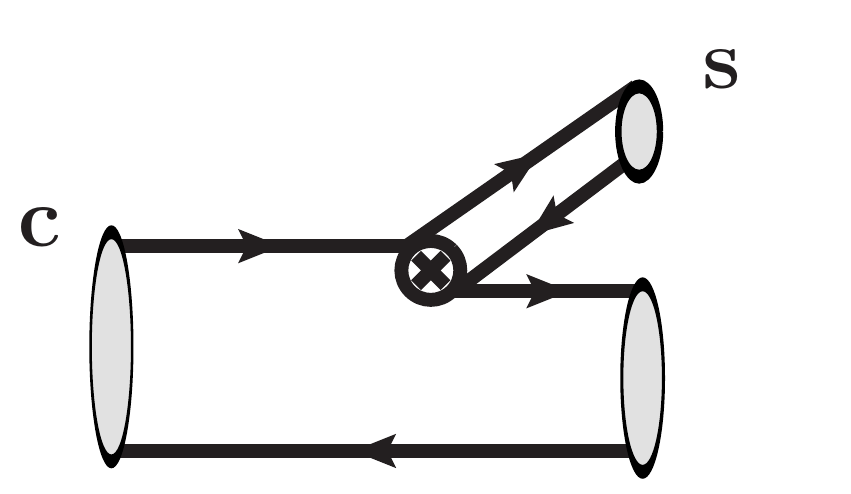}} $-$ 
		 \vcenteredhbox{\includegraphics[width=.19\linewidth]{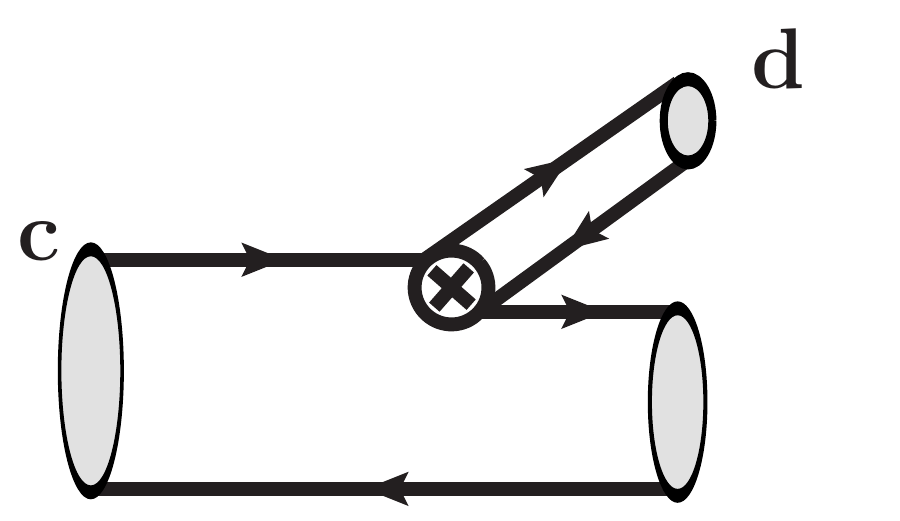}}  & 
		 \vcenteredhbox{\includegraphics[width=.16\linewidth]{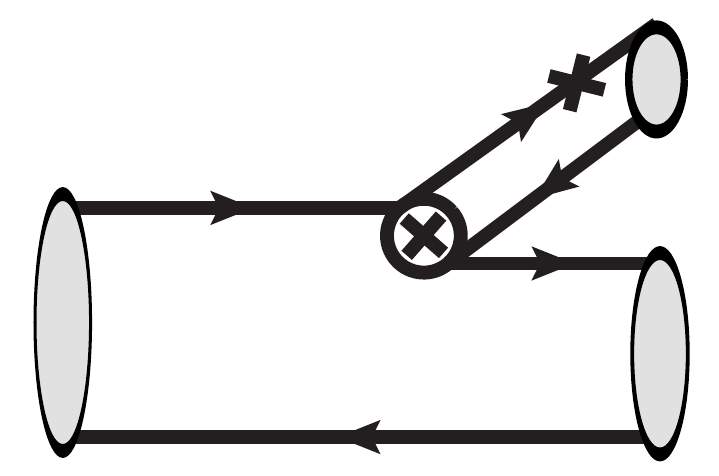}} \\\hline
$C^{(1)}_{2}$ &  \vcenteredhbox{\includegraphics[width=.19\linewidth]{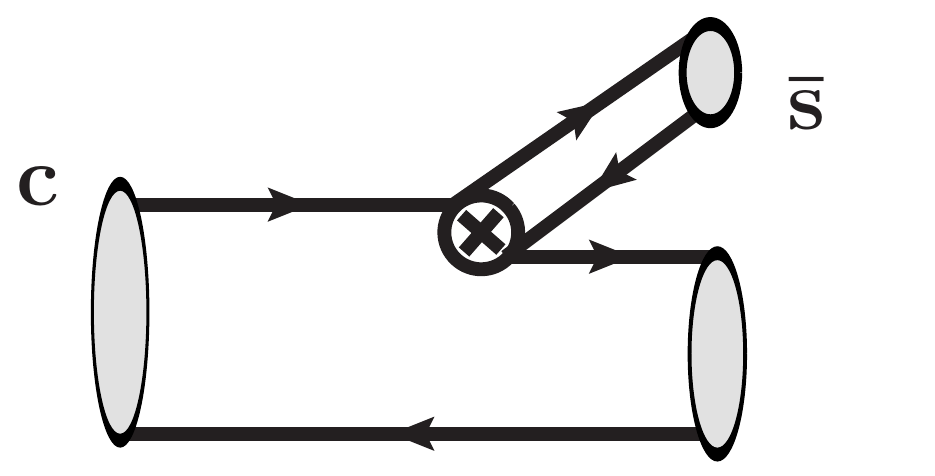}} $-$ 
		 \vcenteredhbox{\includegraphics[width=.19\linewidth]{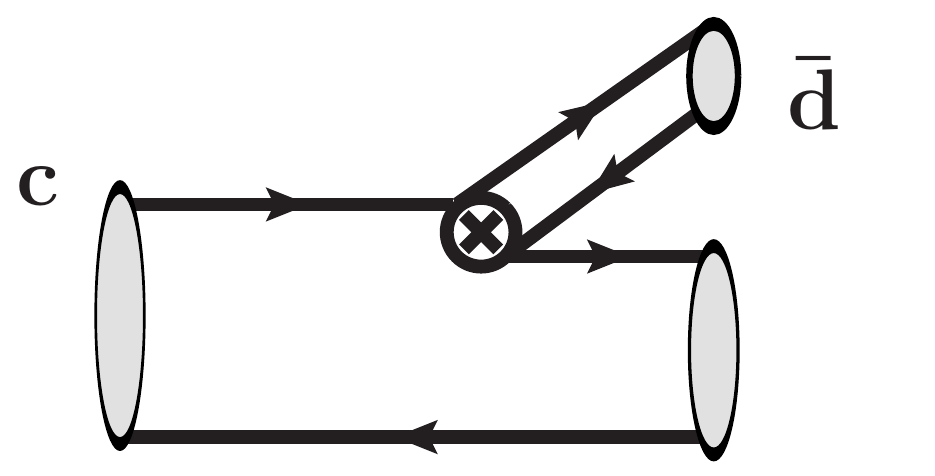}}  & 
		 \vcenteredhbox{\includegraphics[width=.16\linewidth]{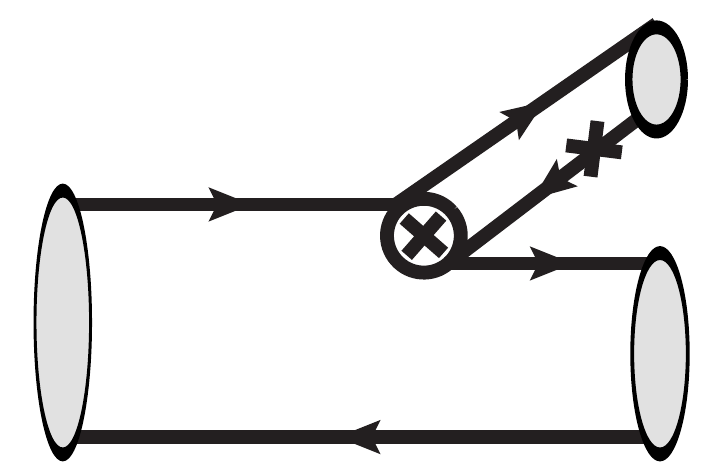}} \\\hline
$C^{(1)}_{3}$ &  \vcenteredhbox{\includegraphics[width=.19\linewidth]{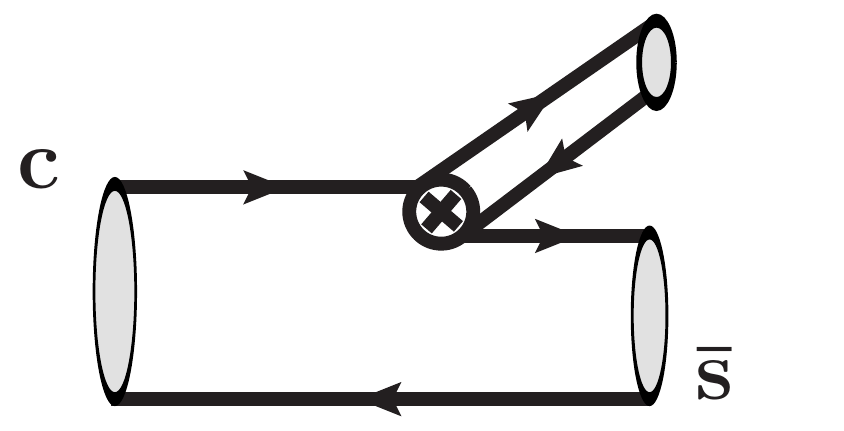}} $-$ 
		 \vcenteredhbox{\includegraphics[width=.19\linewidth]{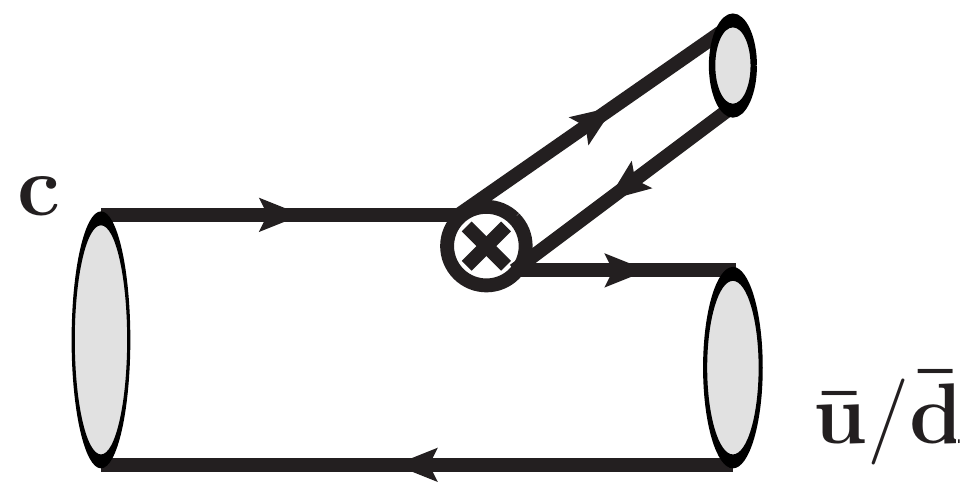}}  & 
		 \vcenteredhbox{\includegraphics[width=.16\linewidth]{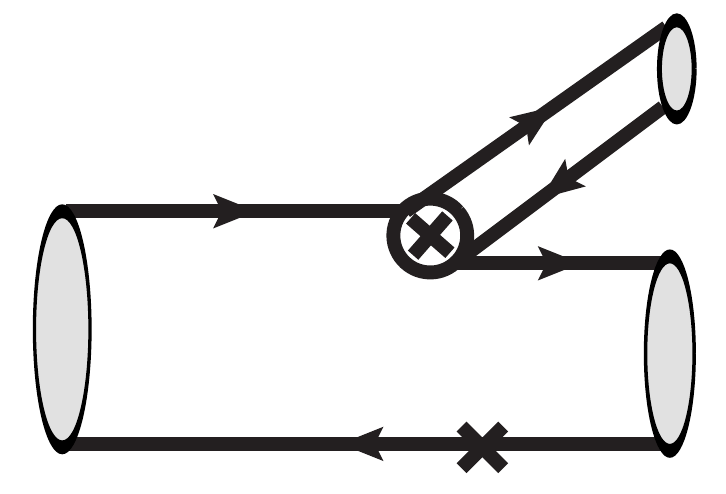}} \\\hline
$E^{(1)}_{1}$ &  \vcenteredhbox{\includegraphics[width=.19\linewidth]{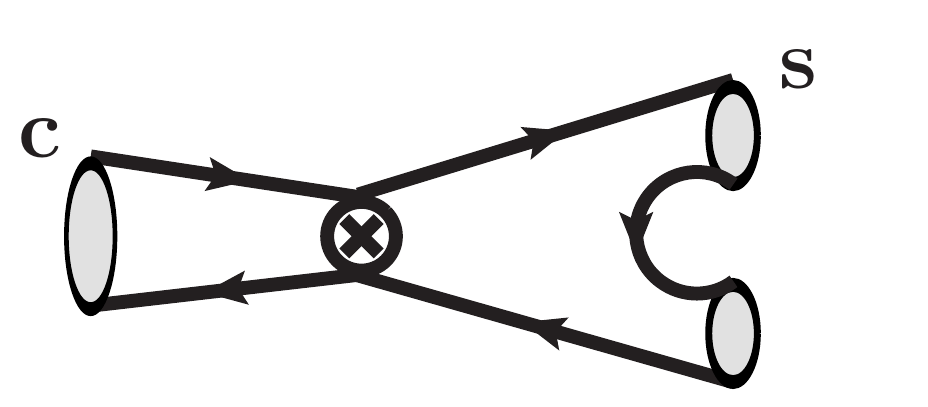}} $-$ 
		 \vcenteredhbox{\includegraphics[width=.19\linewidth]{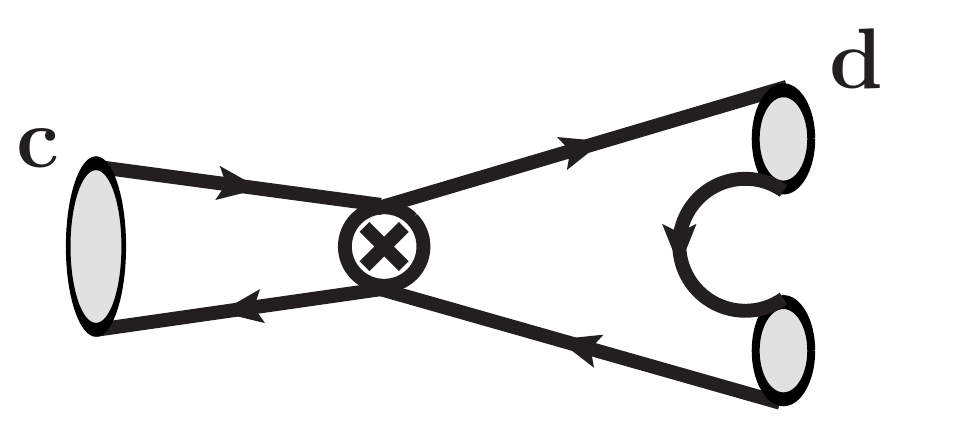}}  & 
		 \vcenteredhbox{\includegraphics[width=.16\linewidth]{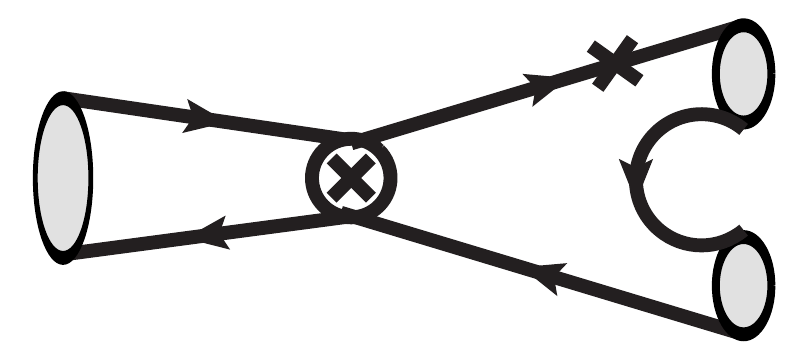}} \\\hline
$E^{(1)}_{2}$ &  \vcenteredhbox{\includegraphics[width=.19\linewidth]{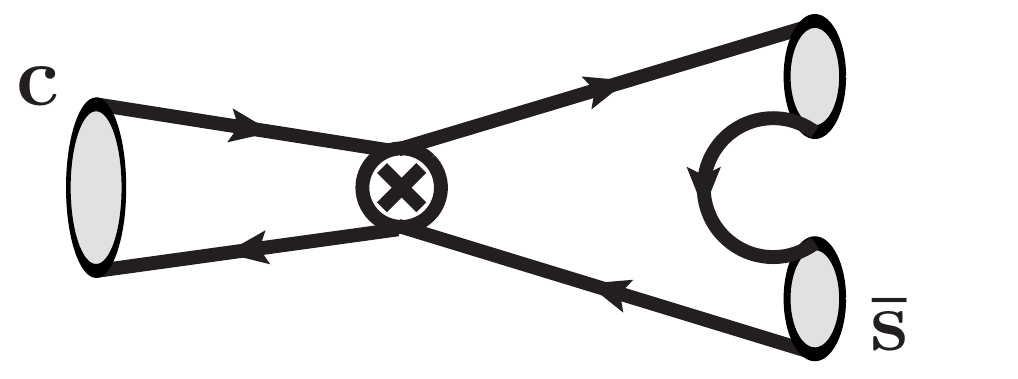}} $-$ 
		 \vcenteredhbox{\includegraphics[width=.19\linewidth]{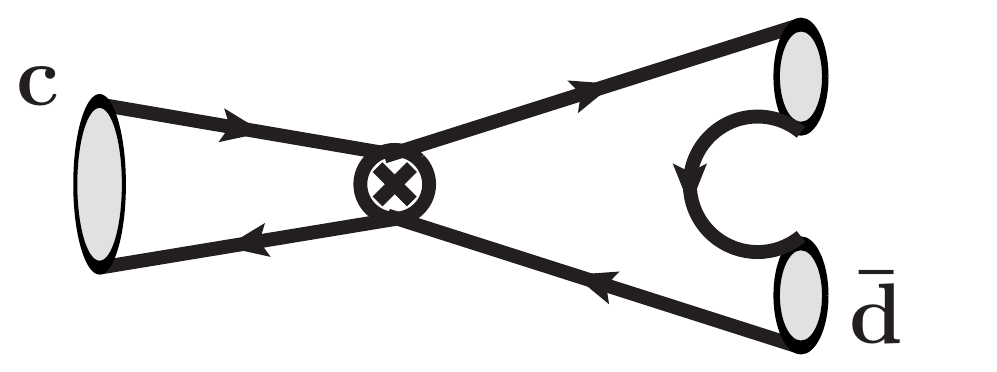}}  & 
		 \vcenteredhbox{\includegraphics[width=.16\linewidth]{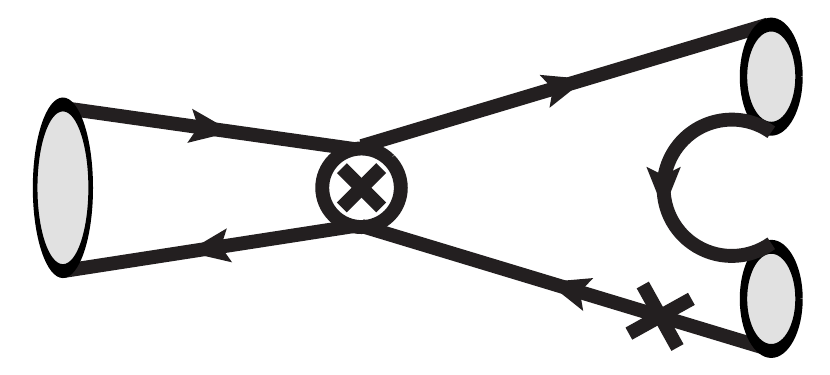}} \\\hline
$E^{(1)}_{3}$ &  \vcenteredhbox{\includegraphics[width=.19\linewidth]{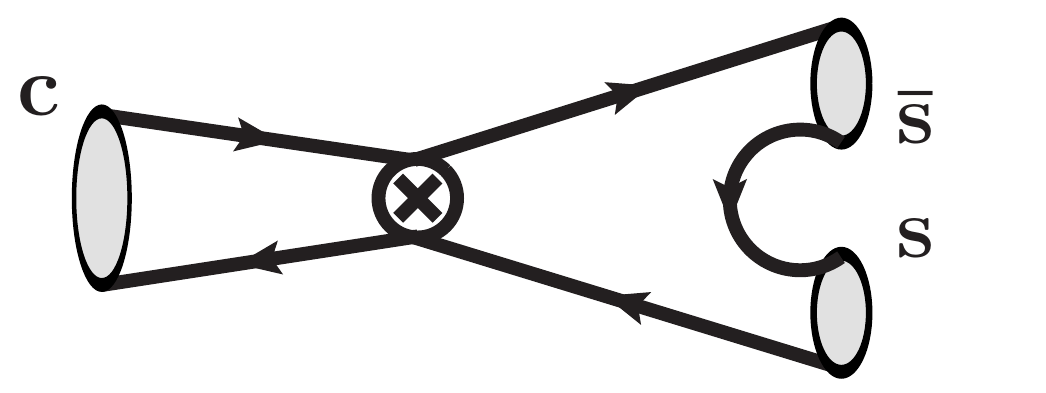}} $-$ 
		 \vcenteredhbox{\includegraphics[width=.19\linewidth]{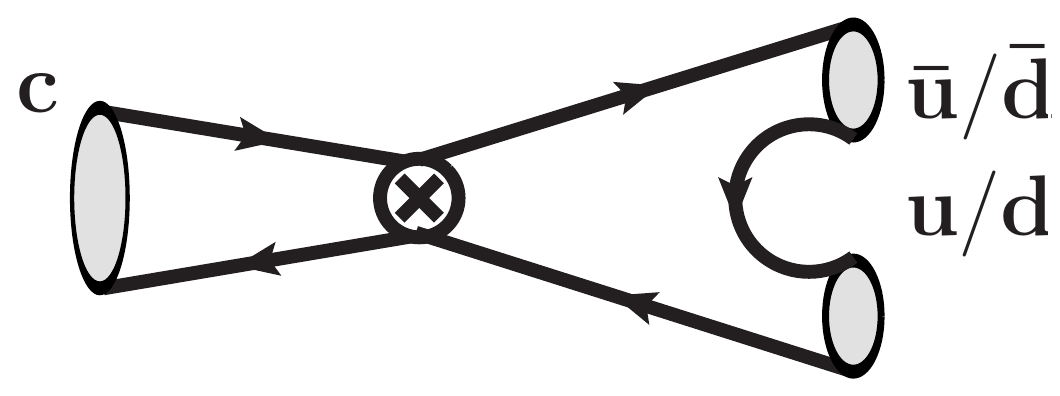}}  & 
		 \vcenteredhbox{\includegraphics[width=.16\linewidth]{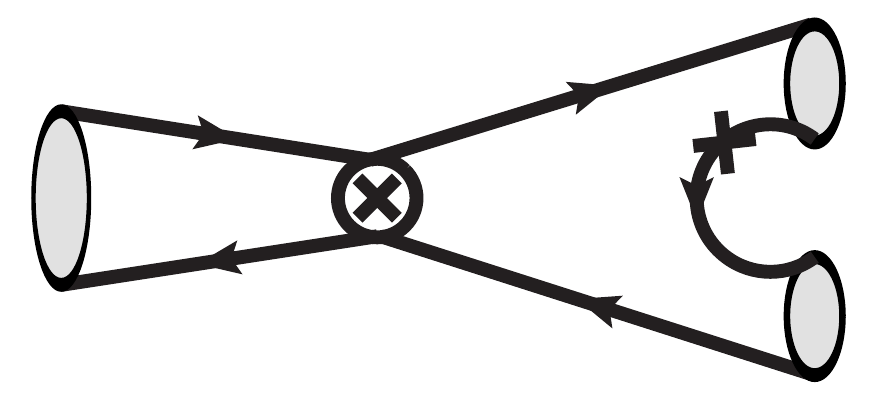}} \\\hline
$P_{\mathrm{break}}$ & \vcenteredhbox{\includegraphics[width=.17\linewidth]{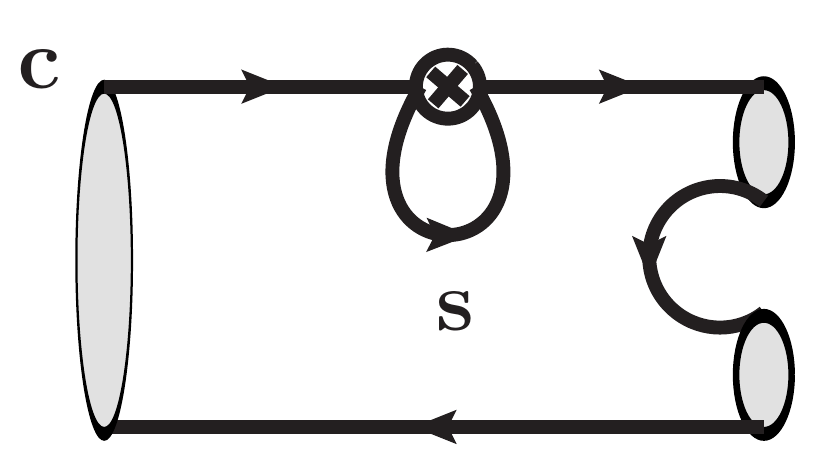}} $-$ 
		       \vcenteredhbox{\includegraphics[width=.17\linewidth]{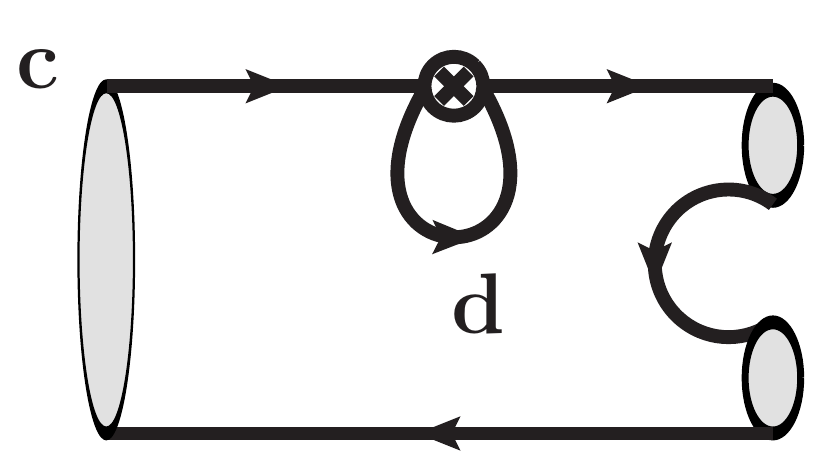}}  & 
		       \vcenteredhbox{\includegraphics[width=.15\linewidth]{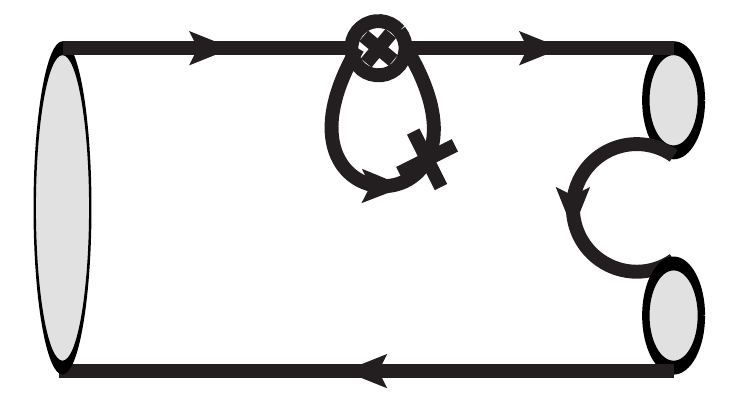}} \\\hline
\end{tabular}
\caption{SU(3)$_F$-breaking {topological amplitudes corresponding to 
the amplitudes} in Tab.~\ref{tab:topoparametrization}. 
{The Feynman rule for $H_{\cancel{\mathrm{SU(3)}}_F}$ is the cross placed
on an $s$ line. \label{tab:su3breaking-diagrams}}}
\end{center}
\end{table*}

\begin{table*}[tp!]
\begin{center}
\begin{tabular}{c|c|c|c|c|c|c|c|c|c|c|c|c|c|c|c|c|c}
\hline \hline
Decay ampl. {$\mathcal{A}(d)$} & $T$ & $T_1^{(1)}$ & $T_2^{(1)}$ & $T_3^{(1)}$  & $A$ & $A_1^{(1)}$ & $A_2^{(1)}$ & $A_3^{(1)}$ & $C$ & $C_1^{(1)}$ & $C_2^{(1)}$ & $C_3^{(1)}$ & $E$  & $E_1^{(1)}$ & $E_2^{(1)}$ & $E_3^{(1)}$ & $P_{\mathrm{break}}$  \\\hline\hline
\multicolumn{18}{c}{SCS} \\\hline\hline
$\mathcal{A}(D^0\rightarrow K^+ K^-)$  	&	1 & 1 & 1 & 0 & 0 & 0 & 0 & 0 & 0 & 0 & 0 & 0 & 1 & 1 & 1 & 0 & 1 \\\hline
$\mathcal{A}( D^0\rightarrow \pi^+ \pi^- )$ 	&	$-1$ & 0 & 0 & 0 & 0 & 0 & 0 & 0 & 0 & 0 & 0 & 0 & $-1$ & 0 & 0 & 0 & 1 \\\hline
$\mathcal{A}( D^0\rightarrow \bar{K}^0 K^0)$ 	&	0 & 0 & 0 & 0 & 0 & 0 & 0 & 0 & 0 & 0 & 0 & 0 & 0 & $-1$ & $-1$ & $1$ & 0 \\\hline
$\mathcal{A}( D^0 \rightarrow \pi^0 \pi^0)$ 	&	0 & 0 & 0 & 0 & 0 & 0 & 0 & 0 & $-\frac{1}{\sqrt{2}}$ & 0 & 0 & 0 & $\frac{1}{\sqrt{2}}$ & 0 & 0 & 0 & $-\frac{1}{\sqrt{2}}$ \\\hline
$\mathcal{A}( D^+ \rightarrow \pi^0 \pi^+)$  	&	$-\frac{1}{\sqrt{2}}$ & 0 & 0 & 0 & 0 & 0 & 0 & 0 & $-\frac{1}{\sqrt{2}}$ & 0 & 0 & 0 & 0 & 0 & 0 & 0 & 0 \\\hline
$\mathcal{A}( D^+ \rightarrow \bar{K}^0 K^+)$ &	1 & 1 & 1 & 0 & $-1$ & 0 & 0 & $-1$ & 0 & 0 & 0 & 0 & 0 & 0 & 0 & 0 & 1 \\\hline
$\mathcal{A}( D_s^+ \rightarrow  K^0 \pi^+)$ 	&	$-1$ & 0 & 0 & $-1$ & 1 & 1 & 1 & 0 & 0 & 0 & 0 & 0 & 0 & 0 & 0 & 0 & 1 \\\hline
$\mathcal{A}( D_s^+ \rightarrow  K^+ \pi^0)$ 	&	0 & 0 & 0 & 0 & $-\frac{1}{\sqrt{2}}$ & $-\frac{1}{\sqrt{2}}$ & $-\frac{1}{\sqrt{2}}$ & 0 & $-\frac{1}{\sqrt{2}}$ & 0 & 0 & $-\frac{1}{\sqrt{2}}$ & 0 & 0 & 0 & 0 & $-\frac{1}{\sqrt{2}}$ \\\hline\hline
\multicolumn{18}{c}{CF} \\\hline\hline
$\mathcal{A}( D^0\rightarrow K^- \pi^+)$   	&	1 & 1 & 0 & 0 & 0 & 0 & 0 & 0 & 0 & 0 & 0 & 0 & 1 & 1 & 0 & 0 & 0 \\\hline
$\mathcal{A}( D^0\rightarrow \bar{K}^0 \pi^0)$ & 	0 & 0 & 0 & 0 & 0 & 0 & 0 & 0 & $\frac{1}{\sqrt{2}}$ & $\frac{1}{\sqrt{2}}$ & 0 & 0 & $-\frac{1}{\sqrt{2}}$ & $-\frac{1}{\sqrt{2}}$ & 0 & 0 & 0 \\\hline
$\mathcal{A}( D^+ \rightarrow \bar{K}^0 \pi^+)$ &	1 & 1 & 0 & 0 & 0 & 0 & 0 & 0 & 1 & 1 & 0 & 0 & 0 & 0 & 0 & 0 & 0 \\\hline
$\mathcal{A}( D_s^+ \rightarrow \bar{K}^0 K^+)$ &	0 & 0 & 0 & 0 & 1 & 1 & 0 & 1 & 1 & 1 & 0 & 1 & 0 & 0 & 0 & 0 & 0 \\\hline\hline
\multicolumn{18}{c}{DCS} \\\hline\hline
$\mathcal{A}( D^0 \rightarrow K^+ \pi^-)$ 	&	1 & 0 & 1 & 0 & 0 & 0 & 0 & 0 & 0 & 0 & 0 & 0 & 1 & 0 & 1 & 0 & 0 \\\hline
$\mathcal{A}( D^0 \rightarrow K^0 \pi^0)$ 	&	0 & 0 & 0 & 0 & 0 & 0 & 0 & 0 & $\frac{1}{\sqrt{2}}$ & 0 & $\frac{1}{\sqrt{2}}$ & 0 & $-\frac{1}{\sqrt{2}}$ & 0  & $-\frac{1}{\sqrt{2}}$ & 0 & 0 \\\hline
$\mathcal{A}( D^+ \rightarrow K^0\pi^+ )$ 	&	0 & 0 & 0 & 0 & 1 & 0 & 1 & 0 & 1 & 0 & 1 & 0 & 0 & 0 & 0 & 0 & 0 \\\hline
$\mathcal{A}( D^+ \rightarrow K^+ \pi^0)$ 	&	$\frac{1}{\sqrt{2}}$ & 0 & $\frac{1}{\sqrt{2}}$ & 0 & $-\frac{1}{\sqrt{2}}$ & 0 & $-\frac{1}{\sqrt{2}}$ & 0 & 0 & 0 & 0 & 0 & 0 & 0 & 0 & 0 & 0 \\\hline
$\mathcal{A}( D_s^+ \rightarrow K^0 K^+ )$ 	&	1 & 0 & 1 & 1 & 0 & 0 & 0 & 0 & 1 & 0 & 1 & 1 & 0 & 0 & 0 & 0 & 0 \\\hline\hline
\end{tabular}
\caption{The coefficients of the decomposition 
of the {physical amplitudes (including SU(3)$_F$ breaking) in 
terms of the topological amplitudes as in   
\eqsto{eq:amps-cf}{eq:amps-dcs}. The table entries are the elements of 
 the coefficient matrix $M$ in \eq{eq:mpa}.  
\label{tab:topoparametrization}}}
\end{center}
\end{table*}

\subsection{SU(3)$_F$-breaking}\label{sec:topobreak}
{Any perturbative treatment starts with a subdivision of the
  {Hamiltonian} $H=H_0+H_1$ into a piece $H_0$ treated without
  approximation and the perturbation $H_1$. The S-matrix element 
  of the transition $i\to f$ triggered by $H_1$ is}
\begin{align}
 &\bra{f} \mathcal{T} e^{-i \int d^4 x H_1(x)} \ket{i}\, .
 \label{eq:fti}
\end{align}
In our case $H_0$ is the QCD {Hamiltonian} with $m_u$ and $m_s$ 
set equal to $m_d$. 
$H_1$ consists of the weak $|\Delta C|=1$ {Hamiltonian} 
$H_W$ and the SU(3)$_F$-breaking {Hamiltonian}
\begin{align}
H_{\cancel{\mathrm{SU(3)}}_F} &= (m_s-m_d) \ov s s, \label{eq:hsd}
\end{align}
where isospin breaking is neglected. {With our choice of $H_0$} 
the asymptotic states $i,f$ are eigenstates of $H_0$ which are $D^+$ or
$D^0$ mesons or two-pion states. To first order in $H_W$ and zeroth and
first order in $H_{\cancel{\mathrm{SU(3)}}_F}$ the transition amplitude in \eq{eq:fti}
becomes 

\begin{align}
  &\bra{f} -i \int d^4 x H_W(x) \ket{i} + \nn\\ 
&\qquad
 \bra{f} -\frac{1}{2} \int\!\!\!\!\int d^4 x d^4y \mathcal{T} H_W (x)
  H_{\cancel{\mathrm{SU(3)}}_F}(y) \ket{i}\,. \label{eq:feynmanrule}  
\end{align}
The second piece accounts for the differences of amplitudes involving a
$D_s^+$ in the initial state or one or two kaons in the final state from
their unflavored counterparts. The Feynman rule of
$H_{\cancel{\mathrm{SU(3)}}_F}$ is an $\ov s s $ vertex which we denote
by a cross on the $s$-quark line.  This approach is essentially
identical to the one of Ref.~\cite{Gronau:1995hm}, {where $B$ decays have
been considered.}
{$H_{\cancel{\mathrm{SU(3)}}_F}$ also leads to $\eta$--$\eta^\prime$
  mixing. Using an $\eta$--$\eta^\prime$ mixing angle in our diagrammatic
  method may lead to a double-counting of SU(3)$_F$-breaking effects and
  we do not consider final states with $\eta^{(\prime)}$'s in the final
  state in this paper.}  The corresponding topological amplitudes are
collected in Tab.~\ref{tab:su3breaking-diagrams}. {We {combine} our
  topological amplitudes {into} a vector
\begin{align}
\mathbf{p} &\equiv \left( T  , 
			  T_1^{(1)} , 
			  T_2^{(1)} , 
			  T_3^{(1)}, 
			  A, 
			  A_1^{(1)}, 
			  A_2^{(1)}, 
			  A_3^{(1)},\right.\nn\\
		  &\left. C , 
			  C_1^{(1)} , 
			  C_2^{(1)} , 
			  C_3^{(1)} , 
			  E  , 
			  E_1^{(1)}, 
			  E_2^{(1)}, 
			  E_3^{(1)}, 
			  P_{\mathrm{break}} \right)^T\, . 
  \label{eq:parametervector}
\end{align}
Then we can write 
\begin{align}
 {M \mathbf{p}} &= {\mathbfcal{A}} \label{eq:mpa} 
\end{align}
with a $17\times 17$ coefficient matrix $M$ and $\mathbfcal{A}= \left(
  \mathcal{A}(D^0\to K^+K^-), \ldots, \mathcal{A}(D_s^+\to K^0
  K^+)\right)^T$ subsuming the decay amplitudes. The $i$-th column of
$M$ contains the coefficients $c_i^d$ of \eqsto{eq:amps-cf}{eq:amps-dcs}.
Tab.~\ref{tab:topoparametrization} shows $\mathbfcal{A}$ in the first
column and lists the elements of $M$ as table entries.}  We remark that
the only final state with two identical mesons is $\ket{\pi^0\pi^0}$. In
$D^0(p_D)\to\pi^0(p_1)\pi^0(p_2)$ two effects must be taken into
account: first, each topological amplitude appears twice (with $p_1$ and
$p_2$ interchanged, leading to a proper Bose-symmetrized state).
Second, in the subsequent phase space integration one integrates the
azimuthal angle over the interval $[0,\pi]$ rather the usual $[0,2\pi]$,
because the two pions are indistinguishable. The resulting factor of 1/2
in the decay rate (compared to the other listed decay rates) is
accommodated through a factor of $1/\sqrt{2}$ on the amplitude level in
Tab.~\ref{tab:topoparametrization}. For example, the factor of $1/\sqrt{2}$
multiplying $E$ is the result of the mentioned factors of 2 and
$1/\sqrt{2}$ and two factors of $1/\sqrt{2}$ stemming from the
$\ket{\pi^0}$ state in \eq{eq:states-3}.  {Note that it would be unwise
  to define the SU(3)$_F$ limit from some average of $s$ and $d$
  diagrams, since {with this choice} the asymptotic states
  constructed from $H_0$ would not correspond to physical
  mesons. Furthermore, there would be far fewer zeros among the
  coefficients in Tab.~\ref{tab:topoparametrization} which would further
  complicate the analysis.}

{There is one more SU(3)$_F$-breaking topological amplitude, the
  penguin annihilation amplitude $PA_{\mathrm{break}} \equiv PA_s -
  PA_d$ depicted in Fig.~\ref{fig:su3limit-penguin-annihilation}.
  While the dynamics described by this amplitude is different from the 
  ones discussed so far,  $PA_{\mathrm{break}}$ enters the decay
  amplitudes in such a way that it can be absorbed into other amplitudes. 
  Thus it is a redundant fit parameter, as explained in the following 
section.}  

\subsection{Redundancies\label{sec:redundancies}}
{The relationship between physical and topological amplitudes is not
  one-to-one. If no other dynamical information on the latter is used,
  the determination of $\mathbf{p}$ from $\mathbfcal{A}$ in \eq{eq:mpa}
  yields an infinite set of solutions describing the data equally
  well. A priori this feature renders fitted numerical values of
  $T,\ldots, P_{\mathrm{break}}$ meaningless and {obscures} the comparison
  of different analyses in the literature. There are two ways to address
  this problem: one can simply remove redundant parameters and
  quote numbers for the linear combinations of the topological
  amplitudes which are in one-to-one correspondence with the physical
  ones. Or one can use further theoretical (and experimental) input to
  constrain the topological amplitudes.   
  We determine redundancies among  
  $T,\ldots, P_{\mathrm{break}}$ in this section and relegate the second
  approach to Sec.~\ref{sec:theoryinput}.}
\begin{figure}[t]
\begin{center}
        \includegraphics[width=0.31\textwidth]{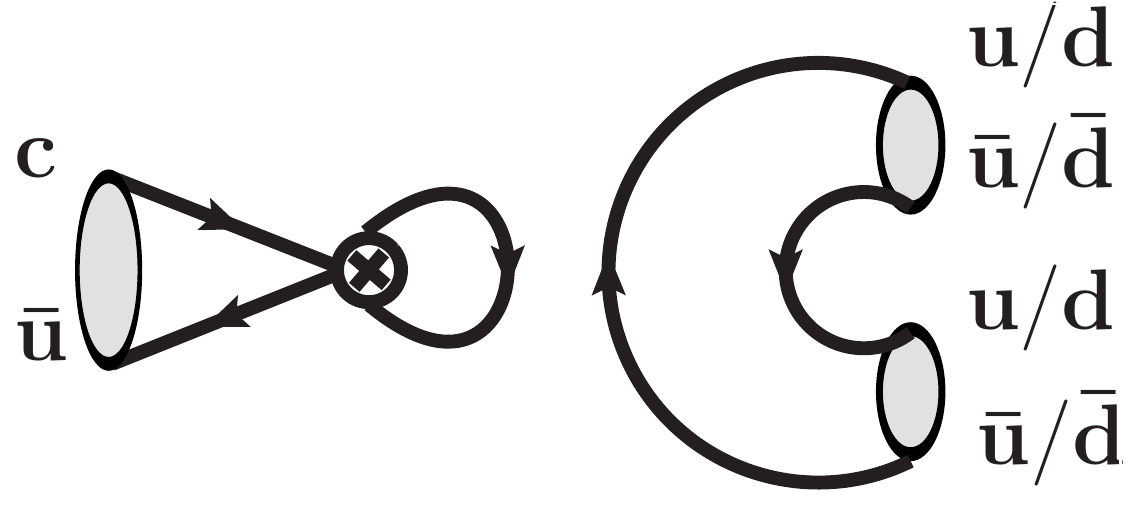}
\end{center}
\caption{Penguin annihilation diagram.
\label{fig:su3limit-penguin-annihilation}
}
\end{figure}

\begin{table}[t]
\begin{center}
\begin{tabular}{c|c}
\hline \hline
Decay $d$ & $PA_{\mathrm{break}}$  \\\hline\hline
\multicolumn{2}{c}{SCS} \\\hline\hline
$D^0\rightarrow K^+ K^-$  	& 1	 \\\hline
$D^0\rightarrow \pi^+ \pi^- $ 	& 1	 \\\hline
$D^0\rightarrow \bar{K}^0 K^0$ 	& $-1$	 \\\hline
$D^0 \rightarrow \pi^0 \pi^0$ 	& $-\frac{1}{\sqrt{2}}$ \\\hline
$D^+ \rightarrow \pi^0 \pi^+$  	& 0	 \\\hline
$D^+ \rightarrow \bar{K}^0 K^+$ & 0	 \\\hline
$D_s^+ \rightarrow  K^0 \pi^+$ 	& 0	 \\\hline
$D_s^+ \rightarrow  K^+ \pi^0$ 	& 0	 \\\hline\hline
\end{tabular}
\caption{The coefficients of the topological amplitude
  $PA_{\mathrm{break}}$ which is absorbed into $E$, $E_{1,2,3}^{(1)}$ in 
 Tab.~\ref{tab:topoparametrization} {as explained in 
Sec.~\ref{sec:redundancies}}. 
\label{tab:topopenguinanni}}
\end{center}
\end{table}

{The first redundancy is related to $PA_{\mathrm{break}}$ of
  Fig.~\ref{fig:su3limit-penguin-annihilation},} which appears in SCS
decays with the coefficients in Tab.~\ref{tab:topopenguinanni}. The
{listed} column of coefficients is linearly dependent on the
four columns of coefficients of $E$, $E_{1,2,3}^{(1)}$ {in
  Tab.~\ref{tab:topoparametrization}}.  I.e.\ we can absorb
$PA_{\mathrm{break}}$ into the exchange amplitudes by redefining $E =
\widehat{E} - PA_{\mathrm{break}}$, $E_{1,2,3}^{(1)} =
\widehat{E}_{1,2,3}^{(1)} + PA_{\mathrm{break}}$.  In
Tab.~\ref{tab:topoparametrization} this redefinition is implicitly
already performed, so that $PA_{\mathrm{break}}$ is not shown there
anymore. {The physical meaning of $E$, $E_{i}^{(1)}$ changes
  accordingly, to be read as $\widehat{E}$, $\widehat{E}_{i}^{(1)}$ with
  the penguin annihilation mechanism included. However,
  $|PA_{\mathrm{break}}|$ is expected to be negligibly small: The
  corresponding Wilson coefficient in $H_W$ is small and the momentum
  flowing through the penguin loop is large (of order of the $D^0$ mass)
  so that the Glashow-Iliopoulos-Maiani (GIM) \cite{Glashow:1970gm} suppression will be
  effective.}

{Further redundancies are related to the fact that our coefficient
  matrix $M$ in \eq{eq:mpa} does not have maximal rank. Considering
  first the} SU(3)$_F$ limit {ignoring
  $T_i^{(1)},A_i^{(1)},C_i^{(1)}$, and $E_i^{(1)}$ one observes that the
  remaining matrix} in Tab.~\ref{tab:topoparametrization} {linking
  $T$, $C$, $A$ and $E$ to the physical amplitudes has only} rank
three. I.e.\ one of {$T$, $C$, $A$ and $E$ is redundant}.

Redundancies of the diagrammatic approach in the SU(3)$_F$ limit are
also discussed in {Ref.}~\cite{Wang:1980ac}, comparing to the
SU(3)$_F$ parametrization in {Ref.}~\cite{Wang:1979dx}. The
corresponding matching for $B$ decays is done in
{Ref.}~\cite{Zeppenfeld:1980ex}.  Note that the redundancies
{change} when taking $\eta^{(\prime)}$ final states into account
\cite{Bhattacharya:2009ps, Hiller:2012xm}, {leading to} more
parameters but also additional sum rules \cite{Grossman:2012ry}.

 Including SU(3)$_F$ breaking, {the 17$\times$17 matrix
  $M$} in Tab.~\ref{tab:topoparametrization} has rank 11.  Consequently
{$\mathbf{p}$ in \eq{eq:parametervector}} contains six redundant
complex parameters. 
The remaining parametric redundancy contained in
Tab.~\ref{tab:topoparametrization} can be systematically found and
removed as follows. It is encoded in the six-dimensional kernel of the
coefficient matrix. The 17-dimensional basis vectors of the kernel are
given in {columns 2 to 7 of} Tab.~\ref{tab:kernel}.  
\begin{table}[t]
\begin{center}
\begin{tabular}{c|cccccc||c|c|c}
\hline\hline
$\mathbf{p}$ & $\mathbf{n}_1$ & $\mathbf{n}_2$ & $\mathbf{n}_3$ & $\mathbf{n}_4$ & $\mathbf{n}_5$ & $\mathbf{n}_6$ & 
$\mathbf{p} - P_{\mathrm{brk}} \mathbf{n}_1$ & $\mathbf{p} - A \mathbf{n}_4$  & $\mathbf{p} - A_3^{(1)} \mathbf{n}_5$ 	 \\\hline 
 $T$                  &  1 & 1 & 0 & 1 & 0 & 0       & $T - P_{\mathrm{brk}}$            &  $T-A$                  &  $T$                           \\   
 $T_1^{(1)}$          &  -1 & -1 & 1 & 0 & 0 & 0     & $T_1^{(1)} + P_{\mathrm{brk}}$    &  $T_1^{(1)}$            &  $T_1^{(1)}$                   \\  
 $T_2^{(1)}$          &  -1 & 0 & -1 & 0 & 0 & 0     & $T_2^{(1)} + P_{\mathrm{brk}}$    &  $T_2^{(1)}$            &  $T_2^{(1)}$                   \\  
 $T_3^{(1)}$          &  0 & 0 & 0 & 0 & 0 & -1      & $T_3^{(1)}$                         &  $T_3^{(1)}$            &  $T_3^{(1)}$                   \\   
 $A$                  &  0 & 0 & 0 & 1 & -1 & 0      & $A$                                 &  $0$                    &  $A + A_3^{(1)}$                           \\ 
 $A_1^{(1)}$          &  0 & 0 & 1 & 0 & 0 & -1      & $A_1^{(1)}$                         &  $A_1^{(1)}$            &  $A_1^{(1)}$                   \\   
 $A_2^{(1)}$          &  0 & 1 & -1 & 0 & 1 & 0      & $A_2^{(1)}$                         &  $A_2^{(1)}$            &  $A_2^{(1)} - A_3^{(1)}$                   \\   
 $A_3^{(1)}$          &  0 & 0 & 0 & 0 & 1 & 0       & $A_3^{(1)}$                         &  $A_3^{(1)}$            &  $0$                   \\    
 $C$         	      &  -1 & -1 & 0 & -1 & 0 & 0    & $C +  P_{\mathrm{brk}}$           &  $C+A$         	     &  $C$         	               \\   
 $C_1^{(1)}$ 	      &  1 & 1 & -1 & 0 & 0 & 0      & $C_1^{(1)} - P_{\mathrm{brk}}$    &  $C_1^{(1)}$ 	     &  $C_1^{(1)}$ 	               \\   
 $C_2^{(1)}$ 	      &  1 & 0 & 1 & 0 & 0 & 0       & $C_2^{(1)} - P_{\mathrm{brk}}$    &  $C_2^{(1)}$ 	     &  $C_2^{(1)}$ 	               \\    
 $C_3^{(1)}$ 	      &  0 & 0 & 0 & 0 & 0 & 1       & $C_3^{(1)}$                         &  $C_3^{(1)}$ 	     &  $C_3^{(1)}$ 	               \\    
 $E$         	      &  0 & -1 & 0 & -1 & 0 & 0     & $E$                                 &  $E+A$         	     &  $E$         	               \\    
 $E_1^{(1)}$ 	      &  0 & 1 & -1 & 0 & 0 & 0      & $E_1^{(1)}$                         &  $E_1^{(1)}$ 	     &  $E_1^{(1)}$ 	               \\   
 $E_2^{(1)}$ 	      &  0 & 0 & 1 & 0 & 0 & 0       & $E_2^{(1)}$                         &  $E_2^{(1)}$ 	     &  $E_2^{(1)}$ 	               \\    
 $E_3^{(1)}$ 	      &  0 & 1 & 0 & 0 & 0 & 0       & $E_3^{(1)}$                         &  $E_3^{(1)}$ 	     &  $E_3^{(1)}$ 	               \\    
 $P_{\mathrm{brk}}$ &  1 & 0 & 0 & 0 & 0 & 0       & $0$                   		   &  $P_{\mathrm{brk}}$   &  $P_{\mathrm{brk}}$ 	 \\\hline\hline
\end{tabular}
\caption{The parameter vector $\mathbf{p}$ as defined in 
\eq{eq:parametervector}, 
vectors $\mathbf{n}_i$ {spanning} the kernel of the 
coefficient matrix {$M$} 
in Tab.~\ref{tab:topoparametrization}, and several redefined
parameter vectors, see {\eqsto{eq:redef}{eq:os}}.
\label{tab:kernel}}
\end{center}
\end{table}
If we {redefine $\mathbf{p}$ in \eq{eq:parametervector} as}
\begin{align}
  \mathbf{p}^{\mathrm{new}} &\equiv \mathbf{p} + \sum_i c_i
  \mathbf{n}_i\,,\quad c_i \in \mathbb{C}\,,
\label{eq:redef}
\end{align}
{this will} not change {$M\mathbf{p}$ in \eq{eq:mpa}}, i.e.\ the
$\mathbf{n}_i$ {define} the \lq\lq{}flat directions\rq\rq{} in
parameter space which correspond to the same {$\mathbfcal{A}$}.  One
  can remove this redundancy by redefining the topological
  {amplitudes and choosing 11 of them as new independent
    parameters}. For example we can set
\begin{align}
\mathbf{\tilde{p}}^{\mathrm{new}} &\equiv 
   \mathbf{p} - P_{\mathrm{break}} \mathbf{n}_1 \label{eq:removePbreak}
\end{align}
which gives the result {in the first column after the double line in} 
Tab.~\ref{tab:kernel}.  Subsequently,
we can redefine the parameters in order to eliminate
$P_{\mathrm{break}}$.  In order to remove all redundancies in one step
one can {choose}
\begin{align}
\mathbf{\hat{p}}^{\mathrm{new}} &\equiv \mathbf{p} - P_{\mathrm{break}}\mathbf{n}_1 
				        - E_3^{(1)} \mathbf{n}_2
					- E_2^{(1)} \mathbf{n}_3
					-\nn\\&\qquad A \mathbf{n}_4
					- A_3^{(1)} ( \mathbf{n}_4 + \mathbf{n}_5 )
					- C_3^{(1)} \mathbf{n}_6\,
\label{eq:os}
\end{align}
and then perform redefinitions of the other parameters in order to remove 
\begin{align}
 P_{\mathrm{break}}\,,
 E_3^{(1)} \,,
 E_2^{(1)} \,,
 A \,,
 A_3^{(1)} \,,
 C_3^{(1)} \, \label{eq:rem}
\end{align}
from the parameterization.  Note the special form of $\mathbf{n}_4$ which
{encodes} the redundancy {present in the} SU(3)$_F$ limit.
{$\mathbf{n}_4$ forces us to eliminate one of $T$, $C$, $E$, or $A$,
  while the other five eliminations involve  SU(3)$_F$-breaking amplitudes
 (e.g.\ those in \eq{eq:rem}).}  The elimination of $A$ only is also shown in
Tab.~\ref{tab:kernel}.  Additionally, from $\mathbf{n}_5$ we see that
the coefficient vector of $A_3^{(1)}$ is linear{ly} dependent on the
other annihilation coefficient vectors. Consequently, $A_3^{(1)}$ can be
absorbed by redefining annihilation amplitudes only, as shown also
explicitly in Tab.~\ref{tab:kernel}.

Note further that the $\mathbf{n}_i$ are linearly independent also when
removing all but the first six elements.  This means it is not possible
to perform redefinitions without touching the tree or annihilation
diagrams.  Equivalently, the submatrix obtained by removing tree and
annihilation diagrams from Tab.~\ref{tab:topoparametrization} has rank
nine, which in this case {equals} the number of {remaining}
parameters,~i.e.\ {the lower {nine} components of $\mathbf{p}$.  This
  observation guides us to the approach of Sec.~\ref{sec:theoryinput}:
  calculating tree and annihilation amplitudes will also remove the
  redundancies}.

{After absorbing some topological amplitudes (e.g.\ those in
  \eq{eq:rem}) into others the new amplitudes have lost their original
  meaning in terms of QCD dynamics. An important question in charm
  physics is the level of GIM cancellation between an $s$ and $d$
  loop. In this paper we encounter $P_{\mathrm{break}}$ as a quantity
  probing the GIM mechanism. A naive quark-level calculation involves a
  suppression factor of $m_s^2/m_c^2$ and renders $P_{\mathrm{break}}$
  negligibly small. Thus any information on the actual size of
  $|P_{\mathrm{break}}|$ may give insight into a possible
  non-perturbative enhancement of GIM-suppressed amplitudes. However, 
  as shown above and exemplified in Tab.~\ref{tab:kernel} the fit to 
  topological amplitudes alone cannot give this information, because 
  $P_{\mathrm{break}}$ cannot be separated from the other parameters
  fitted from the data.}

As we have seen above, the calculation of the kernel gives a method to
remove redundant parameters.  In the same way, the cokernel of $M$ gives
us information on \lq\lq{}redundant\rq\rq{} amplitudes,~i.e.\ {six}
sum rules {fulfilled by the latter},
{all of which} were found in Ref.~\cite{Grossman:2012ry}.  {In
  other words: if one did a Gaussian elimination to determine
  $\mathbf{p}$ from \eq{eq:mpa}, one would end up with a 6$\times$17
  block of zeros in the transformed coefficient matrix $M$ and linear
  combinations of physical amplitudes in the corresponding six entries
  of $\mathbfcal{A}$. These linear combinations vanish by the SU(3)$_F$
  sum rules of Ref.~\cite{Grossman:2012ry}. Thus the discussed
  redundancies are not the consequence of missing experimental
  information but of the symmetry relations underlying these sum rules.}
{It is instructive to rederive these sum rules with our diagrammatic
  method, which is particularly straightforward and intuitive. We do
  this in} Appendix~\ref{sec:sumrules}.

{We checked that} after the removal of all redundancies, the
diagrammatic parameterization and the {common expansion in terms of
  SU(3)$_F$ representations} can be mapped onto each other, i.e.~one can
calculate one set of parameters when given the other one. The mapping
can be obtained explicitly by inverting either the {reduced}
coefficient matrix $M$ or its counterpart in the SU(3)$_F$ method.  Note
that in the SU(3)$_F$ parameterization unphysical degrees of freedom are
present in the very same way.  Analogously, it is possible to redefine
SU(3)$_F$ matrix elements in order to obtain a physical basis
\cite{Hiller:2012xm}.  In Appendix~\ref{sec:matching} we give the
inverse of the SU(3)$_F$ coefficient matrix of \cite{Hiller:2012xm} and
show the result of the extraction of the corresponding SU(3)$_F$ matrix
elements for an example fit point of our diagrammatic {analysis}.  {So
  far our discussion of redundancies has assumed that the amplitudes in
  $\mathbfcal{A}$ are known. In practice, there is no information on
  most of their complex phases (and not all of them are physical). This
  feature introduces additional flat directions in the space of our fit
  parameters and is equally present in the SU(3)$_F$ method.}

\begin{figure}[t]
\begin{center}
        \includegraphics[width=0.31\textwidth]{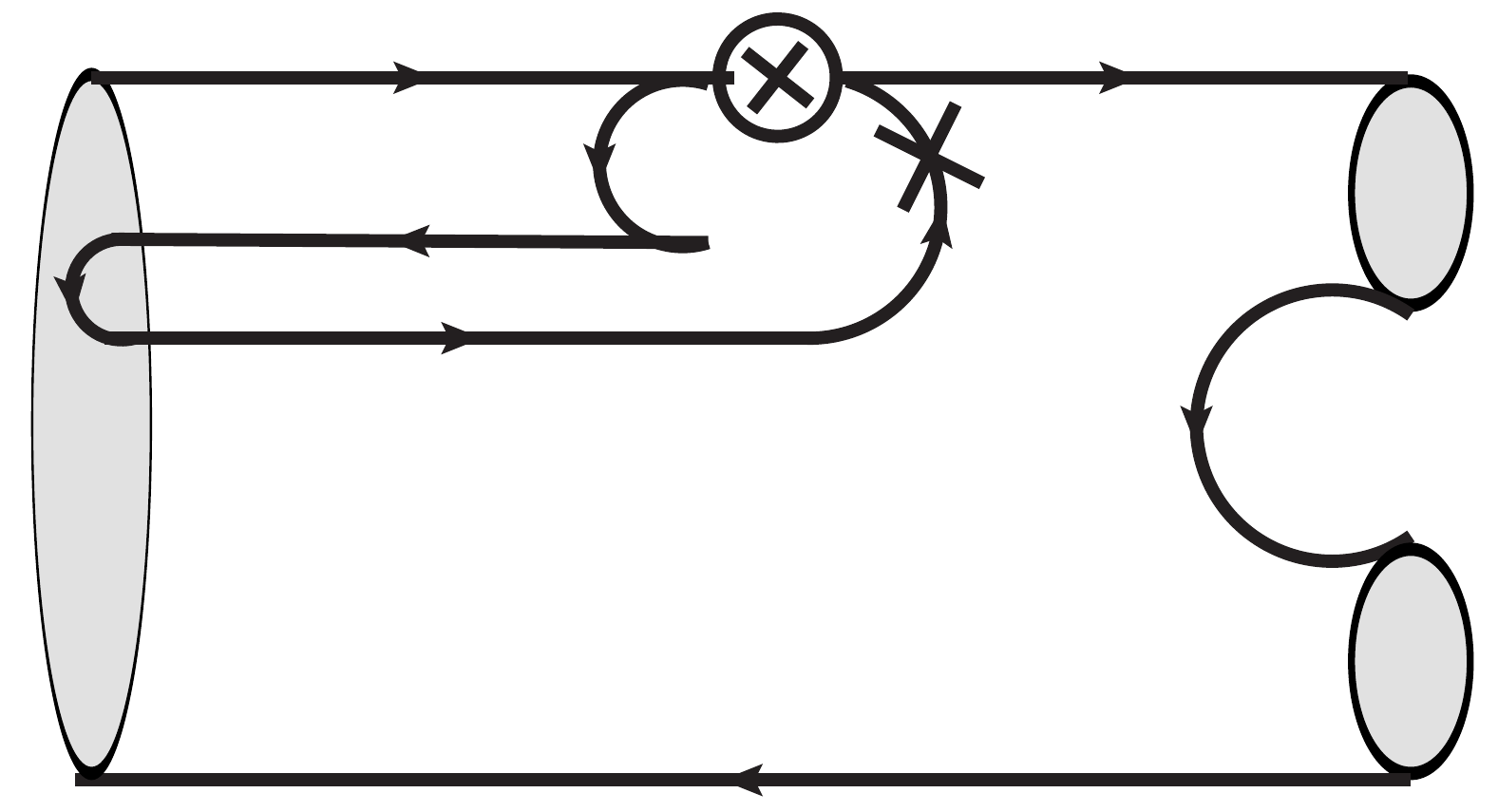}
\end{center}
\caption{Example for a SU(3)$_F$-breaking diagram involving sea quarks 
which can be absorbed into $P_{\mathrm{break}}$, see 
Tab.~\ref{tab:su3breaking-diagrams}.\label{fig:sea-su3break-penguin}
}
\end{figure}

{The discussion above has made clear that the topological-amplitude
  method is complete in the sense that it contains the full information
  contained in an SU(3)$_F$ analysis including SU(3)$_F$ breaking to
  linear order. It is also worthwhile to study this question from the
  viewpoint of QCD dynamics: are there any dynamical mechanisms which
  cannot be mapped onto topological amplitudes? As a first topic we
  discuss final-state rescattering, i.e.\ decays $D\to f^\prime \to f$
  passing through an on-shell intermediate state $f^\prime$. The flavor
  flow for such a rescattering process is always a deformation of a
  diagram in Tab.~\ref{tab:su3limit-diagrams} or
  Fig.~\ref{fig:su3limit-penguin-annihilation} and is therefore included
  in the corresponding topological amplitude. Rescattering effects
  cannot be isolated from the ``direct'' $D\to f$ decay, because the
  dispersive part of $\mathcal{A}(D\to f^\prime \to f)$ cannot be
  separated from that of $\mathcal{A}(D\to f)$ in a meaningful
  way. (Neglecting CP violation we can choose phase conventions such
  that the dispersive and absorptive parts of some amplitude equal its
  real and imaginary parts, respectively.) By the optical theorem the
  absorptive part of $\mathcal{A}(D\to f)$ can be related to
  $\mathcal{A}(D\to f^\prime)$ and the $f^\prime \to f$ scattering
  amplitude, with summation over all intermediate states
  $f^\prime$. This feature holds true for the topological amplitudes as
  well. The imaginary parts of the topological amplitudes found  in
  our fit in Sec.~\ref{sec:fits} are therefore a measure of the size
  of rescattering. The second topic of QCD dynamics addresses the proper
  description of meson states. The state of e.g.\ an energetic kaon can 
   be expanded as   
\begin{align}
  \ket{K^0} &= \ket{d\bar{s}} + \ket{d\bar{s}g} + \ket{d\bar{s}q\bar{q}}
  + \dots\,, \label{eq:fse}
\end{align}
where the notation implicitly contains convolution integrals over the
kaon momentum fraction carried by the indicated partons. Our graphical
description of the topological amplitudes only catches the first term
in \eq{eq:fse}.  The higher Fock states $ \ket{d\bar{s}g},
\ket{d\bar{s}u\bar{u}},\ldots $ are suppressed with powers of the kaon
energy, but in view of the small energy release in $D$ decays this
suppression is unlikely to be realized numerically. We may wonder
whether the contributions with additional $q\bar{q}$ pairs in
\eq{eq:fse} will require the introduction of further amplitude
topologies, with extra quark lines connected with ``sea'' quarks
in the mesons. An example is shown in
Fig.~\ref{fig:sea-su3break-penguin}. However, it is easy to see that
such diagrams are always obtained by forking a quark line of one of the
topological amplitudes considered so far. For instance, the diagram in 
Fig.~\ref{fig:sea-su3break-penguin} is contained in
$P_{\mathrm{break}}$.} 

\section{Theoretical Input on diagrammatic 
SU(3)$_F$ breaking\label{sec:theoryinput}} 

The great advantage of the flavor-flow parameterization over the plain
SU(3)$_F$ approach is the opportunity to {add dynamical input to
  constrain individual topologies.}  We use two different such inputs
which are presented below.

\subsection{$1/N_c$ counting\label{sec:countingtopologies}}

\begin{table*}[tp!]
\begin{center}
\begin{tabular}{c|c|c|c|c|c|c|c|c|c|c}
  \hline \hline
  Decay ampl. {$\widetilde{\mathcal{A}}(d)$} & $\delta_T-\delta_A$ &
  $\widetilde{C}\equiv C+\delta_A$ & $C_1^{(1)}$ & $C_2^{(1)}$ &
  $C_3^{(1)}$ & 
  $\widetilde{E}\equiv E+\delta_A$  & $E_1^{(1)}$ & $E_2^{(1)}$ & 
  $E_3^{(1)}$ & $P_{\mathrm{break}}$  \\\hline\hline
\multicolumn{11}{c}{SCS} \\\hline\hline
$\widetilde{\mathcal{A}}(D^0\rightarrow K^+ K^-)$  		&1 & 0 & 0 & 0 & 0 & 1 & 1 & 1 & 0 & 1 \\\hline
$\widetilde{\mathcal{A}}( D^0\rightarrow \pi^+ \pi^- )$ 	&$-1$ & 0 & 0 & 0 & 0 & $-1$ & 0 & 0 & 0 & 1 \\\hline
$\widetilde{\mathcal{A}}( D^0\rightarrow \bar{K}^0 K^0)$ 	&0 & 0 & 0 & 0 & 0 & 0 & $-1$ & $-1$ & $1$ & 0 \\\hline
$\widetilde{\mathcal{A}}( D^0 \rightarrow \pi^0 \pi^0)$ 	&0 & $-\frac{1}{\sqrt{2}}$ & 0 & 0 & 0 & $\frac{1}{\sqrt{2}}$ & 0 & 0 & 0 & $-\frac{1}{\sqrt{2}}$ \\\hline
$\widetilde{\mathcal{A}}( D^+ \rightarrow \pi^0 \pi^+)$  	&$-\frac{1}{\sqrt{2}}$  & $-\frac{1}{\sqrt{2}}$ & 0 & 0 & 0 & 0 & 0 & 0 & 0 & 0 \\\hline
$\widetilde{\mathcal{A}}( D^+ \rightarrow \bar{K}^0 K^+)$ &	1 & 0 & 0 & 0 & 0 & 0 & 0 & 0 & 0 & 1 \\\hline
$\widetilde{\mathcal{A}}( D_s^+ \rightarrow  K^0 \pi^+)$ 	&	$-1$ & 0 & 0 & 0 & 0 & 0 & 0 & 0 & 0 & 1 \\\hline
$\widetilde{\mathcal{A}}( D_s^+ \rightarrow  K^+ \pi^0)$ 	&	0 & $-\frac{1}{\sqrt{2}}$ & 0 & 0 & $-\frac{1}{\sqrt{2}}$ & 0 & 0 & 0 & 0 & $-\frac{1}{\sqrt{2}}$ \\\hline\hline
\multicolumn{11}{c}{CF} \\\hline\hline
$\widetilde{\mathcal{A}}( D^0\rightarrow K^- \pi^+)$   	&	1 & 0 & 0 & 0 & 0 & 1 & 1 & 0 & 0 & 0 \\\hline
$\widetilde{\mathcal{A}}( D^0\rightarrow \bar{K}^0 \pi^0)$ & 	0 & $\frac{1}{\sqrt{2}}$ & $\frac{1}{\sqrt{2}}$ & 0 & 0 & $-\frac{1}{\sqrt{2}}$ & $-\frac{1}{\sqrt{2}}$ & 0 & 0 & 0 \\\hline
$\widetilde{\mathcal{A}}( D^+ \rightarrow \bar{K}^0 \pi^+)$ &	1 & 1 & 1 & 0 & 0 & 0 & 0 & 0 & 0 & 0 \\\hline
$\widetilde{\mathcal{A}}( D_s^+ \rightarrow \bar{K}^0 K^+)$ &	0 & 1 & 1 & 0 & 1 & 0 & 0 & 0 & 0 & 0 \\\hline\hline
\multicolumn{11}{c}{DCS} \\\hline\hline
$\widetilde{\mathcal{A}}( D^0 \rightarrow K^+ \pi^-)$ 	&	1 & 0 & 0 & 0 & 0 & 1 & 0 & 1 & 0 & 0 \\\hline
$\widetilde{\mathcal{A}}( D^0 \rightarrow K^0 \pi^0)$ 	&	0 & $\frac{1}{\sqrt{2}}$ & 0 & $\frac{1}{\sqrt{2}}$ & 0 & $-\frac{1}{\sqrt{2}}$ & 0  & $-\frac{1}{\sqrt{2}}$ & 0 & 0 \\\hline
$\widetilde{\mathcal{A}}( D^+ \rightarrow K^0\pi^+ )$ 	&	0 & 1 & 0 & 1 & 0 & 0 & 0 & 0 & 0 & 0 \\\hline
$\widetilde{\mathcal{A}}( D^+ \rightarrow K^+ \pi^0)$ 	&	$\frac{1}{\sqrt{2}}$ &  0 & 0 & 0 & 0 & 0 & 0 & 0 & 0 & 0 \\\hline
$\widetilde{\mathcal{A}}( D_s^+ \rightarrow K^0 K^+ )$ 	&	1 & 1 & 0 & 1 & 1 & 0 & 0 & 0 & 0 & 0 \\\hline\hline
\end{tabular}
\caption{{Coefficients of the parameters $(\delta_T-\delta_A,\ldots,
    P_{\mathrm{break}})$ for the amplitudes
    $\widetilde{\mathcal{A}}(d)$, which are obtained from
    $\mathcal{A}(d)$ by subtracting the factorized part, see
    \eq{eq:defat}.  The table entries are the elements of the
    coefficient matrix $\widetilde{M}$ in \eq{eq:mpa2}.}
\label{tab:topoparametrization1Nc}}
\end{center}
\end{table*}
{The $1/N_c$ expansion \cite{'tHooft:1973jz} has first been applied
  to charm physics in Ref.~\cite{Buras:1985xv}. We will apply  $1/N_c$
  counting to the tree and annihilation topologies, which are leading 
in  $1/N_c$. Here we exemplify the method for $T$:
\begin{align}
T = & {T^{\mathrm{fac}} + \delta_T}, \label{eq:fact}\\
{T^{\mathrm{fac}} \equiv} & { \frac{G_F}{\sqrt{2}} a_1 f_{\pi} (m_D^2 -
  m_{\pi}^2 ) F_0^{D\pi}(m_\pi^2)\,.} \label{eq:calctreefac}
\end{align}
Here} $a_1=C_2+C_1/N_c=1.06$ {in terms of the usual Wilson coefficients
$C_{1,2}$ of $H_W$ and the quoted value corresponds to next-to-leading
order in the NDR scheme at a scale of} $1.5\,$GeV.  {It is important to
note that the color exchange between the two quark lines in the $T$
diagram in Tab.~\ref{tab:su3limit-diagrams} is penalized by \emph{two}\
powers of $1/N_c$. {We parameterize this $1/N_c^2$ correction by the
complex parameter $\delta_T$ in \eq{eq:fact}.}  Also the renormalization scale and scheme 
dependences of $a_1$ are suppressed by $1/N_c^2$.  By using \eq{eq:fact}
and the equivalent formulae for the other tree amplitudes
$T+T_1^{(1)},\ldots$ we trade four parameters for a single parameter
$\delta_T$ {with $|\delta_T/T^{\mathrm{fac}}|\leq 0.15$}.
SU(3)$_F$ breaking in this small parameter is neglected, because it is
smaller than the neglected second-order SU(3)$_F$-breaking effects.  $
F_0^{D\pi}(m_\pi^2) $ {entering $T^{\mathrm{fac}}$} is measured in
semileptonic $D$ decays, therefore the $1/N_c$ method uses  
additional experimental input, {too}.  {Also the $A$ amplitudes factorize up
  to corrections of order $1/N_c^2$.  The factorization formulae for all
  {tree and annihilation amplitudes} can be found in
  {Appendices}~\ref{sec:fac-tree-amps} and \ref{sec:fac-anni},
  respectively.  In analogy to $\delta_T$ we define {the complex parameter}
\begin{align}
 \delta_A &= A-A^{\mathrm{fac}} \label{eq:defda}
\end{align}
for the ${\cal O}(1/N_c^2)$ corrections.}  $A^{\mathrm{fac}}$
{depends} on the form factor $F_0^{K\pi}(m_{D_{(s)}}^2)$, see
Appendix~\ref{sec:fac-anni} for details.

$E$, $C$, and $P_{\mathrm{break}}$ are formally suppressed by one power
of $1/N_c$ with respect to $T$. However, $E$ and $C$ are enhanced by
short-distance QCD effects residing in the Wilson coefficients: we write
$H_W\propto C_1Q_1 +C_2 Q_2 = (C_1+C_2/N_c) Q_1 + 2 C_2 Q_8$ with the 
octet$\times$octet operator $Q_8\equiv \overline{u}\gamma_\mu T^ac
\overline{q}^\prime \gamma^\mu T^aq$ and note that $\bra{PP^\prime}
Q_8\ket{D}$ is $1/N_c$ suppressed. However, the Wilson coefficient} $2
C_2=2.4$ {almost exactly offsets the $1/N_c$ suppression, so that $E$ and
$C$ can be almost as large as $T$. We therefore do not place a numerical
constraint on $|E|$, $|C|$, or $|P_{\mathrm{break}}|$ in our fit but
rather keep them general.}

{With the added $1/N_c$ input the diagrammatic analysis becomes more
  constrained compared to the} plain SU(3)$_F$ approach.  
Factorization {fixes the sizes} of the tree and annihilation
{amplitudes within roughly $\sim 15\%$ of $T^{\mathrm{fac}}$, i.e.\
  the size of the $1/N_c^2$ corrections. In the case of $A$ the
  $1/N_c^2$ corrections quantified by $\delta_A$ include final-state
  rescattering effects \cite{Franco:2012ck, Sorensen:1981vu,
    Buccella:1994nf, Buccella:2013tya, Neubert:1997wb, Smith:1998nu},
  which are not proportional to the decay constant $f_D$ which enters
  $A^{\mathrm{fac}}$. We therefore do not normalize $\delta_A$ to
  $A^{\mathrm{fac}}$, but instead allow $|\delta_A|$ to be as large as
  $|\delta_T|$. Factorization has also been used in
  Refs.~\cite{Bhattacharya:2012ah, Bhattacharya:2012pc}, but only to 
  estimate SU(3)$_F$ breaking. We instead use it to constrain the
  overall sizes of $T$ and $A$. 
  Note that we treat $T^{\mathrm{fac}}$ and   $A^{\mathrm{fac}}$
  beyond  linear SU(3)$_F$ breaking, so that these factorized amplitudes}  
  violate the Grossman-Robinson SU(3)$_F$ sum rules
  \cite{Grossman:2012ry}.

The parameters $\delta_T$ and $\delta_A$ {replace} the
first eight entries of $\mathbf{p}$ in \eq{eq:parametervector} as fit
parameters. {We use}
\begin{align}
\mathbf{p}^{\prime} \equiv & \left( \delta_T, \delta_A, 
		  C , 
			  C_1^{(1)} , 
			  C_2^{(1)} , 
			  C_3^{(1)} ,\right.\nn\\ &\quad \left. 
			  E  , 
			  E_1^{(1)}, 
			  E_2^{(1)}, 
			  E_3^{(1)}, 
			  P_{\mathrm{break}} \right)^T\,, 
  \label{eq:newparametervector}
\end{align}
{comprising} 11 parameters {in total}.  
{We next derive the equivalent of \eq{eq:mpa} for this new 
set of parameters. To this end we
define}
\begin{align}
  \widetilde{\mathcal{A}}(d) &\equiv \mathcal{A}(d) - 
   \mathcal{A}^{\mathrm{fac}}(d)\,,\label{eq:defat} \\
  \mathcal{A}^{\mathrm{fac}}(d) &\equiv T^{\mathrm{fac}}(d) +
  A^{\mathrm{fac}}(d)\, . 
\end{align}
The $17\times 11$ coefficient matrix {linking $\mathbf{p}^{\prime}$
  to $ \mathbfcal{\widetilde{A}}=\left(\mathcal{\widetilde{A}}(
    D^0\to K^+K^-), \ldots \mathcal{\widetilde{A}}(D_s^+\to K^0
    K^+)\right)^T $} has {only} rank~10. This has two implications:
First, {there is still a redundant parameter. Second, there is a
  new sum rule among the physical amplitudes.  Addressing the first
  point,} the kernel has the 11-dimensional basis vector
\begin{align}
\mathbf{n} = \left(-1, -1, 1, 0, 0, 0, 1, 0, 0, 0, 0 \right)^T\,, \label{eq:1NCnullspace}
\end{align}
where the order of the entries is the same as in
Eq.~(\ref{eq:newparametervector}). The redefinition
\begin{align}
\mathbf{p}^{\prime\prime} &\equiv \mathbf{p}^{\prime}  + \delta_A \, \mathbf{n} 
\end{align}
with $\mathbf{n}$ as in Eq.~(\ref{eq:1NCnullspace}) {
  absorbs $\delta_A$ into $C$, $E$ and $\delta_T$: setting} 
\begin{align}
\widetilde{C} &= C + \delta_A\,,\label{eq:redefineC}\\ 
\widetilde{E} &= E + \delta_A\,,\label{eq:redefineE}
\end{align}
{one observes that the physical amplitudes only depend on $\widetilde{C}
  $, $\widetilde{E} $, and $\delta_T-\delta_A$.}   
{Writing 
\begin{align}
\widetilde{\mathbf{p}} 
   \equiv &\left( \delta_T- \delta_A, 
		  \widetilde C , 
			  C_1^{(1)} , 
			  C_2^{(1)} , 
			  C_3^{(1)} ,\right.\nn  \\ & 
   \quad\left. 
	           \widetilde E  , 
			  E_1^{(1)}, 
			  E_2^{(1)}, 
			  E_3^{(1)}, 
			  P_{\mathrm{break}} \right)^T\,, 
  \label{eq:newparametervector2}
\end{align}
the desired equivalent of
\eq{eq:mpa} reads 
\begin{align}
 \widetilde M \widetilde{\mathbf{p}} &= \widetilde{\mathbfcal{A}} \,, \label{eq:mpa2}  
\end{align}
with the amplitudes of \eq{eq:defat} on the RHS.  The resulting
{$17\times 10$} coefficient matrix $\widetilde M$ with rank 10 is
shown in Tab.~\ref{tab:topoparametrization1Nc}.  Addressing the second
point, we find the new sum rule}
\begin{align}
  \widetilde{\mathcal{A}}(D^+\rightarrow \bar{K}^0K^+) & - 
  \widetilde{\mathcal{A}}(D_s^+\rightarrow K^0\pi^+) - \nn\\
  &2\sqrt{2}\widetilde{\mathcal{A}}(D^+\rightarrow K^+\pi^0) =
  0\,, \label{eq:1Ncrule1}
\end{align}
{from the cokernel of $\widetilde M$. It tests
the $1/N_c$ counting and is violated by terms which are linear in 
SU(3)$_F$ breaking, but suppressed by two powers of $1/N_c$.}

{The $1/N_c^2$ corrections parametrised by $\delta_{T,A}$ are varied
  in smaller ranges than the other fit parameters.}  If we {consider
  them fixed,} there remain nine unknown complex parameters in
  Eq.~(\ref{eq:newparametervector}) and a corresponding coefficient
matrix with rank nine, {implying a new sum rule.}
We combine the new rule with the one in \eq{eq:1Ncrule1} as:
\begin{align}
  \widetilde{\mathcal{A}}(D^+\rightarrow \bar{K}^0K^+) &- \widetilde{\mathcal{A}}(D_s^+\rightarrow K^0\pi^+) \nn\\
  &= 2 \left(\delta_T -\delta_A\right)\,, \\
  \widetilde{\mathcal{A}}(D^+\rightarrow K^+\pi^0) &= \frac{1}{\sqrt{2}}
  \left(\delta_T - \delta_A\right) \,.
\end{align}
%

\begin{table*}[tp!]
\begin{center}
\begin{tabular}{c|c|c|c|c|c}
\hline \hline
Example fit point with minimal $\chi^2$:        & Point 0                        & Point 1  & Point 2 & Point 3  &Exp. data \\
Applied conditions:     & None   &$P_{\mathrm{break}}=\delta_T=\delta_A=0$ & $\delta_T=\delta_A=0$ & $P_{\mathrm{break}}=0$ &  \textemdash \\\hline\hline

$\vert P_{\mathrm{break}}/T^{\mathrm{fac}}\vert$ & 0.25  & 0       & 0.54       & 0     &\textemdash\\ 
$\mathrm{arg}(P_{\mathrm{break}})$               & $-1.95$ & 0       & 2.21       & 0     &\textemdash\\
$\vert\delta_T\vert/T^{\mathrm{fac}}$            & 0.11  & 0       & 0          & 0.15  &\textemdash\\
$\mathrm{arg}(\delta_T)$                         & 3.07  & 0       & 0          & $-2.74$  &\textemdash\\
$\vert\delta_A\vert/A^{\mathrm{fac}}$            & 0.09  & 0       & 0          & 0.11  &\textemdash\\
$\mathrm{arg}(\delta_A)$                         & 0.67  & 0       & 0          & $0.00$  &\textemdash\\
$F_0^{D_s K}(0)/F_0^{D \pi}(0)$                  & 0.96  & 1.01    & 0.95       & 0.97  &\textemdash\\ 
$F_0^{D K}(0)$                                   & 0.74  & 0.72    & 0.74       & 0.74  &\textemdash\\ 
$F_0^{D \pi}(0)$                                 & 0.64  & 0.64    & 0.64       & 0.64  &\textemdash\\ 
$\vert F_0^{K\pi}(m^2_{D_{(s)} })\vert$          & 2.39  & 1.99    & 4.50       & 1.62  &\textemdash\\ 
$\mathrm{arg}(F_0^{K\pi}(m^2_{D_{(s)}}))$        & 1.71  & $-1.29$ & $-1.15$    & $-2.36$ &\textemdash\\\hline 
$T^{\mathrm{fac}}/10^{-6}$ GeV                   & 2.52  & 2.52    & 2.52       & 2.52  &\textemdash\\\hline 

$T^{\mathrm{fac}}(D^+\rightarrow \bar{K}^0 K^+)/10^{-6}$ GeV    & 3.40             & 3.34             & 3.40             &   3.40   &\textemdash\\
$T^{\mathrm{fac}}(D_s^+\rightarrow K^0 \pi^+)/10^{-6}$ GeV      & $-2.53$          & $-2.68$          & $-2.51$          &   $-2.57$   &\textemdash\\ 
$T^{\mathrm{fac}}(D^+\rightarrow K^+\pi^0)/10^{-6}$ GeV         & 2.22             & 2.22             & $2.22$           &   2.22   &\textemdash\\ 
$A^{\mathrm{fac}}(D_s^+\rightarrow K^0 \pi^+)/10^{-6}$ GeV      & $-0.18 + i\, 1.22$ & $0.28 - i\, 0.99$  & $0.94 - i\, 2.12$  &  $-0.59 - i\, 0.59$   &\textemdash\\ 
$A^{\mathrm{fac}}(D^+\rightarrow K^+\pi^0)/10^{-6}$ GeV         & $0.10 - i\, 0.68$  & $-0.16 + i\, 0.55$ & $-0.53 + i\, 1.19$ &  $0.33 + i\, 0.33$   &\textemdash\\\hline 
$\mathcal{B}(D^+\rightarrow K_S K^+) / 10^{-3}$                 & 2.83             & 4.04             & 2.85             &  2.83   & $2.83\pm 0.16$ \cite{Beringer:1900zz}\\ 
$\mathcal{B}(D_s^+\rightarrow K_S \pi^+) / 10^{-3}$             & 1.22             & 1.22             & 1.23             & 1.22    & $1.22\pm 0.06$ $^\dagger$\cite{Beringer:1900zz, Zupanc:2013byn, Onyisi:2013bjt} \\ 
$\mathcal{B}(D^+\rightarrow K^+ \pi^0) / 10^{-4}$               & 1.83             & 1.83             & 1.72             &  1.83  & $1.83\pm 0.26$ \cite{Beringer:1900zz} \\\hline
$\chi^2$                     & 0.00         & 63.93   & 0.20   & 0.00   & \textemdash \\ 
$\nu$                        &\textemdash  & 5     & 3    &  2     & \textemdash \\
Significance of rejection    &\textemdash  & $7.0\sigma$      &  $0.03\sigma$    &   $0.0\sigma$  & \textemdash \\\hline\hline
\end{tabular}
\caption{Fits to the branching ratios $\mathcal{B}(D^+\rightarrow K_S K^+)$, $\mathcal{B}(D_s^+\rightarrow K_S \pi^+)$, $\mathcal{B}(D^+\rightarrow K^+ \pi^0)$ only, without taking 
correlations and additional constraints on SU(3)$_F$ breaking, see Sec.~\ref{sec:SU3Xing}, into account. 
The form factors are varied as described in Sec.~\ref{sec:fits} and 
{Appendices}~\ref{sec:fac-tree-amps} and \ref{sec:fac-anni}.
The $\chi^2$ is the one taking into account the three given branching ratios and form factors only.  
$\nu$ are the number of degrees of freedom compared to {the fit scenario of} point 0. The significance of rejection takes point 0 as null hypothesis. $^\dagger$Our average.
\label{tab:1Nc-example}}
\end{center}
\end{table*}


%
{The amplitudes in \eq{eq:1Ncrule1} are
  related to those} with $K_{S,L}$ in the final state as
\begin{align}
  \mathcal{A}(D^+\rightarrow K_{S,L} K^+)    &= {\mp}
 \frac{1}{\sqrt{2}} \mathcal{A}(D^+\rightarrow \bar{K}^0 K^+ )\,, \\
  \mathcal{A}(D_s^+\rightarrow K_{S,L} \pi^+ ) &= \frac{1}{\sqrt{2}}
  \mathcal{A}(D_s^+\rightarrow K^0 \pi^+ )\, .
\end{align}
The corresponding branching ratios read:
\begin{align}
 & \mathcal{B}(D^+\rightarrow {K_{S,L}} K^+) = \vert\lambda_{sd}\vert^2\,\mathcal{P}( D^+, K^0, K^+ ) \times \nn\\
 &\quad \left| \mathcal{A}^{\mathrm{fac}}( D^+\rightarrow \bar{K}^0 K^+) + (\delta_T-\delta_A) + P_{\mathrm{break}}\right|^2 \,, \label{eq:1NcBR1}\\[0.2cm]
&  \mathcal{B}(D_s^+\rightarrow {K_{S,L}}\pi^+) = \vert\lambda_{sd}\vert^2\, \mathcal{P}( D_s^+, K^0, \pi^+ ) \times \nn\\
&\quad \left| \mathcal{A}^{\mathrm{fac}}( D_s^+\rightarrow K^0\pi^+) - (\delta_T-\delta_A) + P_{\mathrm{break}}\right|^2\,, \label{eq:1NcBR2} \\[0.2cm]
&  \mathcal{B}(D^+\rightarrow K^+ \pi^0) = \vert V_{cd}^* V_{us}\vert^2\, \mathcal{P}( D^+, K^+, \pi^0 )\times \nn\\
&\quad \left| \mathcal{A}^{\mathrm{fac}}( D^+\rightarrow K^+ \pi^0) + (\delta_T-\delta_A)\right|^2 \,, \label{eq:1NcBR3}
\end{align}
with $\mathcal{P}(D,P_1,P_2)$ as defined in Eq.~(\ref{eq:phasespace}).
{\eqsto{eq:1NcBR1}{eq:1NcBR3} permit one to probe our combined SU(3)$_F$
  and $1/N_c$ expansion quantitatively, since a too large value of
  $|\delta_T-\delta_A|$ extracted from \eqsto{eq:1NcBR1}{eq:1NcBR3}
  would falsify the method. Furthermore, the size of
  $|P_{\mathrm{break}}|$ gives insight into an important issue of QCD
  dynamics, the size of the GIM suppression in the difference between
  strange and down loops.}  

In Tab.~\ref{tab:1Nc-example} we show example fits to the branching
ratios Eqs.~(\ref{eq:1NcBR1})--(\ref{eq:1NcBR3}) only, testing the
dependence of the fit result on the $1/N_c^2$ corrections and the broken
penguin.

In the first place, we illustrate that the data can easily be
accommodated for realistic values of $\delta_T$, $\delta_A$ and
$P_{\mathrm{break}}$ (point 0). Taking these parameters out of the
fit (point 1) results in a bad description of the data which is rejected
at $7\sigma$. It is possible to describe the data with an enhanced
broken penguin only (point 2) but also with $P_{\mathrm{break}} = 0$ and
adjusting $\delta_T$ and $\delta_A$ (point 3). A better knowledge of the
form factor $F_0^{K\pi}(m_{D_{(s)}}^2)$, see
Appendix~\ref{sec:fac-anni}, is crucial in order to disentangle an
  enhanced penguin from $1/N_c^2$ corrections. This could be
  provided by future high statistics measurements of $\tau$ decays
\cite{Epifanov:2007rf, Kuhn:1992nz}.

\subsection{Measuring diagrammatic SU(3)$_F$ breaking\label{sec:SU3Xing}}

In order to describe SU(3)$_F$ breaking in the framework of the diagrammatic approach, we introduce the following measures 
{in analogy to Ref.~\cite{Hiller:2012xm}}. 
We define 
\begin{align}
\delta^{\prime, \mathcal{T}}_X &\equiv \mathrm{max}_d \left|\frac{\mathcal{A}_X^{\mathcal{T}}(d)}{\mathcal{A}(d)}\right|\,, \label{eq:su3X-topo-single}
\end{align}
where $\mathcal{T} = C, E, P_{\mathrm{break}}$ and $\mathcal{A}_X^{\mathcal{T}}(d)$ is the part of the amplitude of decay $d$ stemming 
from the corresponding SU(3)$_F$-breaking parameter(s) only. $\mathcal{A}(d)$ denotes the full amplitude of decay $d$. 
The parameters defined in Eq.~(\ref{eq:su3X-topo-single}) give 
a measure for the {maximal} SU(3)$_F$-breaking contribution to the full amplitude from each topology.

{A measure of the maximal SU(3)$_F$ breaking residing in any of
  the topologies $C$, $E$ and $P_{\mathrm{break}}$ is therefore}
\begin{align}
  \delta^{\prime, \mathrm{topo}}_X &\equiv \mathrm{max}_d
  \left|\frac{\sum_{\mathcal{T}}
      \mathcal{A}_X^{\mathcal{T}}(d)}{\mathcal{A}(d)}\right|\,. 
  \label{eq:su3X-topo}
\end{align}
Note that the SU(3)$_F$ breaking stemming from our calculation of the $T$ and $A$ topologies using factorization is not included in the definition Eq.~(\ref{eq:su3X-topo}).\\

Further{more}, we quantify the relative SU(3)$_F$ breaking of $C$ and
$E$ topologies by the measures
\begin{align}
\delta_X^{C_i^{(1)}/\widetilde{C}} &\equiv \left| \frac{C_i^{(1)}}{\widetilde{C}}\right| \,, &
\delta_X^{E_i^{(1)}/\widetilde{E}} &\equiv \left| \frac{E_i^{(1)}}{\widetilde{E}}\right| \,,
\end{align}
respectively.  In the fit we always demand all the above measures to be
$\leq 50\%$. In $\delta^{\prime, E}_X$ and $\delta^{\prime,
  \mathrm{topo}}_X$ we {ignore} $\mathcal{B}(D^0\rightarrow \bar{K}^0
K^0)$ {when} taking the maximum, {because this} branching ratio
{vanishes in the SU(3)$_F$ limit.}  Note that $E_3^{(1)}$ appears in
the {omitted} channel $D^0\rightarrow \bar{K}^0 K^0$ only,
{therefore $\delta^{\prime, E}_X=0$ is insensitive to the size of}
$E_3^{(1)}\neq 0$.  Furthermore, in case an amplitude vanishes at some
point in parameter space we {also exclude it from the calculation of}
the maxima in Eqs.~(\ref{eq:su3X-topo-single}) and
(\ref{eq:su3X-topo}).

\section{Fit to Branching ratio measurements\label{sec:fits}}
{In our global fit we use the available measured branching fractions 
  and the strong phase difference $\delta_{K^+\pi^-}$ and impose} the
theoretical constraints quoted in Sec.~\ref{sec:theoryinput}. {18 of the fit
  parameters are related to topological amplitudes:}
\begin{center}
$\vert \widetilde{C}/T^{\mathrm{fac}}\vert$, 
$\quad\mathrm{arg}(\widetilde{C})$,
$\quad\vert \widetilde{E}/T^{\mathrm{fac}}\vert$, 
$\quad\mathrm{arg}(\widetilde{E})$,\\
$\vert C_i^{(1)}\vert$, 
$\quad\mathrm{arg}(C_i^{(1)})$,
$\quad\vert E_i^{(1)}\vert$, 
$\quad\mathrm{arg}(E_i^{(1)})$,\\
$\vert P_{\mathrm{break}}/T^{\mathrm{fac}}\vert$,
$\quad\mathrm{arg}(P_{\mathrm{break}})$,
\end{center}
with $i=1,2,3$ and $T^{\mathrm{fac}}$ is calculated from
Eq.~(\ref{eq:calctreefac}).
{We normalize to $T^{\mathrm{fac}}$ rather than
  $T=T^{\mathrm{fac}}+\delta_T$, because our fit is only} sensitive
{to the combination $\delta_T-\delta_A$ and therefore leaves
  $\delta_T$ undetermined.}
These 18 quantities are supplied by four parameters measuring the $1/N_c^2$
corrections to the tree and annihilation diagrams: \begin{center}
    $\vert\delta_T\vert/T^{\mathrm{fac}}$, $\quad\mathrm{arg}(\delta_T)$,
    $\quad\vert \delta_A\vert/T^{\mathrm{fac}}$, $\quad\mathrm{arg}(\delta_A)$\,. \end{center}
In addition we need five parameters related to form factors:
\begin{center}
$F_0^{D_s K}(0)/F_0^{D\pi}(0)$, 
$\quad F_0^{D K}(0)$, 
$\quad F_0^{D \pi}(0)$,\\ 
$\quad\vert F_0^{K\pi}(m_D^2)\vert$,
$\quad\mathrm{arg}(F_0^{K\pi}(m_D^2))$\,,
\end{center}
and set $F_0^{K\pi}(m_{D_s}^2) = F_0^{K\pi}(m_D^2)$.
\begin{table*}
\begin{center}
\begin{tabular}{c|c|c|c}
\hline \hline
Hypothesis 					     & Significance of rejection & $\Delta \chi^2$ & dof \\\hline\hline
$P_{\mathrm{break}} = 0$                                         &   $0.7\sigma$  &  1.3      & 2 \\ 
$P_{\mathrm{break}} = E_i^{(1)} = C_i^{(1)} = 0 \, \forall\, i$  &   $>5\sigma$   &  431.4    & 14 \\ 
$E_i^{(1)} = 0\, \forall\, i$                                    &   $3.0\sigma$  &  20.3     & 6 \\ 
$E = E_i^{(1)} = 0\, \forall\, i$                                &   $>5\sigma$   &  156.4    & 8 \\ 
$C_i^{(1)} = 0\, \forall\, i$                                    &   $4.3\sigma$  &  31.6     & 6 \\ 
$C = C_i^{(1)} = 0\, \forall\, i$                                &   $>5\sigma$   &  $267\cdot 10^3$  & 8 \\\hline\hline 
\end{tabular}
\caption{Results of several likelihood ratio tests. Shown are the
obtained $\chi^2$, the relative degrees of freedom (\emph{dof}) of the hypothesis compared to the {null hypothesis, which is the} full fit, and 
the significance at which the hypothesis can be rejected. 
\label{tab:likelihoodratiotests}}
\end{center}
\end{table*}
Altogether these are 27 real parameters, which are fitted to 16 measured
branching ratios and one strong phase.  The experimental input values,
including the respective correlations, are listed in Appendix
\ref{eq:Inputdata}.  The number of parameters is larger than the number
of observables. However, the 27 parameters are subject to 10
{constraints on the maximal size of} linear SU(3)$_F$ breaking, see
Sec.~\ref{sec:SU3Xing}, {and the bounds $|\delta_{T,A}|\leq
  0.15\, T^{\rm fac}$.}  At the global minimum we obtain $\chi^2 =
0.0$,~i.e.,~the parameterization and theoretical input is in
perfect agreement with the data.  {Thus} the data {are both
  compatible with our chosen bound on} SU(3)$_F$ breaking {(i.e.\ all
  measures {defined} in Sec.~\ref{sec:SU3Xing} are smaller than 50\%)} and
the six Grossman-Robinson SU(3)$_F$ sum rules \cite{Grossman:2012ry}.

In order to study the relative importance of the topological amplitudes
{for the description of the data}, we perform likelihood ratio tests.
We look at several scenarios where some of the parameters of our fit are
fixed. In order to keep the fit simple, we assume the validity of
Wilks' theorem~\cite{Wilks:1938dza},~i.e., we calculate the
$p$-value according to \cite{Beringer:1900zz, Wiebusch:2012en}
\begin{align}
p &= 1 - P_{\nu/2}\left(\Delta \chi^2/2 \right)\,, \label{eq:calcPvalue}
\end{align}
with the normalized lower incomplete Gamma function $P_{\nu/2}$
{depending on} the number $\nu$ of relatively fixed parameters
compared to the full fit.  For a general discussion of the assumptions
{underlying} Eq.~(\ref{eq:calcPvalue}) see
Ref.~\cite{Wiebusch:2012en}.

The results of our likelihood ratio tests are shown in
Tab.~\ref{tab:likelihoodratiotests}.  {This table} shows at which
significance we can reject a certain hypothesis. For example, we can
reject 
$P_{\mathrm{break}} = 0$ at only {$\sim 0.7\sigma$}, {implying that}
$P_{\mathrm{break}} = 0$ is {well} consistent with the data.
However, the fit shows a clear need for SU(3)$_F$ breaking: the
SU(3)$_F$-limit fit with $P_{\mathrm{break}} = E_i^{(1)} = C_i^{(1)} = 0
\, \forall\, i$ {is} rejected at~$>5\sigma$. Looking at the SU(3)$_F$
breaking in specific {topological amplitudes} we find a slight
tendency toward a stronger SU(3)$_F$ breaking in the color-suppressed
tree than in the exchange diagrams.

\begin{figure*}[tp]
\begin{center}
\subfigure{
        \includegraphics[width=0.43\textwidth]{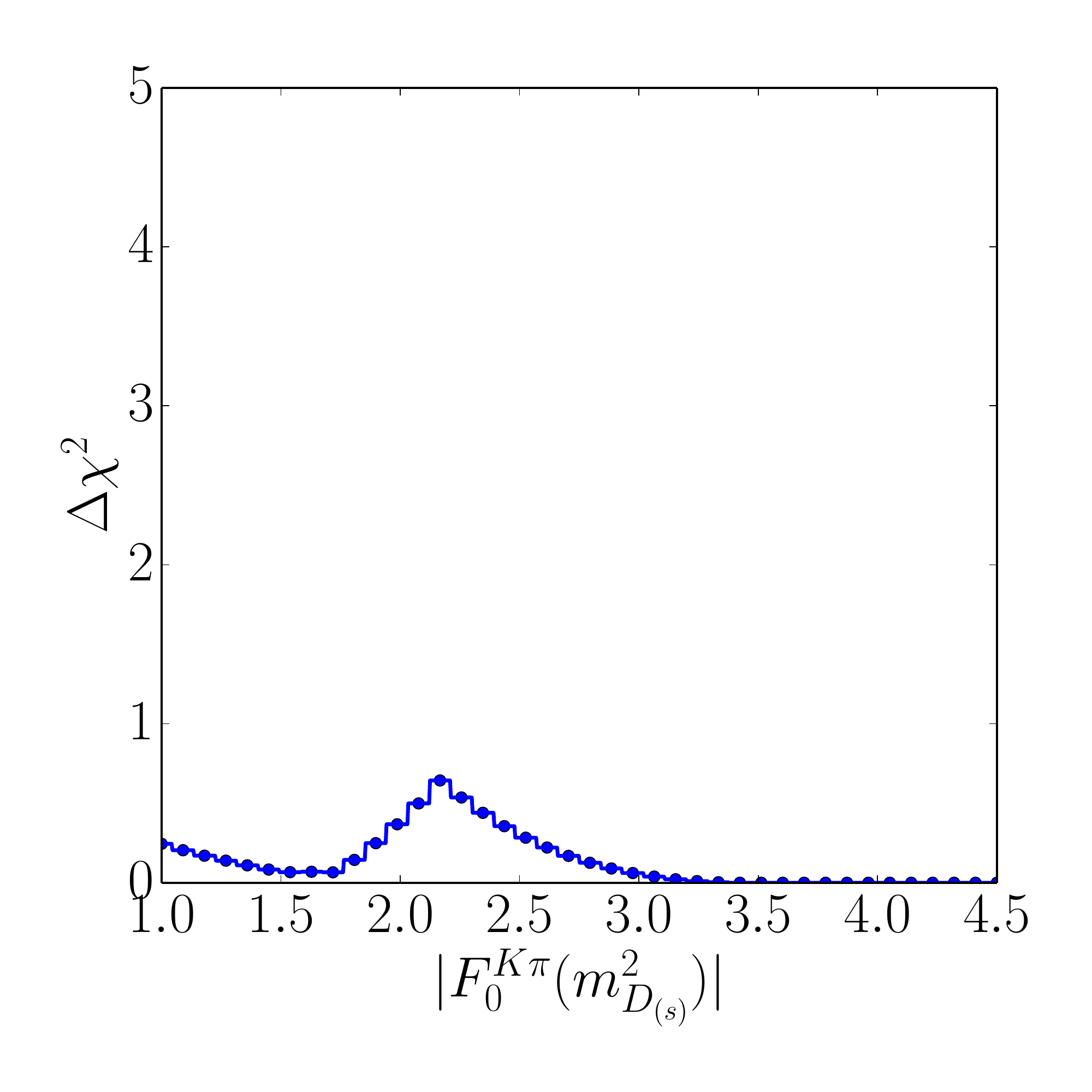}
}
\end{center}
\caption{
The form factor $F_0^{K\pi}(m_{D_{(s)}}^2)$. 
\label{fig:formfactor-Kpi}
}
\end{figure*}

In Figs.~\ref{fig:chi2-su3limit-topo}--\ref{fig:formfactor-Kpi} we show
plots of the fit parameters, measures of SU(3)$_F$ breaking and {fit
  predictions for} observables.  We see that in the multi-parameter
space the {best-fit solutions cover broad regions and typically
  several disconnected best-fit regions exist. Considering that there
  are more parameters than fitted quantities the large degeneracy of the
  best-fit region is not surprising. It is moot to quote best-fit values
  for the parameters, because one can move in a wide valley with $\Delta
  \chi^2=0$. We suspect that the alternative approach of a Bayesian
  analysis would single out a small portion of this $\Delta
  \chi^2=0$ valley as a consequence of the Bayesian prior placed on the 
  fit parameters and the central limit theorem of statistics.}
{Therefore frequentist analyses like ours are more adequate to
  the problem.}
   
The {phase of $\widetilde C$ significantly deviates from 0 and $\pm\pi$
  (see Fig.~{\ref{fig:plot-1d-02}}), which points to large
  rescattering effects.}  The fit results for {$\vert
  \widetilde{C}/T^{\mathrm{fac}}\vert$} and {$\vert
  \widetilde{E}/T^{\mathrm{fac}}\vert$}, see
Fig.~\ref{fig:chi2-su3limit-topo}, show {disconnected regions at
  95\% CL. $ \widetilde{C}$ and $ \widetilde{E}$ are suppressed by
  $1/N_c$ but involve a large Wilson coefficient $\sim 2.4$ as discussed
  in Sec.~\ref{sec:countingtopologies}. Thus only solutions with {$\vert
    \widetilde{C}/T^{\mathrm{fac}}\vert, \vert
    \widetilde{E}/T^{\mathrm{fac}}\vert \lsim 1$} are consistent with
  $1/N_c$ counting, which singles out one of the three regions in the
  {$\vert \widetilde{C}/T^{\mathrm{fac}}\vert$--$\vert
    \widetilde{E}/T^{\mathrm{fac}}\vert$} plane.}

The needed maximum size of total SU(3)$_F$ breaking on the amplitude is
given by $\delta_X^{\prime, \mathrm{topo}}\sim 30\%$ in agreement with
{Ref.}~\cite{Hiller:2012xm}, as can be read off
Fig.~\ref{fig:plot-1d-22}. Note that $\delta_X^{\prime, \mathrm{topo}}$
as well as $\delta^{\prime, \mathcal{T}}_X$ do not {measure the average
  but the maximal} size of SU(3)$_F$ breaking in one of the {17} decay
channels {except for $D^0\rightarrow \bar{K}^0 K^0$}, see
Sec.~\ref{sec:SU3Xing}.  Thus, these measures are very conservative and
{could} in principle be biased by {a single} channel.  However, the need
for SU(3)$_F$ breaking in individual parameters can be considerably
smaller than $30\%$,~\emph{e.g.}, $\vert E_i^{(1)}\vert\sim~0$, $\vert
C_{2,3}^{(1)}\vert\sim~0$ {is well allowed} at $1\sigma$, see
Figs.~\ref{fig:plot-1d-0579} and \ref{fig:plot-1d-111315}, respectively.
Also a $\vert P_{\mathrm{break}}/T\vert$ below $5\%$ already gives very
{good} fits, see Fig.~\ref{fig:plot-1d-17}.  In
Fig.~\ref{fig:plot-1d-21} we see that the same is consistently the case
for $\delta_X^{\prime,P_{\mathrm{break}}}$.

From Figs.~\ref{fig:plot-1d-19} and \ref{fig:plot-1d-20} we again see 
the slight tendency for larger SU(3)$_F$ breaking in the
color-suppressed tree topologies compared to the exchange diagrams.

As illustrated in Tab.~\ref{tab:1Nc-example} {discussed in}
Sec.~\ref{sec:countingtopologies} {around
  \eqsto{eq:1NcBR1}{eq:1NcBR3}},  
the broken penguin
$P_{\mathrm{break}}$ {is correlated with the parameter
  $\delta_T-\delta_A$ quantifying $1/N_c^2$ corrections to factorizable
  amplitudes.} 
This {feature} can be {verified} in Fig.~\ref{fig:plot-2d-16}
{which shows this correlation.}  A vanishing penguin
$P_{\mathrm{break}}\sim 0$ is allowed at the price of $1/N_c$ breaking
corrections of order $\gtrsim 15\%$. Note again that this correlation
heavily depends on the poorly measured form factor
$F_0^{K\pi}(m_{D_{(s)}}^2)$.  Interestingly, the fit result for
$F_0^{K\pi}(m_{D_{(s)}}^2)$, see Fig.~\ref{fig:formfactor-Kpi}, is not
completely flat, showing its nontrivial influence on the branching
ratios of charm decays.  The branching ratio $\mathcal{B}(D^+\rightarrow
K^+ \pi^0)$ depends on no topological parameters besides $\delta_T$ and
$\delta_A$.  Its fit result, which is given in
Fig.~\ref{fig:plot-1d-29}, shows that our assumptions on the ranges for
$\delta_T$ and $\delta_A$ are {loose enough to accommodate the
  measured branching fraction.} However, large fit results for $\mathcal{B}(D^+\rightarrow K^+\pi^0)$ 
are slightly disfavored.

\begin{figure*}[t]
\begin{center}
\subfigure[\label{fig:plot-1d-35} Black: {prediction of 
  Refs.}~\cite{Bigi:1994aw, Rosner:2006bw, Gao:2006nb,
  Gao:2014ena}. {Red (below the black point): experimental error}]{
        \includegraphics[width=0.43\textwidth]{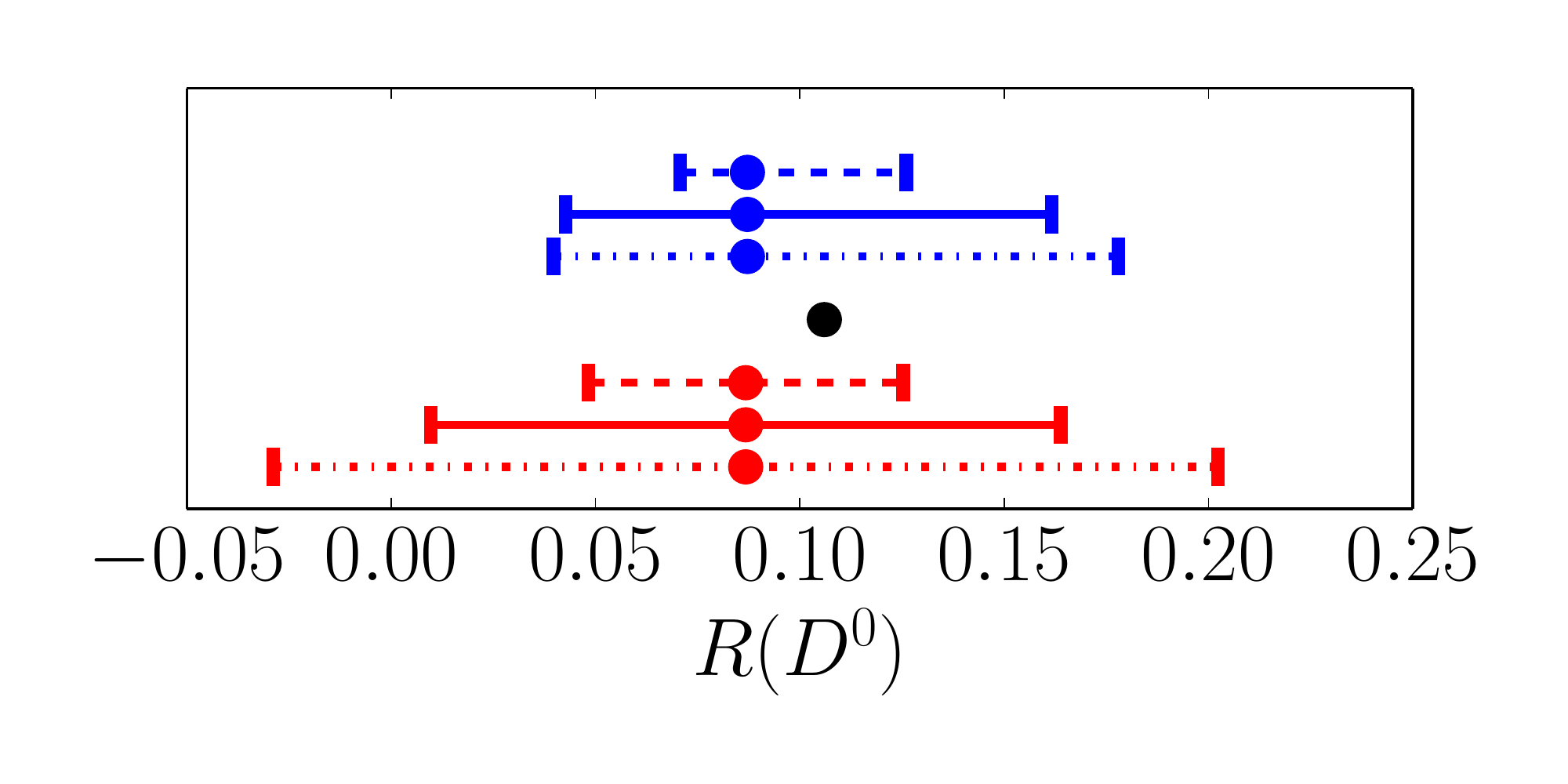}
        }
\hfill
\subfigure[\label{fig:plot-1d-38} Black:  {prediction of 
  Refs.}~\cite{Bigi:1994aw, Rosner:2006bw, Gao:2006nb, Gao:2014ena}]{
        \includegraphics[width=0.43\textwidth]{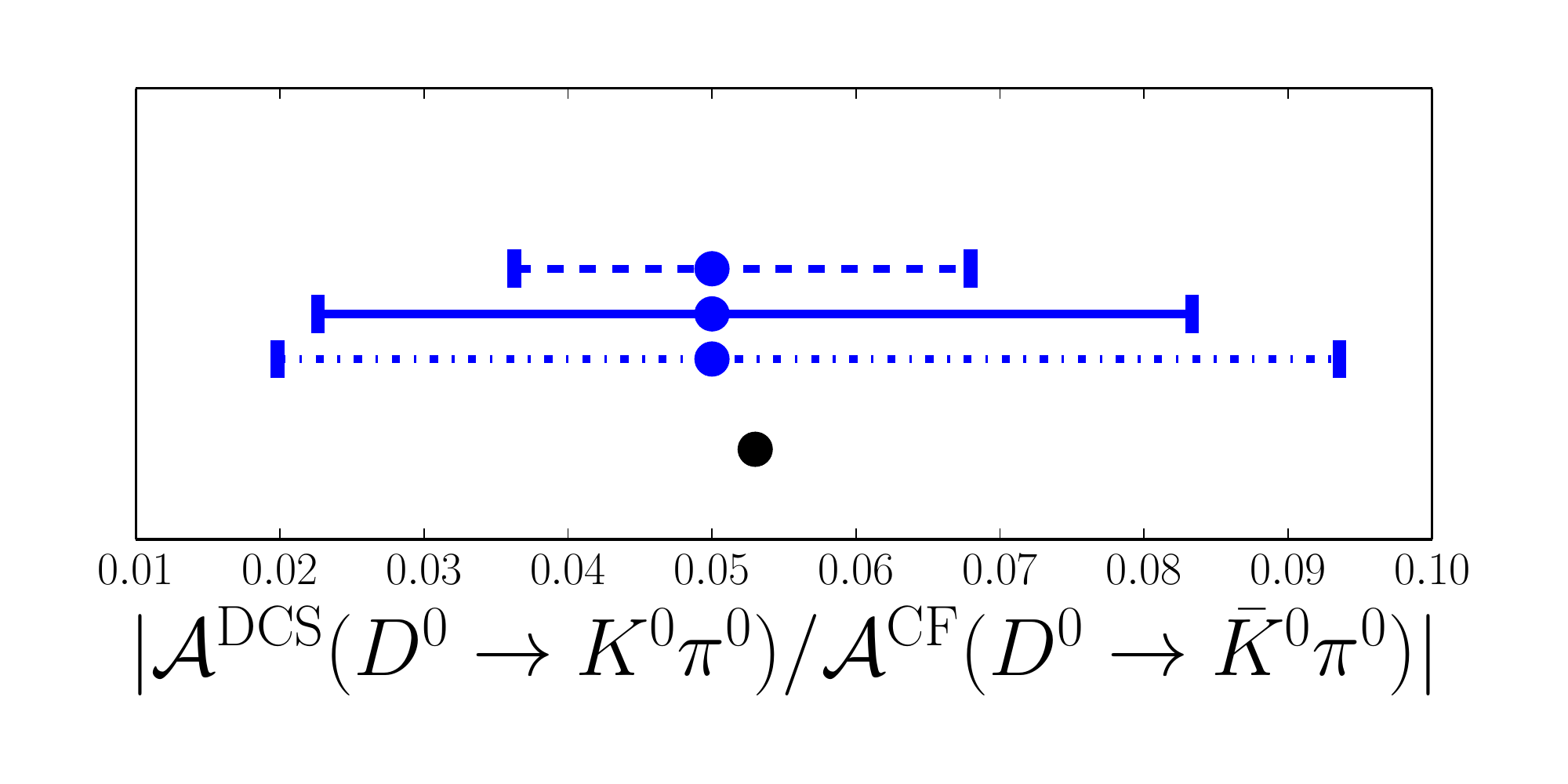}
}

\subfigure[\label{fig:plot-1d-36} Black: {$1\sigma$ range predicted in
   Ref.}~\cite{Gao:2014ena}]{
       \includegraphics[width=0.43\textwidth]{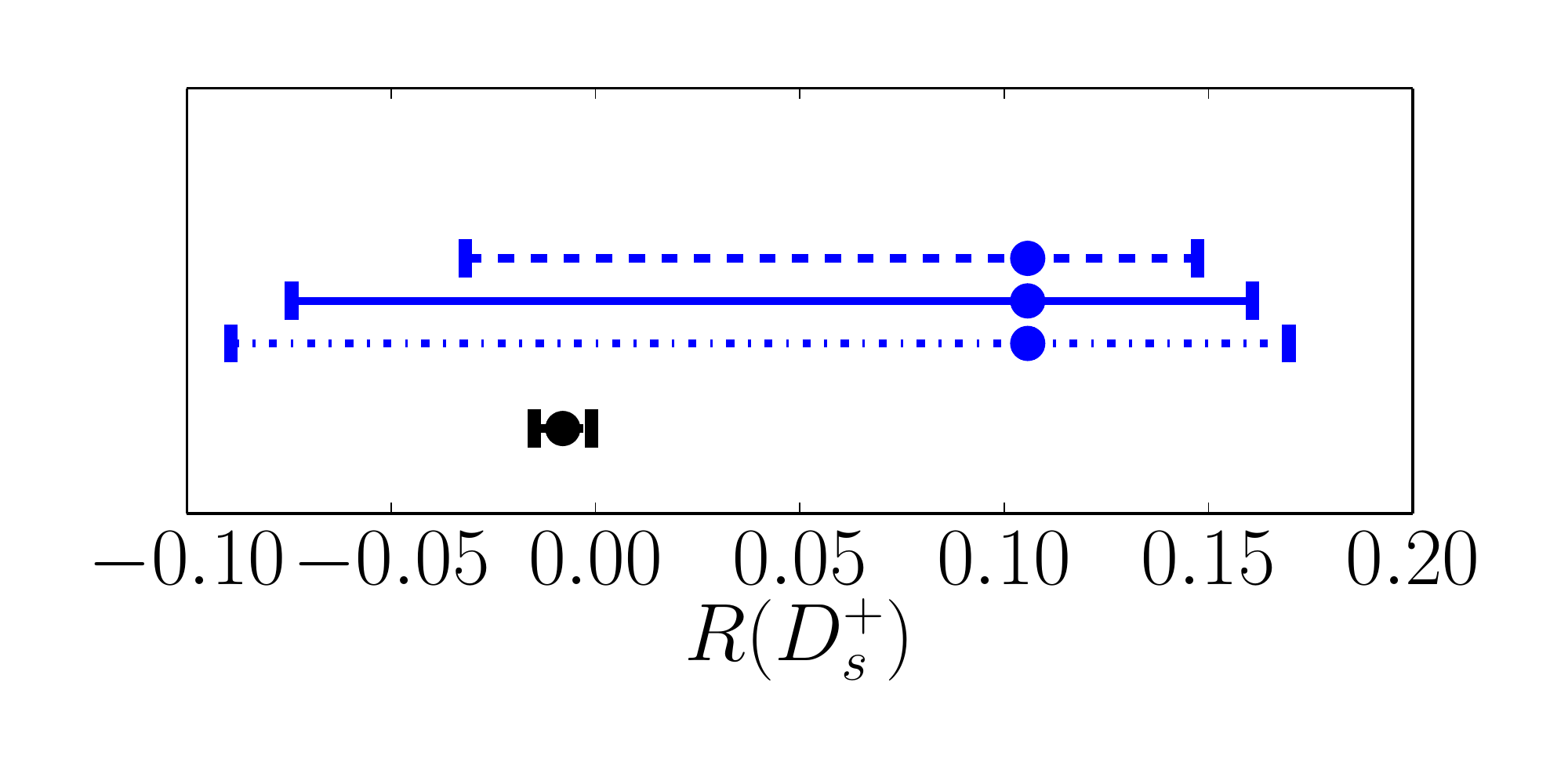}
        }
\hfill %
\subfigure[\label{fig:plot-2d-39} Black: {$1\sigma$ range predicted in 
   Ref.~}\cite{Gao:2014ena}]{
        \includegraphics[width=0.43\textwidth]{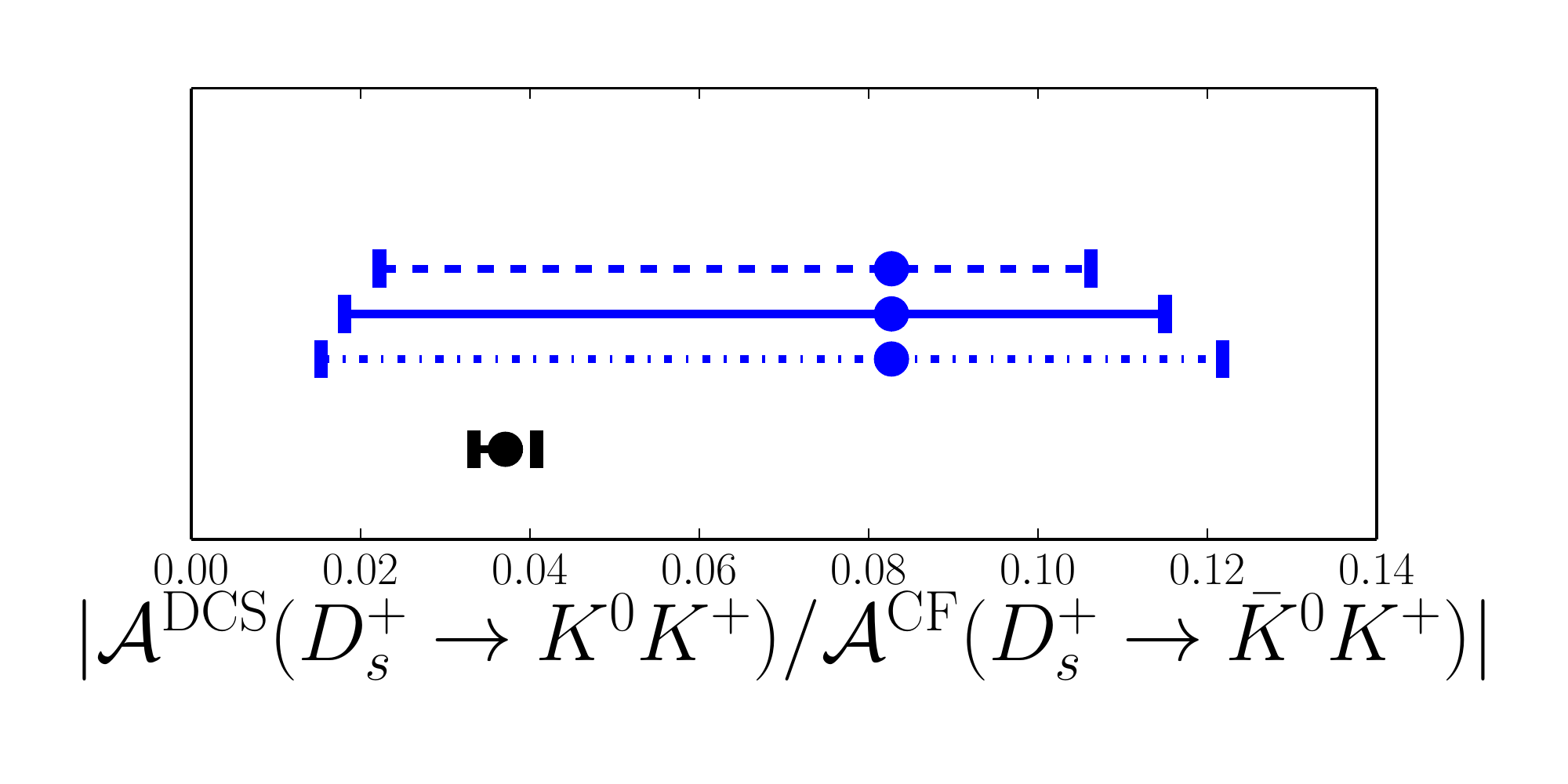}}

\subfigure[\label{fig:plot-1d-37}]{
        \includegraphics[width=0.43\textwidth]{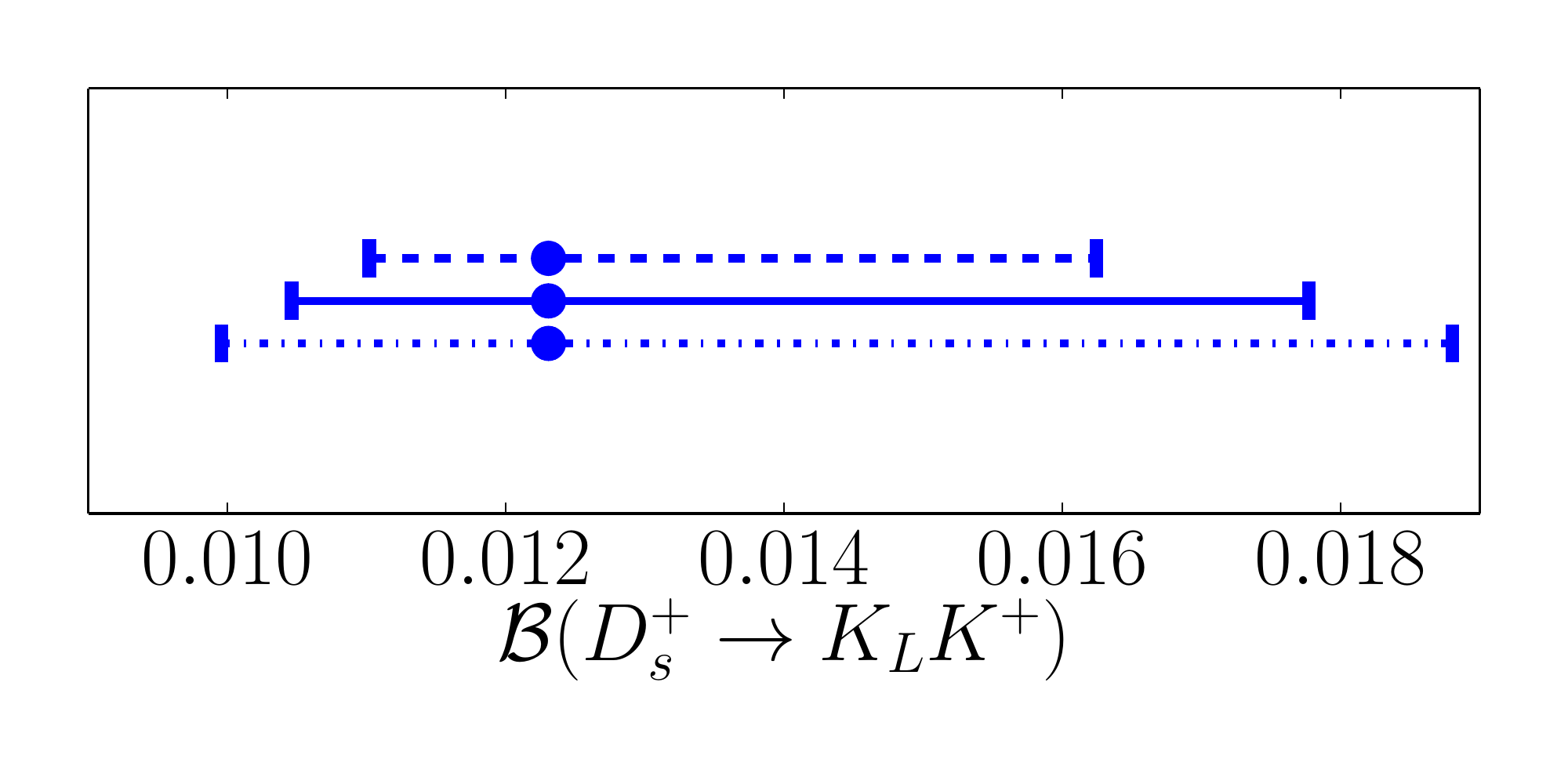}
}

\end{center}
\caption{Blue: Our results for several observables probing doubly
  Cabibbo-suppressed amplitudes (see \eqsand{eq:rd0}{eq:rds}). 
  The lines correspond to
  $1\sigma$ (dashed), $2\sigma$ (solid) and $3\sigma$ (dashed-dotted)
  {confidence intervals}, respectively. The {experimental error
    in} $R(D^0)$
   is obtained by Gaussian error
  propagation from Tab.~\ref{tab:branchingratios}. The 
  results from other groups \cite{Bigi:1994aw, Rosner:2006bw,
    Gao:2006nb, Gao:2014ena} {are shown in black.} 
In case of $R(D^0)$ (a) and
  $\left\vert\frac{\mathcal{A}^{\mathrm{DCS}}(D^0\rightarrow
      K^0\pi^0)}{\mathcal{A}^{\mathrm{CF}}(D^0\rightarrow \bar{K}^0
      \pi^0)} \right\vert$ {(b)} no errors are given in
  Refs.~\cite{Bigi:1994aw, Rosner:2006bw, Gao:2006nb, Gao:2014ena}.
\label{fig:dcs-obs}}
\end{figure*}

{We may next ask whether we can use our fit output to predict
  individual branching fractions better than they are currently measured.
  Our general finding is as in Fig.~\ref{fig:plot-1d-29}, the fit output
  for the $\Delta \chi^2$ profiles essentially tracks the fit input.  To
  find non-trivial predictions for future measurements we must study
  correlations between at least two observables. A nice result is shown}
in Fig.~\ref{fig:plot-2d-19} {revealing} the correlation of
$\mathcal{B}(D^0\rightarrow K_L\pi^0)$ and $\mathcal{B}(D^0\rightarrow
K_S\pi^0)$.  In the SU(3)$_F$ limit the branching ratios are strongly
correlated through their parametric dependence\footnote{{In order to
    find the correct relative signs in \eqsand{eq:dkp1}{eq:dkp2} one
    must define $K_{S,L}$ %
    correctly. \eqsand{eq:states-1}{eq:states-2} comply with $\ket{K^0}=
    C\ket{\Kbar{}^0}=-CP\ket{\Kbar{}^0}$ entailing
    $\ket{K_{S}}\simeq(\ket{K^0} - \ket{\Kbar{}^0})/\sqrt2$. We have
    checked our results by studying the full decay chain $D^0\rightarrow
    \KorKbar [\to \pi^+\pi^-] \pi^0$, from which the $K^0$ sign
    conventions drop out.}}
\begin{align}
  \mathcal{B}(D^0\rightarrow K_S \pi^0) &\sim \vert E-C\vert^2  +2\lambda^2 \vert E-C\vert^2\,, \label{eq:dkp1}\\
  \mathcal{B}(D^0\rightarrow K_L \pi^0) &\sim \vert E-C\vert^2
  -2\lambda^2 \vert E-C\vert^2\,, \label{eq:dkp2}
\end{align}
which {implies} $\mathcal{B}(D^0\rightarrow K_L \pi^0) \lesssim
\mathcal{B}(D^0\rightarrow K_S \pi^0)$. This relation is {a priori
  absent once} SU(3)$_F$-breaking effects {are included, because the
  latter can be larger than $|E-C|$. However, the global fit rejects
  this possibility: in Fig.~\ref{fig:plot-2d-19} the region
  corresponding to 95\% CL entirely satisfies
  $\mathcal{B}(D^0\rightarrow K_L \pi^0) < \mathcal{B}(D^0\rightarrow
  K_S \pi^0)$. Performing a dedicated likelihood ratio test we find that
  $\mathcal{B}(D^0\rightarrow K_L \pi^0) < \mathcal{B}(D^0\rightarrow
  K_S \pi^0)$ holds with a significance of more than $4\sigma$.  Our
  fit} excludes a large region of the $\mathcal{B}(D^0\rightarrow K_S
  \pi^0)$--$\mathcal{B}(D^0\rightarrow K_L \pi^0)$ plane which is still
  allowed by the individual measurements.  {To quantify our findings
  further we define
\begin{align}
  R(D^0) &\equiv \frac{ \mathcal{B}(D^0\rightarrow K_S \pi^0) -
    \mathcal{B}(D^0\rightarrow K_L \pi^0) }{ \mathcal{B}(D^0\rightarrow
    K_S \pi^0) + \mathcal{B}(D^0\rightarrow K_L \pi^0) }\,
 \label{eq:rd0}
\end{align}
and quote the confidence intervals in the first row of
Tab.~\ref{tab:dcs-obs}. The ratio of the magnitudes of the DCS and CF
amplitudes is listed in the third row of this table.
Fig.~\ref{fig:dcs-obs} visualizes these confidence intervals and also
shows the prediction of Refs.~\cite{Bigi:1994aw, Rosner:2006bw,
  Gao:2006nb, Gao:2014ena}, which is the black dot corresponding to
$R(D^0) = 2 \tan^2\theta_C$ (where $\theta_C$ is the Cabibbo angle).
The result is quoted without uncertainty in these papers and
Refs.~\cite{Rosner:2006bw, Gao:2006nb, Gao:2014ena} argue that
corrections from SU(3) breaking to these relations are small.
Refs.~\cite{Gao:2006nb,Gao:2014ena} arrive at this conclusion by
calculating the amplitudes in QCD factorization \cite{Beneke:1999br,
  Beneke:2001ev}, which is a calculational method valid for values of
$m_c$ much larger than the hadronic scale governing the infrared
structure of the decays.  Our fit permits sizable corrections to $R(D^0)
= 2 \tan^2\theta_C$ from the SU(3) breaking contributions, so that future
measurements will give insight into the size of SU(3) breaking and the
viability of QCD factorization in charm physics.}  

Another test of doubly Cabibbo-suppressed contributions involves 
the decays $D_s^+\rightarrow K_{S,L} K^+$. We study 
\begin{align}
  R(D_s^+) &\equiv \frac{ \mathcal{B}(D_s^+\rightarrow K_S K^+) -
    \mathcal{B}(D_s^+\rightarrow K_L K^+) }{
    \mathcal{B}(D_s^+\rightarrow K_S K^+) + \mathcal{B}(D_s^+\rightarrow
    K_L K^+) } \label{eq:rds}
\end{align}
and predict the not yet measured observables $R(D_s^+)$ and
$\mathcal{B}(D_s^+\rightarrow K_L K^+)$, {see Tab.~~\ref{tab:dcs-obs}
  and Fig.~\ref{fig:dcs-obs}. }  Again comparing our result with the
prediction in Ref.~\cite{Gao:2014ena} {we find} much larger
uncertainties. {Thus also in $D_s^+\rightarrow K_{S,L} K^+$ future
  data will test the accuracy} of QCD factorization {assumed} in
Ref.~\cite{Gao:2014ena}.

\begin{table}[b]
\begin{center}
\begin{tabular}{c|c|c|c}
\hline \hline
Observable      &  $\pm1\sigma$ & $\pm2\sigma$ & $\pm3\sigma$ \\\hline\hline
$R(D^0) $ &  
        $0.09^{+0.04}_{-0.02}$  &    
        $0.09^{+0.07}_{-0.04}$  &     
        $0.09^{+0.09}_{-0.05}$  \\ 
$R(D_s^+) $ &  $0.11^{+0.04}_{-0.14}$ &    
               $0.11^{+0.06}_{-0.18}$ &    
               $0.11^{+0.06}_{-0.20}$  \\ 
$\mathcal{B}(D_s^+\rightarrow K_L K^+)$ &  $0.012^{+0.004}_{-0.001}$ &     
                                           $0.012^{+0.005}_{-0.002}$ &      
                                           $0.012^{+0.006}_{-0.002}$    \\ 
$\left\vert\frac{\mathcal{A}^{\mathrm{DCS}}(D^0\rightarrow K^0\pi^0)}{\mathcal{A}^{\mathrm{CF}}(D^0\rightarrow \bar{K}^0 \pi^0)} \right\vert$   &   
                        $0.05^{+0.02}_{-0.01}$ &
                        $0.05^{+0.03}_{-0.03}$ &
                        $0.05^{+0.04}_{-0.03}$ \\ 
$\left\vert\frac{\mathcal{A}^{\mathrm{DCS}}(D_s^+\rightarrow K^0K^+)}{\mathcal{A}^{\mathrm{CF}}(D_s^+\rightarrow \bar{K}^0 K^+)} \right\vert$  &  
                        $0.08^{+0.02}_{-0.06}$  &       
                        $0.08^{+0.03}_{-0.06}$  & 
                        $0.08^{+0.04}_{-0.07}$ \\ \hline\hline 
\end{tabular}
\caption{Fit results for several observables probing doubly
    Cabibbo-suppressed amplitudes. The corresponding plots are shown in
    Fig.~\ref{fig:dcs-obs}.\label{tab:dcs-obs}}
\end{center}
\end{table}

\section{Conclusion}
\begin{figure*}[t]
\begin{center}
\subfigure[\label{fig:plot-1d-01}]{
        \includegraphics[width=0.4\textwidth]{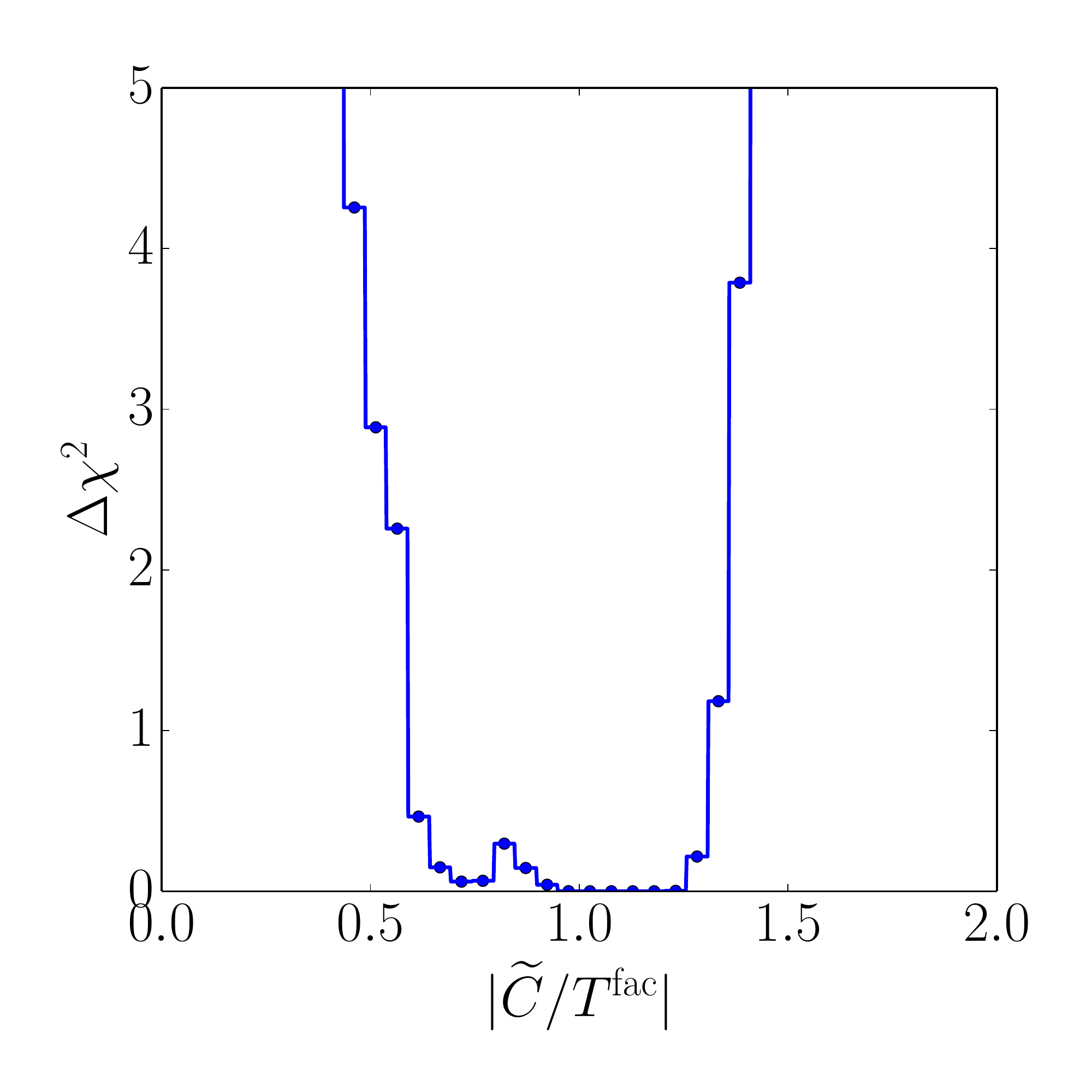}
        }
\hfill
\subfigure[\label{fig:plot-1d-02}]{
        \includegraphics[width=0.4\textwidth]{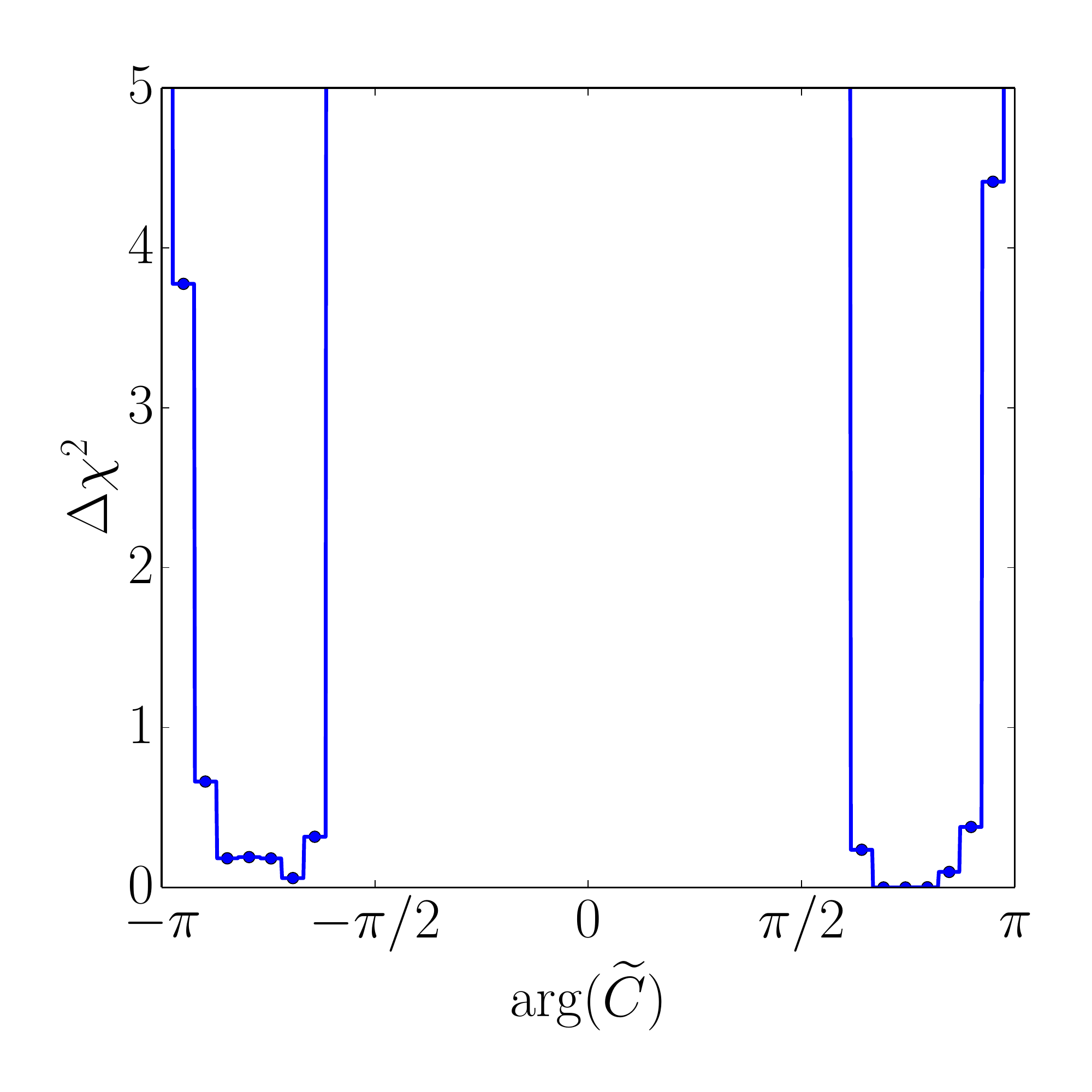}
        }
\hfill %
\subfigure[\label{fig:plot-1d-03}]{
        \includegraphics[width=0.4\textwidth]{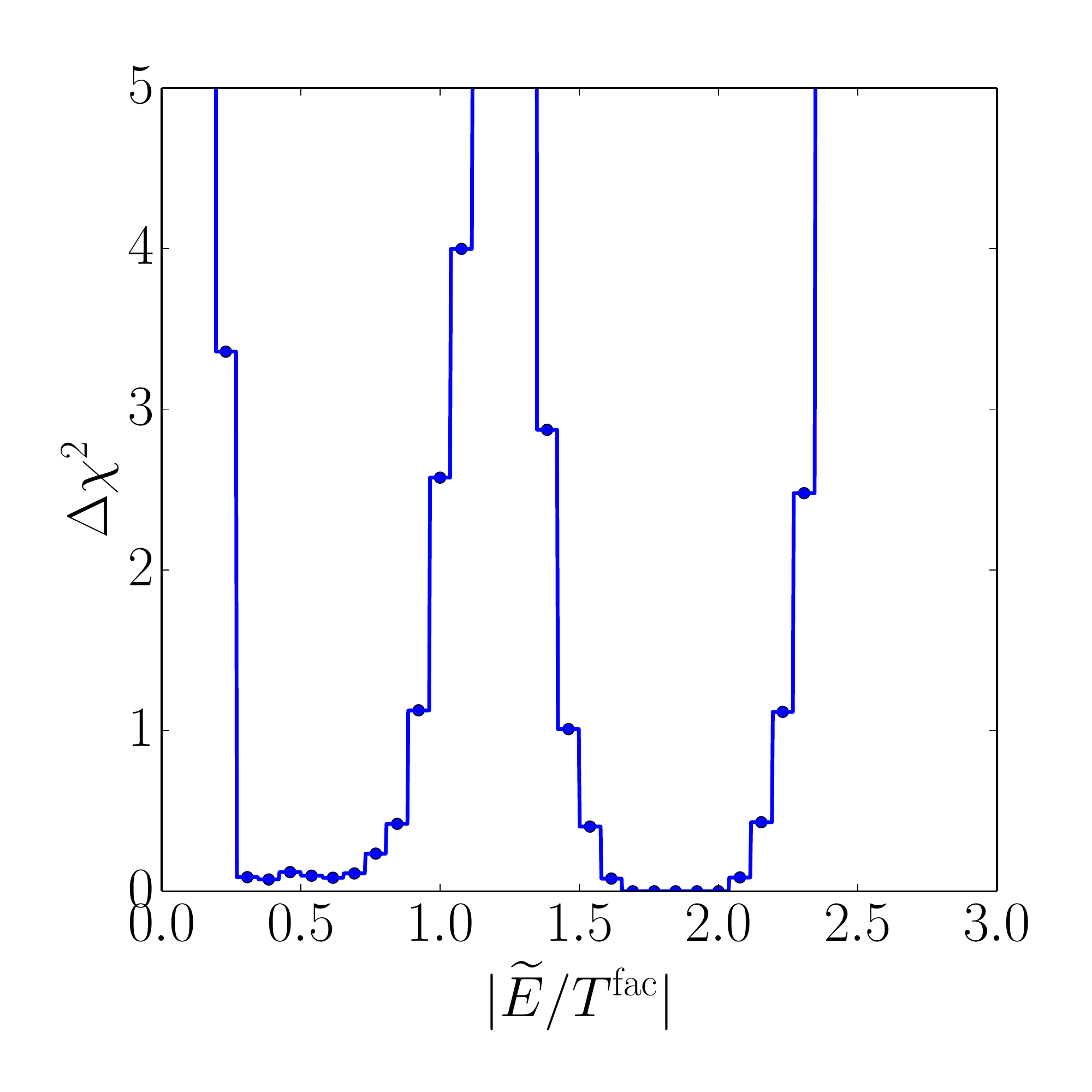}
}
\hfill
\subfigure[\label{fig:plot-1d-04}]{
        \includegraphics[width=0.4\textwidth]{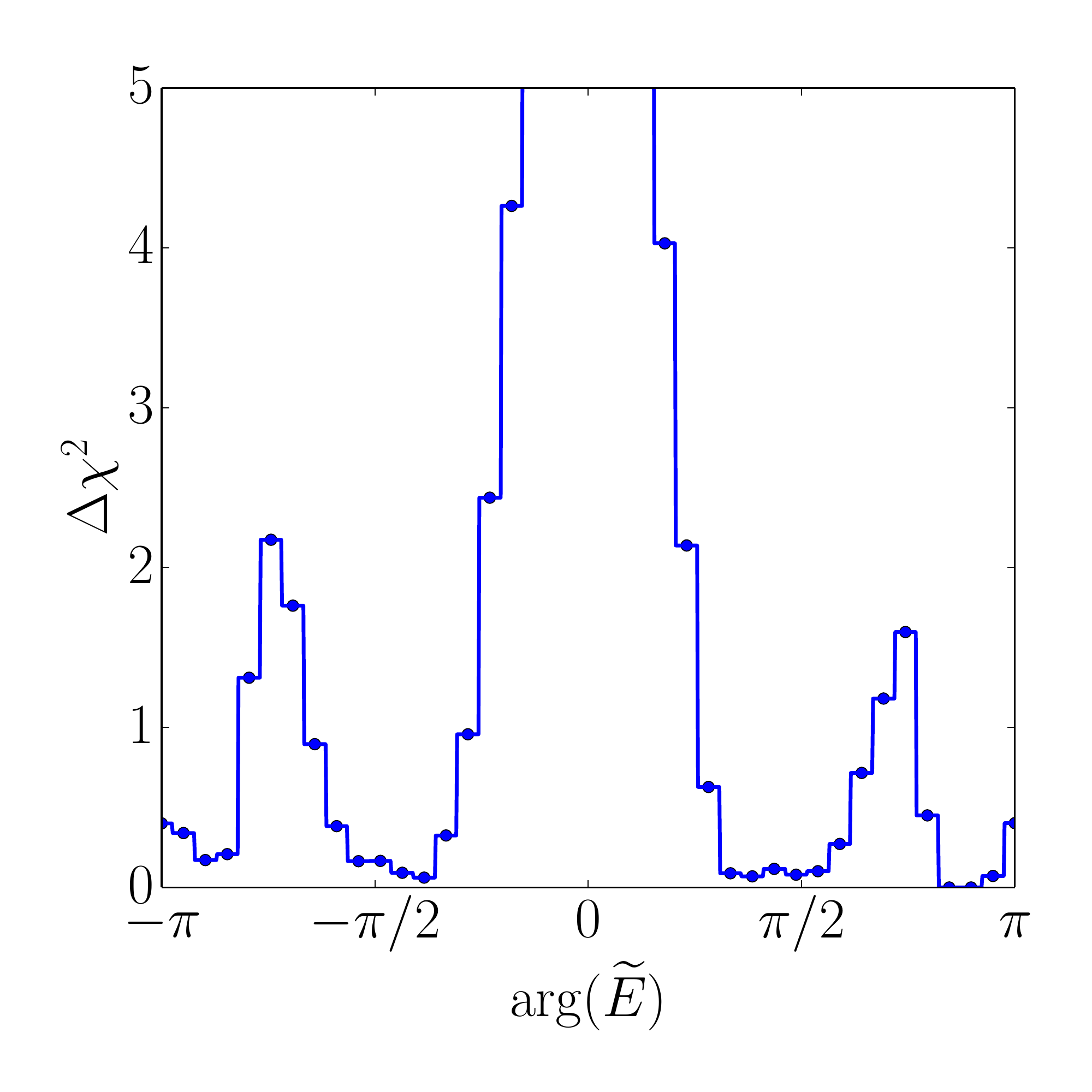}
}
\hfill %
\subfigure[\label{fig:plot-2d-12}]{
        \includegraphics[width=0.4\textwidth]{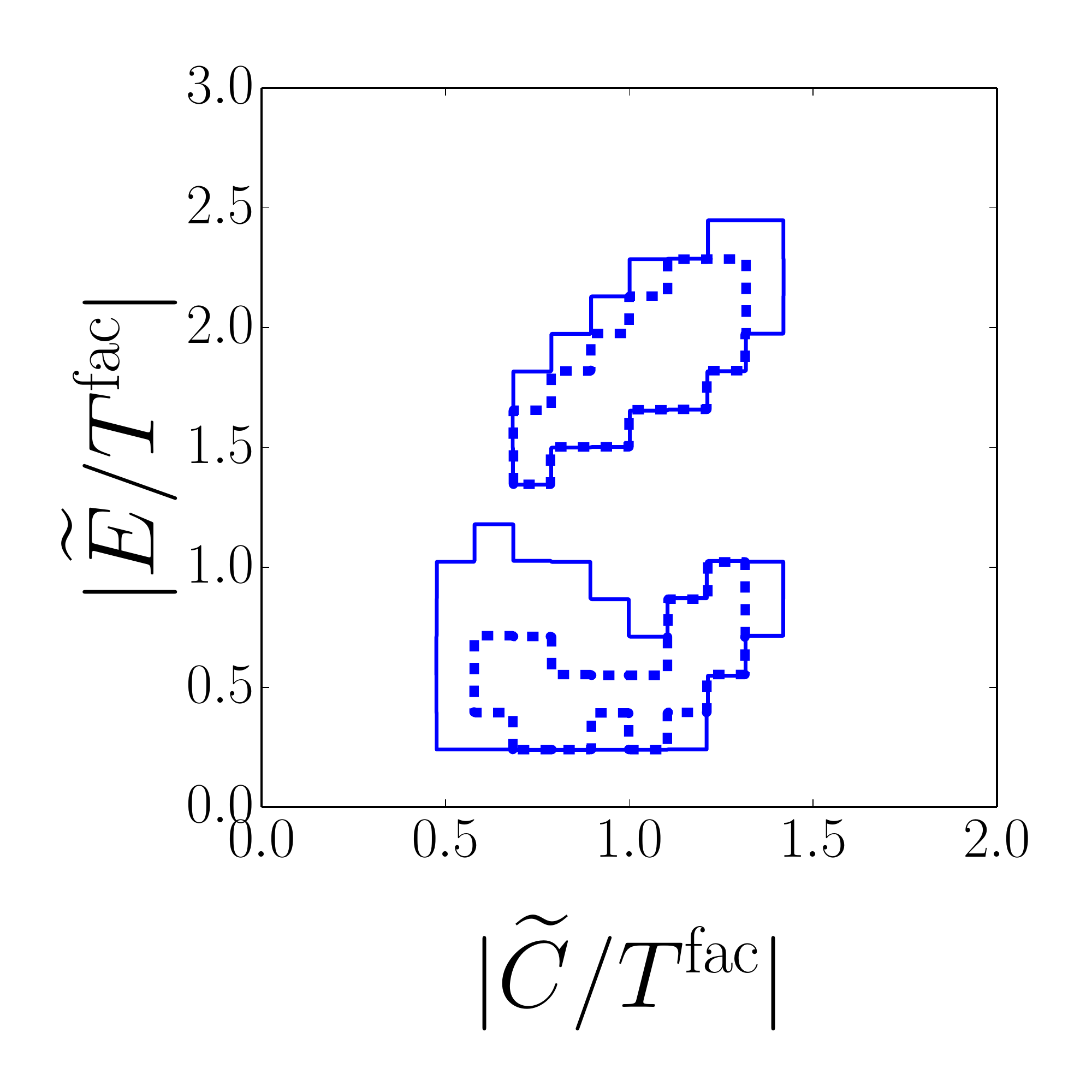}
}
\hfill %
\subfigure[\label{fig:plot-2d-02}]{
        \includegraphics[width=0.4\textwidth]{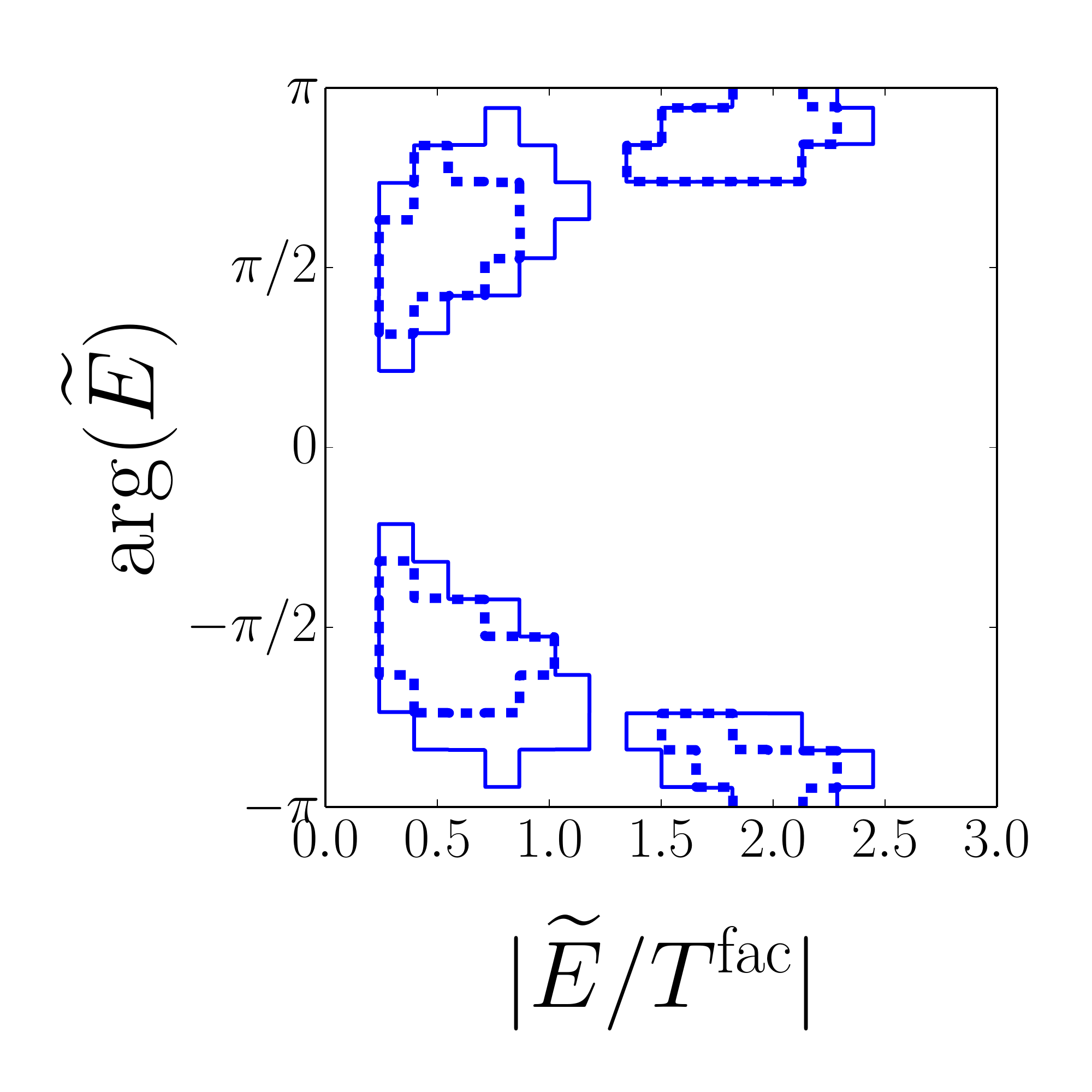}
}
\end{center}
\caption{SU(3)$_F$ limit topologies. In Figs.~{(e)} and (f) the
    dashed (solid) line denotes the 68\% (95\%) C.L. contour.
\label{fig:chi2-su3limit-topo}
}
\end{figure*}

\begin{figure*}[t]
\begin{center}
\subfigure[\label{fig:plot-1d-19}]{
        \includegraphics[width=0.43\textwidth]{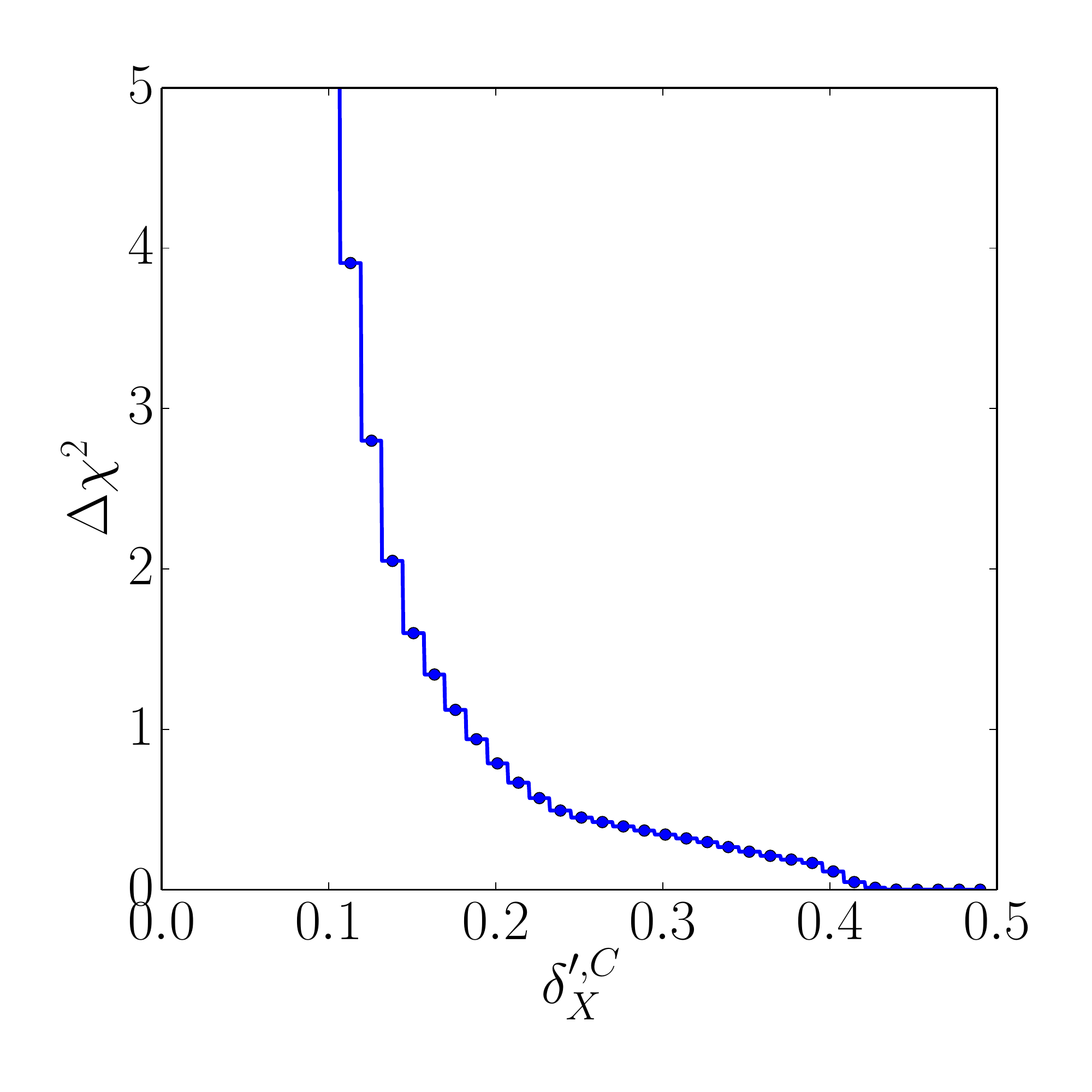}
}
\quad %
\subfigure[\label{fig:plot-1d-20}]{
        \includegraphics[width=0.43\textwidth]{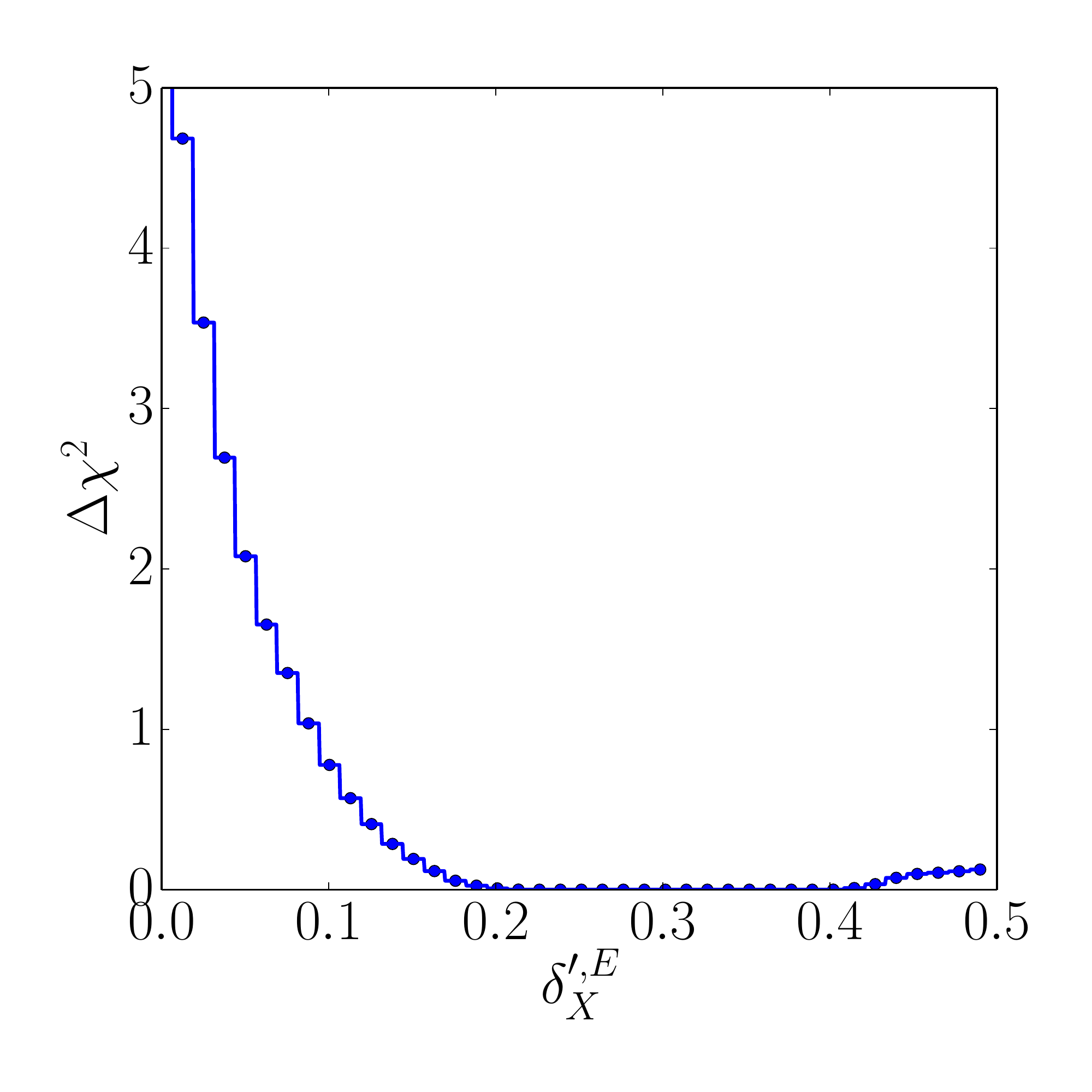}
}

\subfigure[\label{fig:plot-1d-21}]{
        \includegraphics[width=0.43\textwidth]{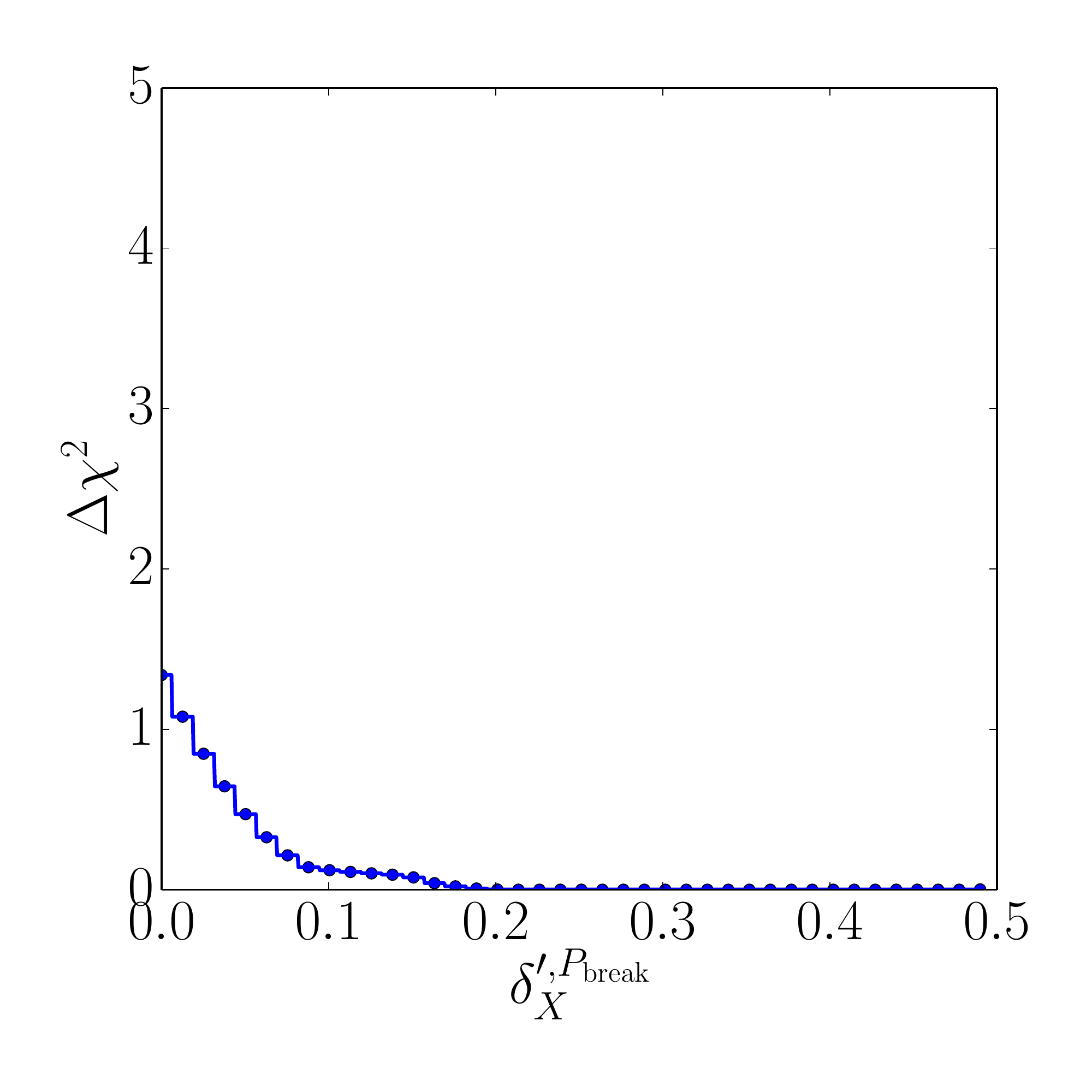}
}
\quad 
\subfigure[\label{fig:plot-1d-22}]{
        \includegraphics[width=0.43\textwidth]{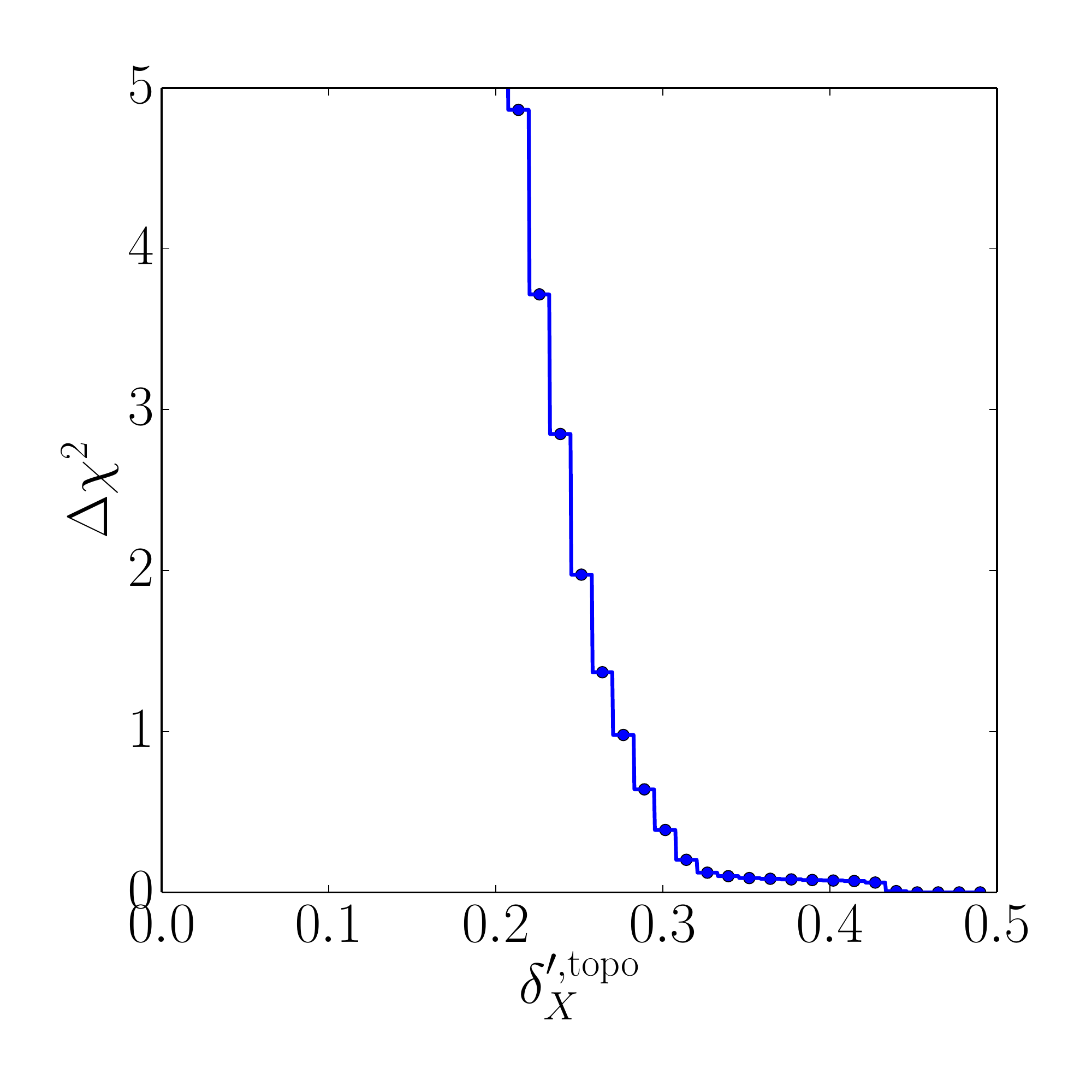}
}
\end{center}
\caption{ $\Delta \chi^2$ {profile of the parameters
    $\delta_X^{\prime,C,E,P_{\mathrm{break}}}$ measuring}
  SU(3)$_F$-breaking in {$C$, $E$, and $P_{\mathrm{break}}$} (a,b,c)
  and {of $\delta_X^{\prime,\mathrm{topo}}$ defined in
    \eq{eq:su3X-topo}, which quantifies the overall SU(3)$_F$-breaking}
  (d).  \label{fig:chi2-measures-SU3break} }
\end{figure*}

\begin{figure*}[t]
\begin{center}
\subfigure[\label{fig:plot-1d-0579}]{
        \includegraphics[width=0.31\textwidth]{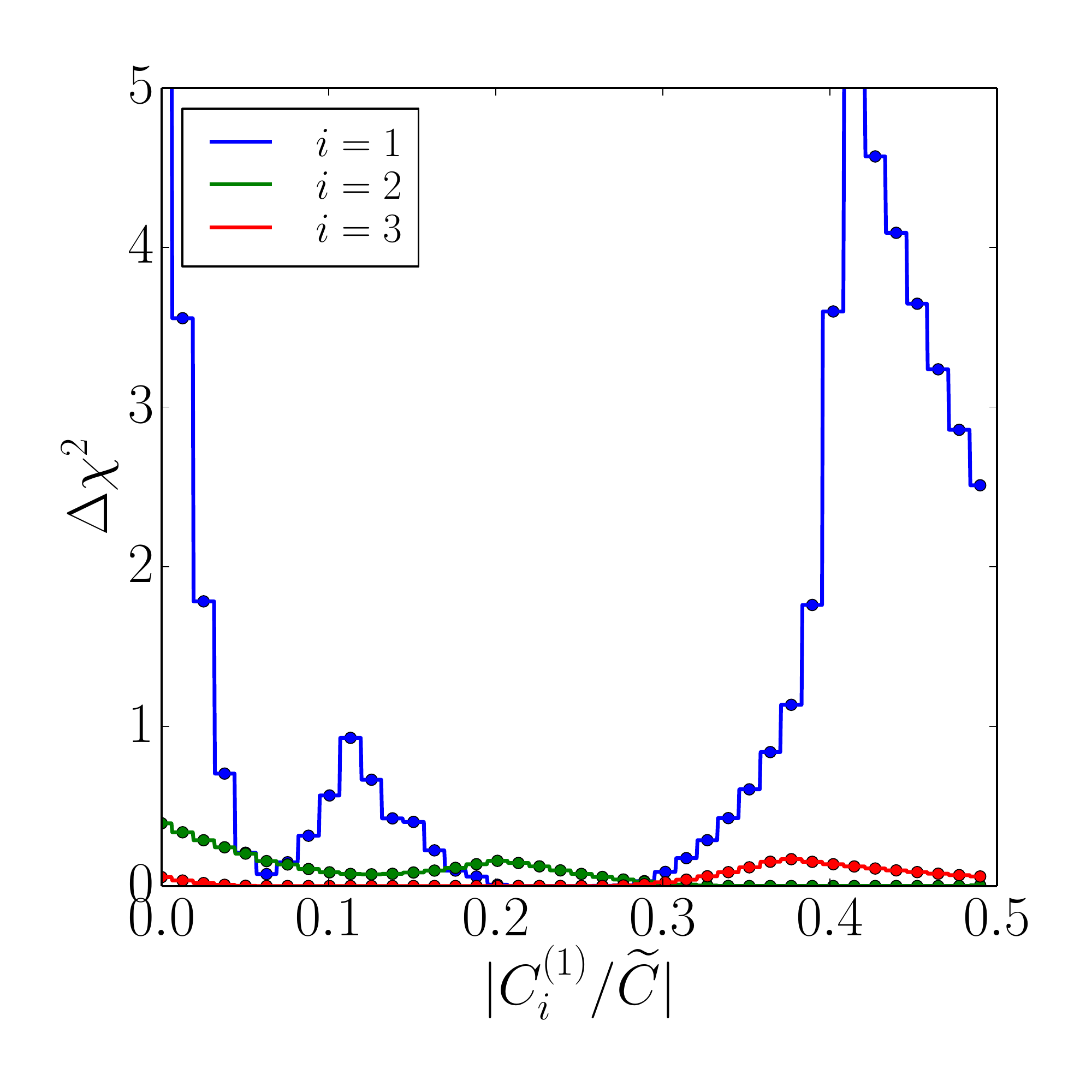}
}
\qquad
\subfigure[\label{fig:plot-1d-06810}]{
        \includegraphics[width=0.31\textwidth]{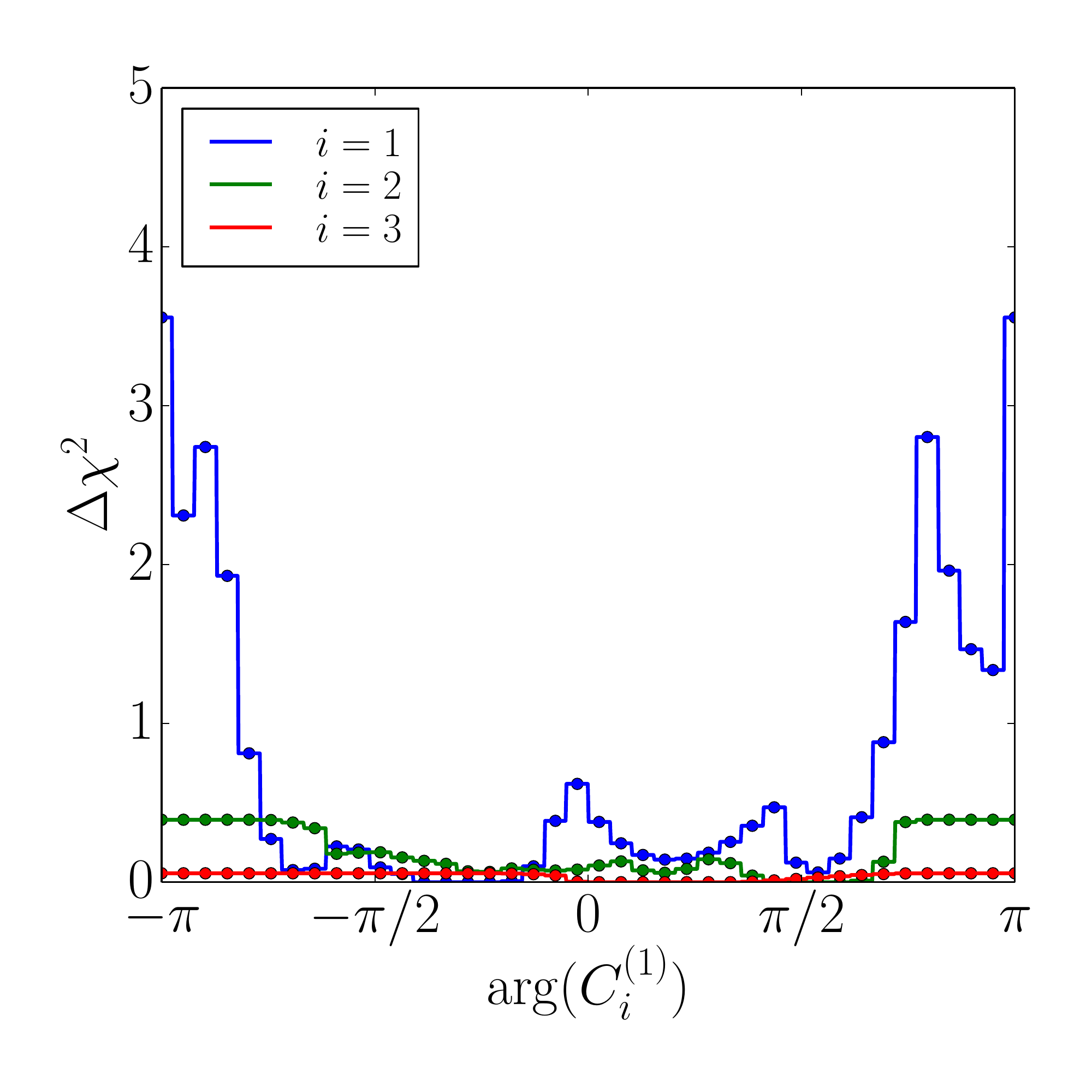}
}

\subfigure[\label{fig:plot-1d-111315}]{
        \includegraphics[width=0.31\textwidth]{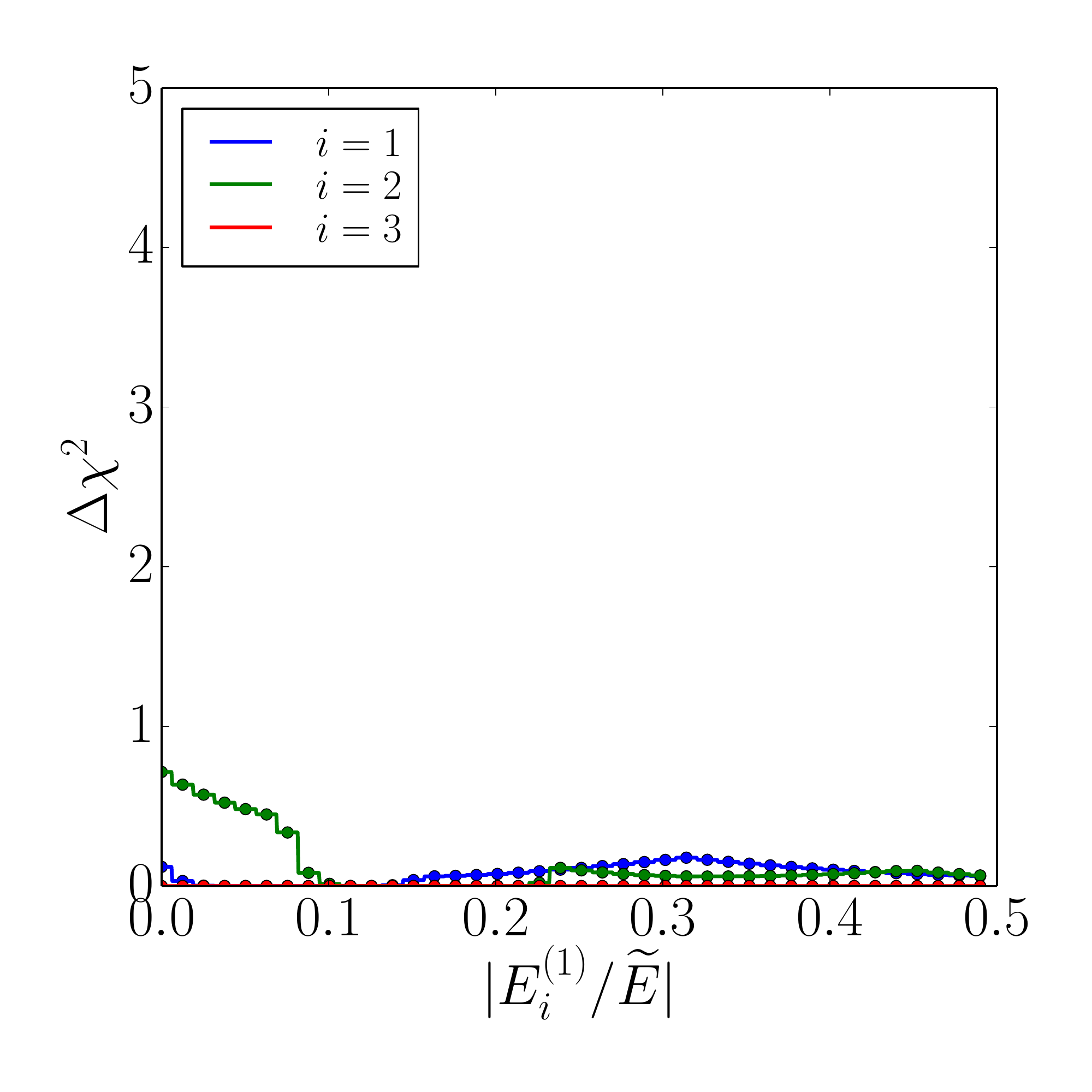}
}
\qquad
\subfigure[\label{fig:plot-1d-121416}]{
        \includegraphics[width=0.31\textwidth]{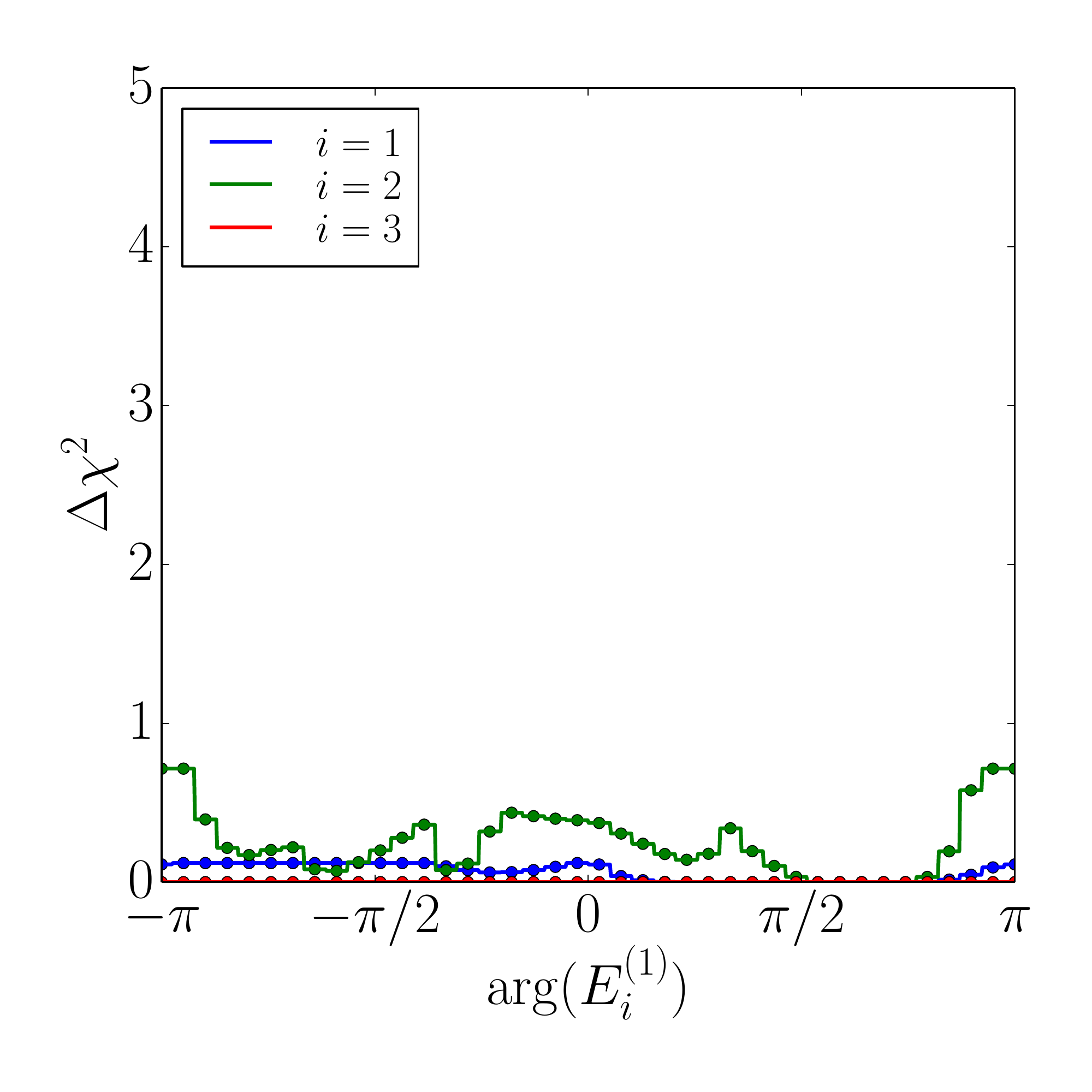}
}
\end{center}
\caption{
SU(3)$_F$-breaking color-suppressed tree (a,b) and exchange (c,d) topologies. 
\label{fig:chi2-su3break-topo}
}
\end{figure*}

\begin{figure*}[t]
\begin{center}
\subfigure[\label{fig:plot-1d-17}]{
        \includegraphics[width=0.31\textwidth]{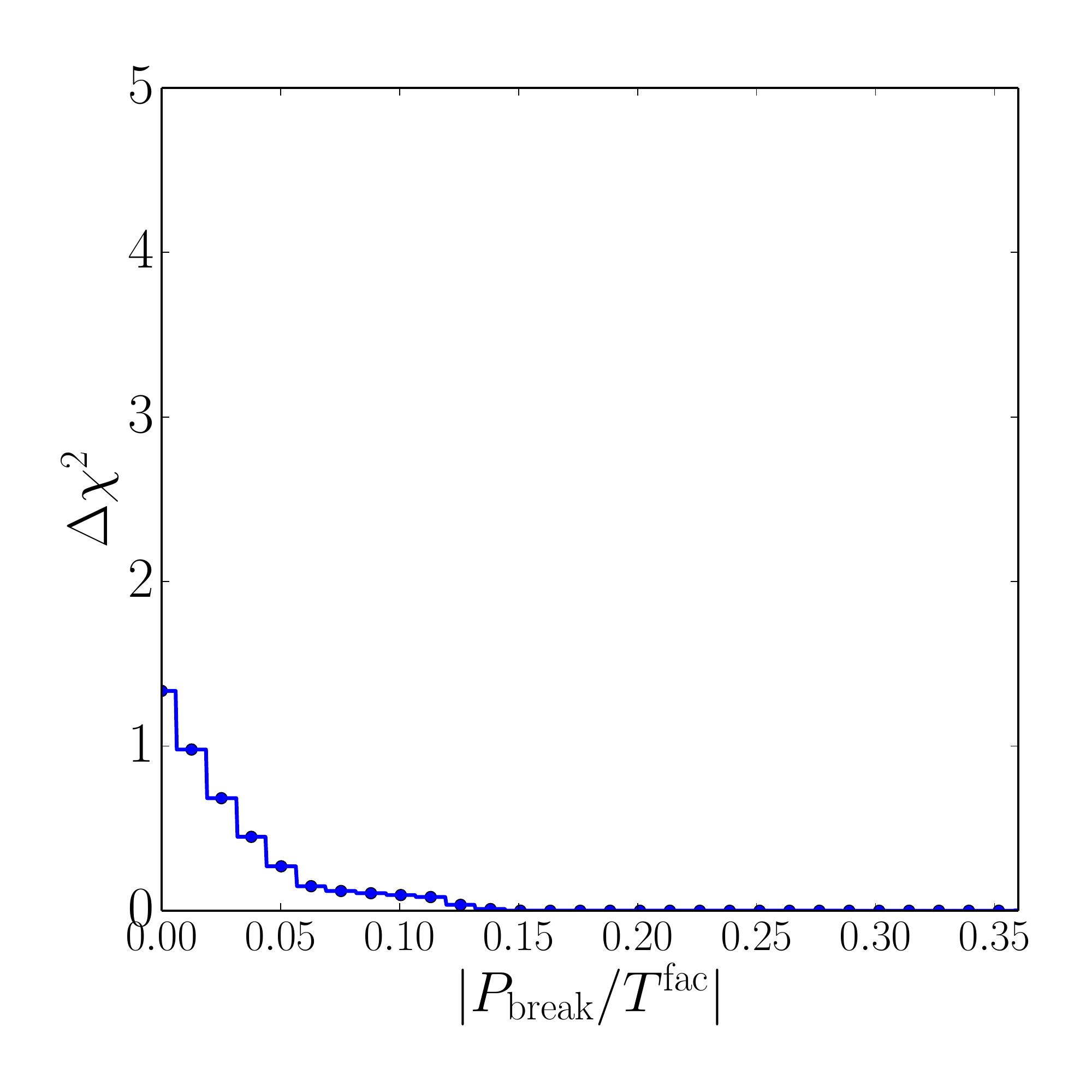}
}
\hfill %
\subfigure[\label{fig:plot-1d-18}]{
        \includegraphics[width=0.31\textwidth]{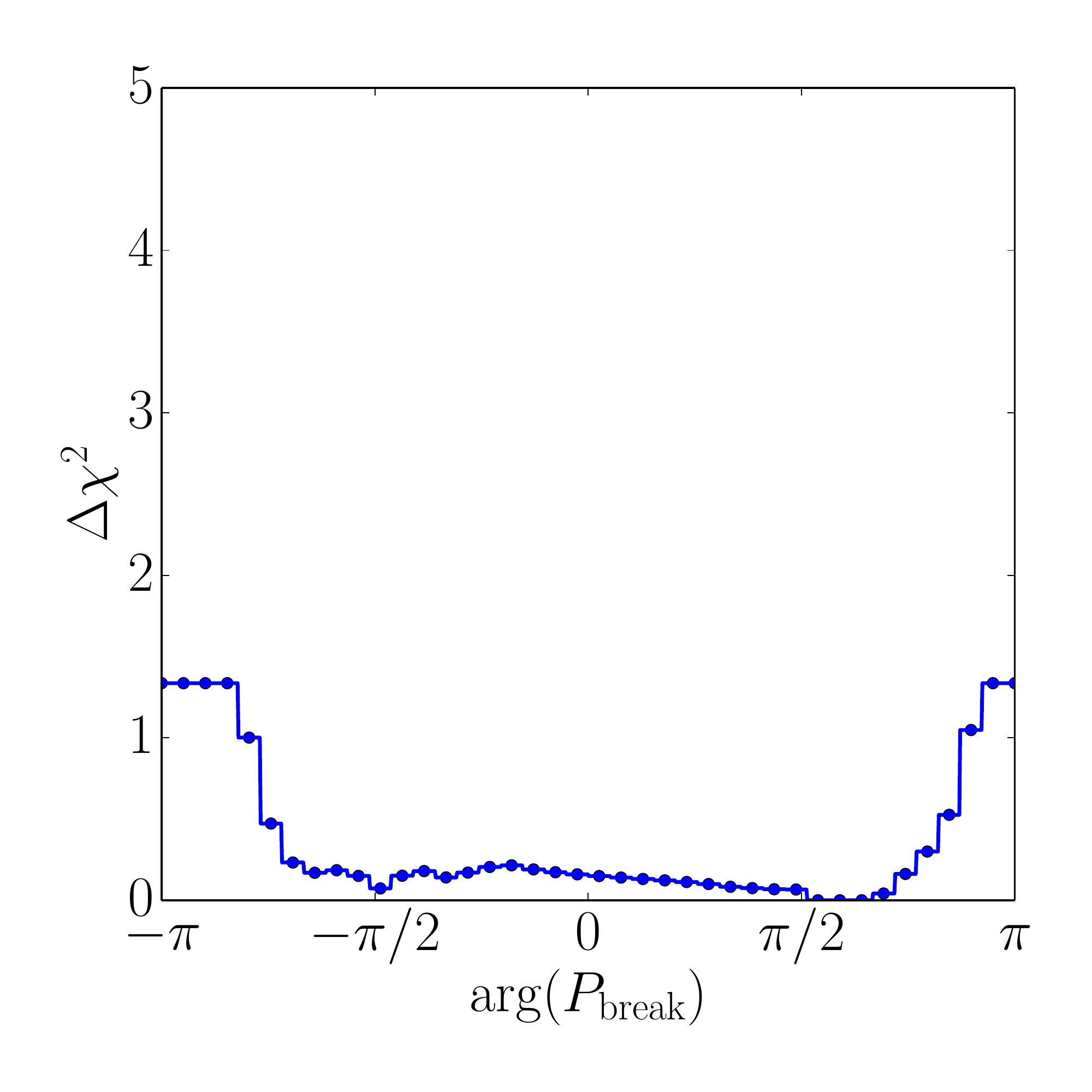}
}
\hfill
\subfigure[\label{fig:plot-2d-16}]{
        \includegraphics[width=0.31\textwidth]{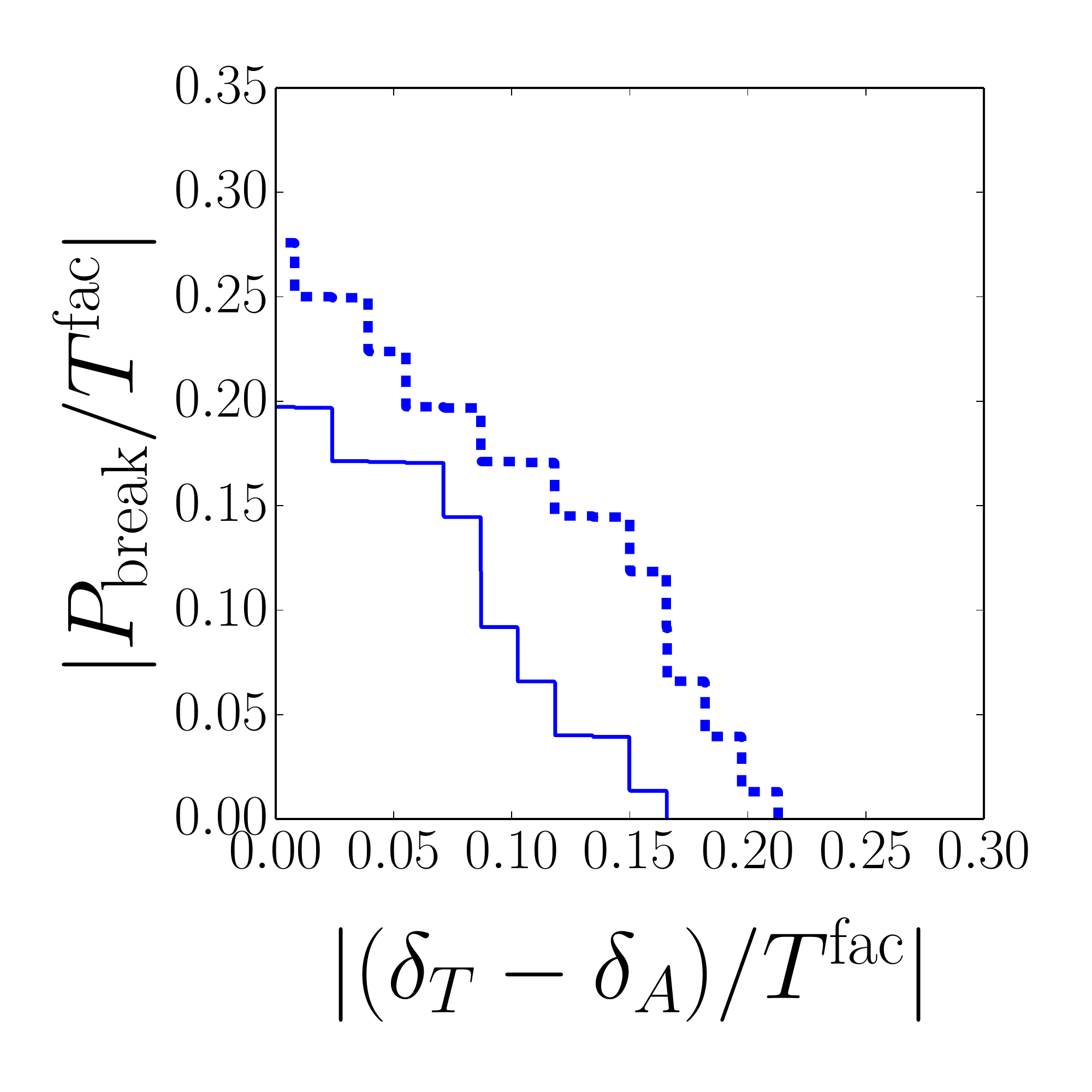}
}
\end{center}
\caption{
The broken penguin (a,b) and its correlation to parameters measuring 
the $1/N_c$ corrections (c).
In Fig.~(c) the dashed (solid) line denotes the 68\% (95\%) C.L. contour 
and the region to the right of the contours is allowed. 
\label{fig:chi2-pingu}
}
\end{figure*}

{We have studied the decay amplitudes of $D$ mesons into two
  pseudoscalar mesons with the topological-amplitude approach. To this
  end we have incorporated linear SU(3)$_F$ breaking into the method and
  have shown that the topological amplitude method can be mapped onto
  the standard decomposition of the decay amplitudes in terms of reduced
  amplitudes characterized by SU(3)$_F$ representations. Unlike plain
  SU(3)$_F$ analyses the topological-amplitude method permits the use of
  a $1/N_c$ expansion to calculate the factorizable tree and
  annihilation amplitudes in terms of form factors and decay constants,
  up to corrections of order $1/N_c^2$. This additional theoretical
  input has led us to a new sum rule between the branching fractions of
  $D^+\rightarrow K_SK^+$, $D_s^+\rightarrow K_S\pi^+$ and
  $D^+\rightarrow K^+\pi^0$. This sum rule correlates the
  non-factorizable $1/N_c^2$ terms with the penguin amplitude
  $P_{\mathrm{break}}$. The latter quantity is of prime interest to
  understand the dynamics of flavour-changing neutral current
  transitions in the charm sector, because $P_{\mathrm{break}}$ is
  suppressed by the GIM mechanism and vanishes in the limit $m_s=m_d$.

We have then performed a global fit using all available branching
ratios and the experimental information on the strong phase difference
$\delta_{K^+\pi^-}$. In our analysis we have included the information on
correlations between experimental errors. It is possible to find a
perfect fit, with a large parameter region satisfying $\chi^2=0$. This
means that current data comply with i) the Grossman-Robinson sum
rules \cite{Grossman:2012ry}, ii) our chosen upper bound of 50\% on
SU(3)$_F$ breaking, and iii) our assumption that the $1/N_c^2$
corrections to the factorizable amplitudes are smaller than 15\% of the
factorized tree amplitude. The main phenomenological results of our
paper are various likelihood ratio tests addressing the sizes of the
topological amplitudes and their SU(3)$_F$ breaking
(Tab.~\ref{tab:likelihoodratiotests} and 
Figs.~\ref{fig:chi2-su3limit-topo}--\ref{fig:chi2-pingu}).
Importantly, }
we find that there is no evidence for an enhanced broken penguin.  The
hypothesis $P_{\mathrm{break}} = 0$ is rejected at below $1\sigma$
only,~i.e.,~insignificantly.  Improvements of
$\mathcal{B}(D^+\rightarrow K_SK^+)$, $\mathcal{B}(D_s^+\rightarrow
K_S\pi^+)$, $\mathcal{B}(D^+\rightarrow K^+\pi^0)$ and especially the
form factor $F_0^{K\pi}(m^2_{D_{(s)}})$ could advance our knowledge of
the GIM mechanism in charm by pinning down the proportions of broken
penguin and $1/N_c^2$ corrections.  The current status is summarized in
Fig.~\ref{fig:plot-2d-16}.  While the SCS branching ratios
$\mathcal{B}(D^+\rightarrow K_SK^+)$ and $\mathcal{B}(D_s^+\rightarrow
K_S\pi^+)$ are known at a precision of $\lesssim 6\%$ the relative
uncertainty of the DCS branching ratio $\mathcal{B}(D^+\rightarrow
K^+\pi^0)$ is about $\sim 14\%$ and {leaves} room for improvement. As the
latter is the only charm decay into kaons and pions which depends on
factorizable contributions only, it is very important to improve its
measurement.  With a simultaneously improved $F_0^{K\pi}(m^2_{D_{(s)}})$
the branching ratio $\mathcal{B}(D^+\rightarrow K^+\pi^0)$ serves as a
test of factorization in charm decays.
 
We observe a slightly larger SU(3)$_F$ breaking in color-suppressed tree
than in exchange diagrams. {In no channel more than $\sim 30\%$}
SU(3)$_F$ breaking is needed to describe the data {(not considering
  $D^0\to K_S K_S$, which is forbidden in the SU(3)$_F$ limit)};
{this finding agrees} with the plain SU(3)$_F$ analysis of
Ref.~\cite{Hiller:2012xm}.  However, as a matter of principle one cannot
decide whether or not the actual SU(3)$_F$ breaking is larger than
this. This can potentially only be achieved by future QCD calculations
on the lattice \cite{Hansen:2012tf}.  In the data, there is no
indication of this to be the case.

\begin{figure*}[t]
\begin{center}
\subfigure[\label{fig:plot-1d-29}]{
        \includegraphics[width=0.43\textwidth]{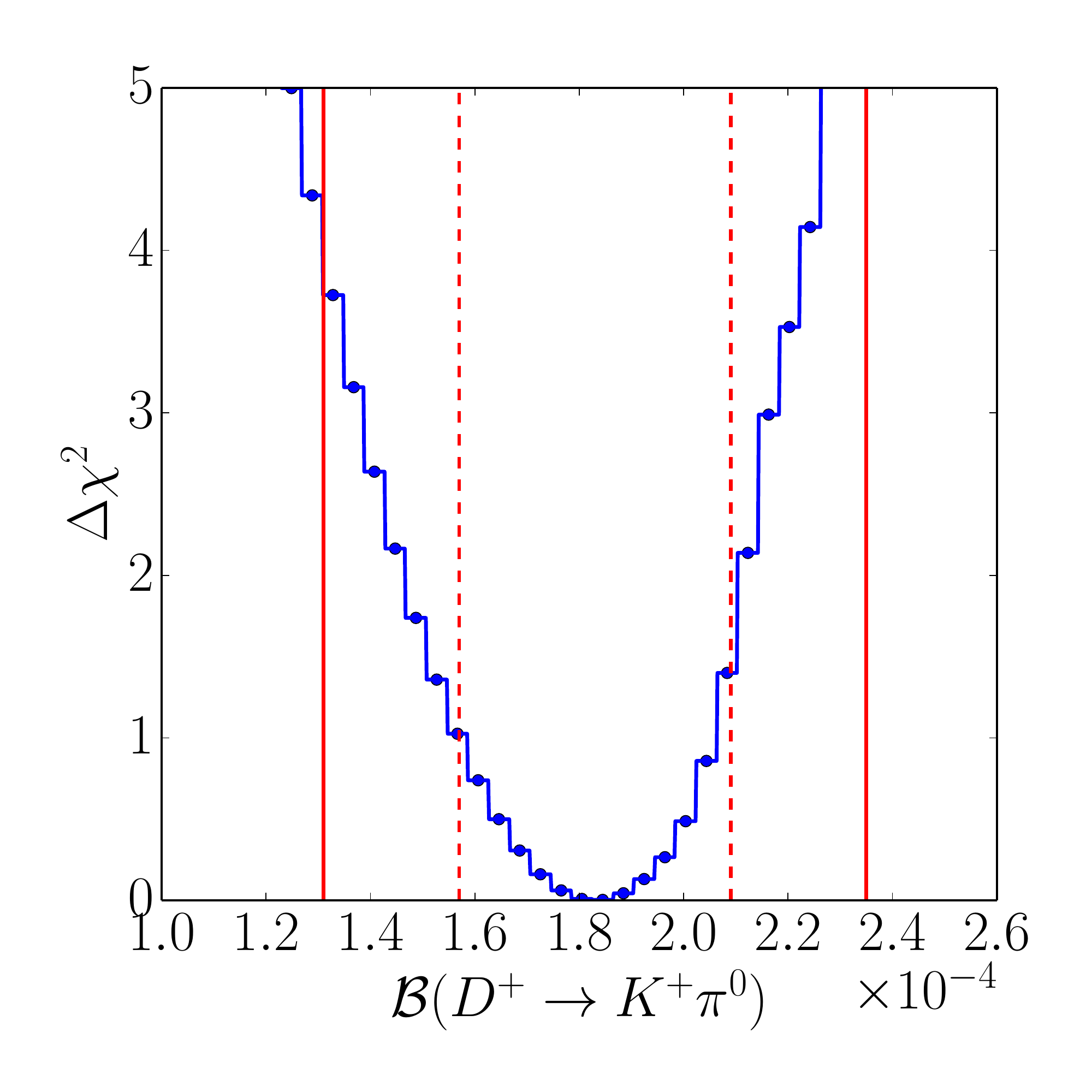}
}
\quad 
\subfigure[\label{fig:plot-2d-19}]{
        \includegraphics[width=0.43\textwidth]{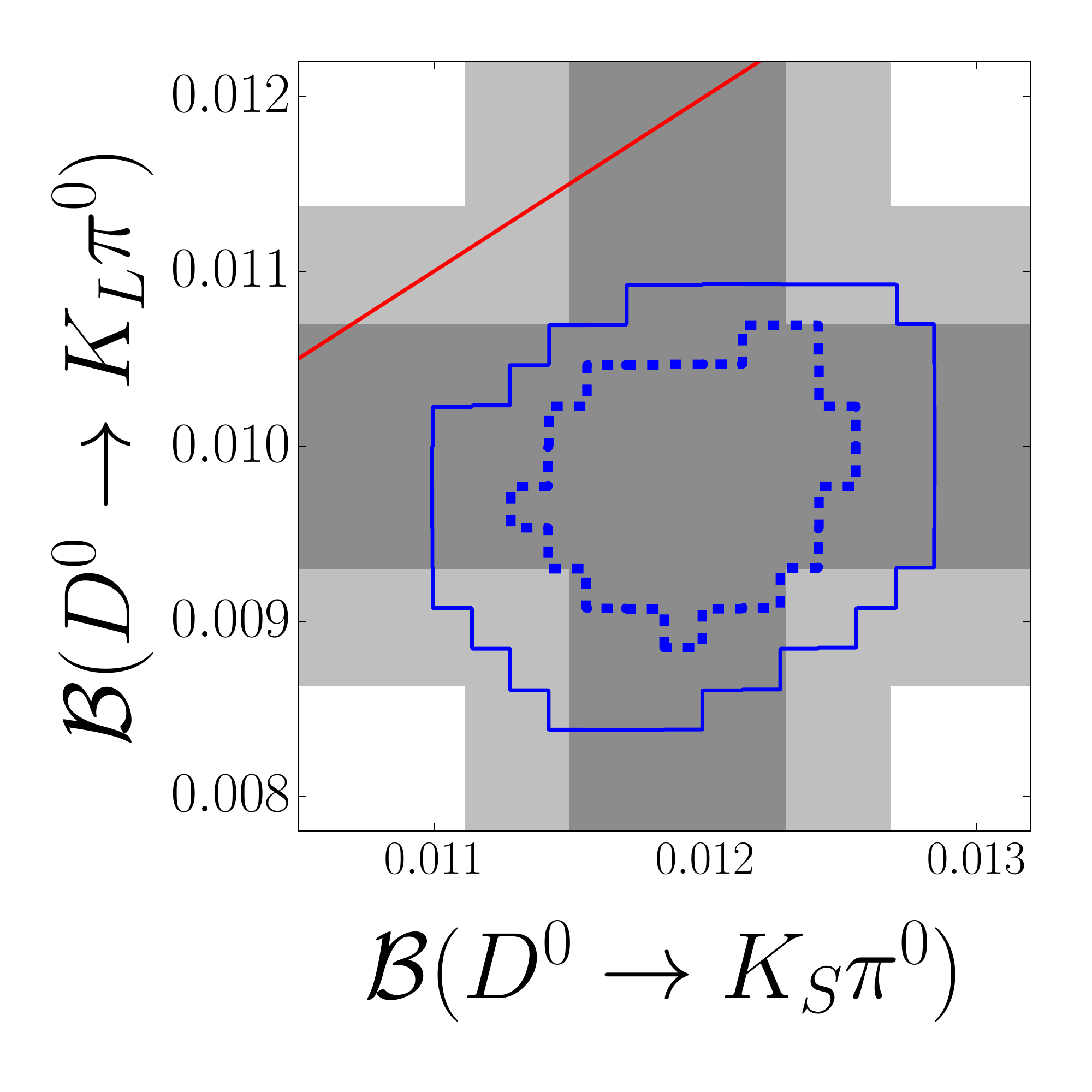}
}
\end{center}
\caption{ $\Delta \chi^2$ {profile of} $\mathcal{B}(B^+\rightarrow
  K^+\pi^0)$ (a) and correlation between 
  $\mathcal{B}(D^0\rightarrow K_L\pi^0)$ and $\mathcal{B}(D^0\rightarrow
  K_S\pi^0)$~(b). In Fig.~(a) the red dashed (solid) line 
  {indicates the $1\sigma$ ($2\sigma$) experimental} error.
  In Fig.~(b) the
  dashed (solid) lines are the 68\% (95\%) C.L.\ contours of our fit and
  the dark (light) gray {shading denotes the  68\% (95\%) C.L.\ 
  region of the measurements}. Here, the solid red line {corresponds to}
  $\mathcal{B}(D^0\rightarrow K_L\pi^0) =
  \mathcal{B}(D^0\rightarrow K_S\pi^0)$.
\label{fig:observables}
}
\end{figure*}

{With our topological-amplitude fit it is further possible to make
  predictions for branching fractions which can be probed by future
  measurements.  Despite our conservative ranges for the SU(3)$_F$
  breaking parameters, we find a correlation between
  $\mathcal{B}(D^0\rightarrow K_L\pi^0)$ and $\mathcal{B}(D^0\rightarrow
  K_S\pi^0)$ probing the doubly Cabibbo-suppressed contributions to
  these modes: Fig.~\ref{fig:observables} entails the prediction
  $\mathcal{B}(D^0\rightarrow K_L\pi^0)<\mathcal{B}(D^0\rightarrow
  K_S\pi^0)$ at more than~$4\sigma$.}

\begin{acknowledgments}
  We thank Philipp Frings, Gudrun Hiller and Martin Jung for useful
  discussions.  The fits are performed using the \texttt{python} version
  of the software package \texttt{myFitter}~\cite{Wiebusch:2012en}.  StS
  thanks Martin Wiebusch for \texttt{myFitter} support and the provision
  of \texttt{myFitter-python-0.2-beta}.  The Feynman diagrams are drawn
  using \texttt{Jaxodraw}
  \cite{Binosi:2003yf,Vermaseren:1994je}. Parts of the computations were performed 
  on the NEMO cluster of Baden-W\"urttemberg (framework program bwHPC). The presented work 
  is supported by 
  BMBF under contract no.~05H12VKF.
\end{acknowledgments}

\clearpage

\appendix

\section{Input Data}
\label{eq:Inputdata}

We give the input data used in the fits, including the correlation
coefficients, in
Tabs.~\ref{tab:branchingratios}--\ref{tab:inputformfactor}.  For details
on the input values for the form factors see Appendices
\ref{sec:fac-tree-amps} and \ref{sec:fac-anni}.

\begin{table}[t]
{\scriptsize
\begin{center}
\begin{tabular}{ccr}
\hline\hline
Observable            &      Measurement &  References  \\\hline
\multicolumn{3}{c}{SCS branching ratios} \\\hline
$\mathcal{B}(D^0\rightarrow K^+ K^-  )   $   & $   \left(3.96\pm  0.08\right)\cdot10^{-3} $ &  \cite{Beringer:1900zz}  \\
$\mathcal{B}(D^0\rightarrow \pi^+ \pi^-) $   & $   \left( 1.402\pm  0.026\right)\cdot10^{-3} $ & \cite{Beringer:1900zz}  \\
$\mathcal{B}(D^0\rightarrow K_S K_S  )   $   & $   \left(0.17\pm  0.04\right)\cdot10^{-3} $ & \cite{Beringer:1900zz} \\
$\mathcal{B}(D^0\rightarrow \pi^0 \pi^0) $   & $   \left(0.820\pm  0.035\right)\cdot10^{-3} $ &  \cite{Beringer:1900zz}  \\
$\mathcal{B}(D^+\rightarrow \pi^0 \pi^+) $   & $   \left(1.19\pm  0.06\right)\cdot10^{-3} $ &  \cite{Beringer:1900zz} \\
$\mathcal{B}(D^+\rightarrow K_S K^+  )   $   & $   \left(2.83\pm  0.16\right)\cdot10^{-3} $ &  \cite{Beringer:1900zz} \\
$\mathcal{B}(D_s^+\rightarrow K_S \pi^+ ) / \mathcal{B}(D_s^+\rightarrow K_S K^+)$ & 
	$   \left(8.12 \pm 0.28\right)\cdot10^{-2} $ &  \cite{Beringer:1900zz} \\
$\mathcal{B}(D_s^+\rightarrow K^+ \pi^0 ) / \mathcal{B}(D_s^+\rightarrow K_S K^+) $  & 
	$   \left(4.2  \pm 1.4 \right) \cdot10^{-2} $ & \cite{Beringer:1900zz} \\\hline
\multicolumn{3}{c}{ CF branching ratios} \\\hline
$\mathcal{B}(D^0\rightarrow K^- \pi^+ )  $   & $   \left(3.88\pm  0.05\right)\cdot10^{-2} $  &  \cite{Beringer:1900zz}  \\
$\mathcal{B}(D^0\rightarrow K_S \pi^0 )  $   & $   \left(1.19\pm  0.04\right)\cdot10^{-2}  $ &  \cite{Beringer:1900zz} \\
$\mathcal{B}(D^0\rightarrow K_L \pi^0 )  $   & $   \left(1.00\pm  0.07\right)\cdot10^{-2}  $ &   \cite{Beringer:1900zz} \\
$\mathcal{B}(D^+\rightarrow K_S \pi^+ )  $   & $   \left(1.47\pm  0.07\right)\cdot10^{-2}  $ &  \cite{Beringer:1900zz} \\
$\mathcal{B}(D^+\rightarrow K_L \pi^+ )  $   & $   \left(1.46\pm  0.05\right)\cdot10^{-2}  $ &  \cite{Beringer:1900zz} \\
$\mathcal{B}(D_s^+\rightarrow K_S K^+  )   $   & $   \left(1.50\pm 0.05\right)\cdot10^{-2}  $ & $^\dagger$\cite{Zupanc:2013byn,Onyisi:2013bjt}  \\\hline
\multicolumn{3}{c}{DCS branching ratios} \\\hline
$\mathcal{B}(D^0\rightarrow K^+ \pi^- ) / \mathcal{B}(D^0\rightarrow K^- \pi^+ ) $   & $   0.00349\pm 0.00004  $ &  \cite{Amhis:2012bh}  \\
$\mathcal{B}(D^+\rightarrow K^+ \pi^0 )  $   & $   \left(1.83\pm  0.26\right)\cdot10^{-4}  $ &  \cite{Beringer:1900zz}  \\\hline\hline
\multicolumn{3}{c}{$K^+ \pi^-$ strong phase difference} \\\hline
$\delta_{K\pi}$   &  $(6.45 \pm 10.65)^{\circ}$    &    $^\ddagger$\cite{Amhis:2012bh}    \\\hline\hline
\end{tabular}
\caption{Input data for charm meson branching ratios and the
  strong phase difference $\delta_{K\pi}$ used in our fit. Note that as
  we {incorporate} the correlations {reported in 
  Ref.}~\cite{Beringer:1900zz}, for
  consistency we do not take into account the
  {experimental} updates of the following branching fractions: 
  $\mathcal{B}(D^0\rightarrow K^-
  \pi^+)$ \cite{Bonvicini:2013vxi} and $\mathcal{B}(D^+\rightarrow K_S
  \pi^+)$ \cite{Bonvicini:2013vxi}.  For the correlation coefficients
  see Tabs.~\ref{tab:correlations-D0}, \ref{tab:correlations-Dp} and
  \ref{tab:correlations-hfag}.  Note that $\mathcal{B}(D_s^+\rightarrow
  K^+\pi^0)$ and $\mathcal{B}(D_s^+\rightarrow K_S\pi^+)$ are not part
  of the PDG fit,~i.e.,~there are no correlation coefficients
  given for these decay modes. We therefore have no correlation matrix
  for $D_s^+$ decays.  The value for $\mathcal{B}(D^0\rightarrow K^+
  \pi^- ) / \mathcal{B}(D^0\rightarrow K^- \pi^+ ) $ is taken from the
  Heavy Flavor Averaging Group (HFAG) in order to take its correlation
  with $\delta_{K\pi}$ into account, see
  Tab.~\ref{tab:correlations-hfag}.  $^\dagger$Our
  average. $^\ddagger$Our symmetrization of uncertainties.
\label{tab:branchingratios} }
\end{center}
}
\end{table}

\begin{table}[t]
\begin{center}
\begin{tabular}{cccccccc}
\hline \hline
 &
$\mathcal{B}_{D^0}^{K^+ K^-}$     &   
$\mathcal{B}_{D^0}^{\pi^+ \pi^-}$ &   
$\mathcal{B}_{D^0}^{K_S K_S}$     &   
$\mathcal{B}_{D^0}^{\pi^0 \pi^0}$ &   
$\mathcal{B}_{D^0}^{K^- \pi^+}$   &   
$\mathcal{B}_{D^0}^{K_S \pi^0}$     \\\hline 
$\mathcal{B}_{D^0}^{K^+ K^-}$     & 1.00  & 0.38  &  0.03 &  0.09 & 0.60  & 0.21     \\
$\mathcal{B}_{D^0}^{\pi^+ \pi^-}$ & 0.38  & 1.00  &  0.03 &  0.09 & 0.62  & 0.22     \\
$\mathcal{B}_{D^0}^{K_S K_S}$     & 0.03  & 0.03  & 1.00  &  0.01 & 0.05  & 0.03     \\
$\mathcal{B}_{D^0}^{\pi^0 \pi^0}$ & 0.09  & 0.09  & 0.01  & 1.00  & 0.14  & 0.05     \\
$\mathcal{B}_{D^0}^{K^- \pi^+}$   & 0.60  & 0.62  & 0.05  &  0.14 & 1.00  & 0.35     \\
$\mathcal{B}_{D^0}^{K_S \pi^0}$   & 0.21  & 0.22  & 0.03  &  0.05 & 0.35  & 1.00     \\\hline\hline 
\end{tabular}
\caption{Correlation coefficients for $D^0$ {branching ratios} 
\cite{Beringer:1900zz} used in our fit. 
We abbreviate $\mathcal{B}_i^f \equiv \mathcal{B}(i\rightarrow f)$.
Note that $\mathcal{B}(D^0\rightarrow K_L \pi^0)$ is not part of the 
PDG fit and is used without correlations to the other modes.
\label{tab:correlations-D0} }
\end{center}
\end{table}

\begin{table}[t]
\begin{center}
\begin{tabular}{cccccc}
\hline \hline
 &
$\mathcal{B}_{D^+}^{K_S K^+}$     &   
$\mathcal{B}_{D^+}^{K_S \pi^+}$   &   
$\mathcal{B}_{D^+}^{K^+ \pi^0 }$   \\\hline 
$\mathcal{B}_{D^+}^{K_S K^+}$     & 1.00  & 0.75  & 0.05 \\
$\mathcal{B}_{D^+}^{K_S \pi^+}$   & 0.75  & 1.00  & 0.06 \\
$\mathcal{B}_{D^+}^{K^+ \pi^0 }$  & 0.05  & 0.06  & 1.00  \\\hline\hline
\end{tabular}
\caption{Correlation coefficients for $D^+$ {branching ratios} \cite{Beringer:1900zz} used in our fit, 
see the caption of Tab.~\ref{tab:correlations-D0} for the notation used. 
Note that $\mathcal{B}(D^+\rightarrow K_L \pi^+)$ and $\mathcal{B}(D^+\rightarrow \pi^0\pi^+)$ are not part of the PDG fit and are 
used without correlations to the other modes.
\label{tab:correlations-Dp}}
\end{center}
\end{table}

\begin{table}[t]
\begin{center}
\begin{tabular}{ccc}
\hline \hline
&
$\delta_{K\pi}$     &
$\frac{\mathcal{B}(D^0\rightarrow K^+ \pi^- ) }{ \mathcal{B}(D^0\rightarrow K^- \pi^+ ) }$    
  \\\hline 
$\delta_{K\pi}$    & 1.000   &  0.404  \\
$\frac{\mathcal{B}(D^0\rightarrow K^+ \pi^- ) }{ \mathcal{B}(D^0\rightarrow K^- \pi^+ ) }$     & 0.404  & 1.000  \\\hline\hline
\end{tabular}
\caption{Correlation between $\mathcal{B}(D^0\rightarrow K^+ \pi^- ) / \mathcal{B}(D^0\rightarrow K^- \pi^+ )$ and 
$\delta_{K\pi}$ \cite{Amhis:2012bh} used in our fit. \label{tab:correlations-hfag}}
\end{center}
\end{table}

\begin{table}[t] 
\begin{center}
\begin{tabular}{c|c|c} \hline\hline
$F_0^{DK}(0)$    & $0.737\pm 0.005$                   & $^\dagger$\cite{CHARM2013:Zupanc, Liu:2012bn, Besson:2009uv, Widhalm:2006wz,Aubert:2007wg}\\\hline  
$F_0^{D\pi}(0)$  & $0.638\pm 0.012$                   & $^\dagger$\cite{CHARM2013:Zupanc, Liu:2012bn,Besson:2009uv, Widhalm:2006wz}\\\hline 
$F_0^{D_s K}(0)$ & $(1\pm 5 \%)\times F_0^{D \pi}(0)$ &          \cite{Koponen:2012di,Koponen:2013tua}\\\hline\hline 
\end{tabular}
\caption{Numerical input for the form factors.
The form factor $F_0^{D_s K}(0)$ is varied {flatly} 
within {the} theory uncertainty \cite{Hocker:2001xe}. 
Table adapted from \cite{Hiller:2014prep}. $^\dagger$Our average.
\label{tab:inputformfactor}}
\end{center}
\end{table}

\section{Mapping of {the} topological on the 
  SU(3)$_F$ parameterization}
\label{sec:matching}

As discussed in Sec.~\ref{sec:redundancies} the topological flavor-flow
parameterization and the linear SU(3)$_F$ expansion can be mapped onto
each other after the removal of redundancies in each
parameterization. 
In Tabs.~\ref{tab:matchingtoposu3-part-1} and
\ref{tab:matchingtoposu3-part-2} we give two {numerical} examples for
the mapping.
Both redundant parameters and 
redundant decay amplitudes have to be removed in order to obtain two
{corresponding} $11\times 11$ regular coefficient matrices.  Then,
the mapping can be calculated by inverting one or the other coefficient
matrix.  {We choose to omit the redundant amplitudes}
\begin{align}
& D^0\rightarrow \pi^0\pi^0\,,\quad D^0\rightarrow K^-\pi^+\,,
\quad D^0\rightarrow K^+\pi^-\,,\\
&D^0\rightarrow K^0\pi^0\,,\quad   D^+\rightarrow K^0\pi^+\,, 
\quad D^+\rightarrow K^+ \pi^0 \,,
\end{align}
using the sum rules presented in Appendix~\ref{sec:sumrules}.  We {next}
calculate the redefined SU(3)$_F$ matrix elements {in terms of the
  remaining decay amplitudes} by inverting the SU(3)$_F$ coefficient
matrix given in Tabs.~I and V of Ref.~\cite{Hiller:2012xm}. The result
for {this} inverse matrix is given in
Tab.~\ref{tab:matchingtoposu3-part-3}.  In order to illustrate how to
read Tab.~\ref{tab:matchingtoposu3-part-3}, we {exemplify}
\begin{align}
  A_{27}^{15} &= \frac{2 \sqrt{2}}{3}\mathcal{A}(D^+\rightarrow
  \bar{K}^0 \pi^+) + \frac{\sqrt{2}}{3} \mathcal{A}(D_s^+\rightarrow
  K^0K^+) \,.
\label{eq:matching-example} 
\end{align}
Inserting {the expansions of} $\mathcal{A}(D^+\rightarrow \bar{K}^0
\pi^+)$ and $\mathcal{A}(D_s^+\rightarrow K^0K^+)$ {in terms of
  topological amplitudes} into \eq{eq:matching-example} gives the
{desired expression of} the SU(3)$_F$ matrix element $A_{27}^{15}$
{in terms of the topological amplitudes}. Note that in the {matching} we implicitly disregard higher
order SU(3)$_F$-breaking effects {which are included in the approximate
  factorization formulas.  Strictly speaking, these invalidate the
  linear SU(3)$_F$ sum rules, see also
  Sec.~\ref{sec:countingtopologies}.}  However, this can be safely
neglected as we only aim at a description of the data at linear
SU(3)$_F$ breaking here.

While the {exemplified} topological-amplitude fit points respect the
SU(3)$_F$ power counting, the SU(3)$_F$ breaking matrix elements can
nevertheless be quite large, like $\vert \widetilde{B}_1^3\vert\sim 0.7$
in Tab.~\ref{tab:matchingtoposu3-part-1} in case of example
point~I. This shows that {several small SU(3)$_F$ breaking parameters
  of the topological-amplitude fit can add up to a larger SU(3)$_F$
  breaking matrix element of the} group-theoretical {approach}.
However, as demonstrated by the example
point II in Tab.~\ref{tab:matchingtoposu3-part-2}, there are also
solutions where both diagrammatic and group theoretic languages give
SU(3)$_F$ breaking $\lesssim 50\%$.  

\begin{table}
\begin{tabular}{c|c||c|c}
\hline\hline
Topological parameter & Value & SU(3)$_F$ matrix element & Value \\\hline
$\vert\delta_A\vert/T^{\mathrm{fac}}$           & 0.14  &  $\vert \widetilde{A}_{27}^{15}\vert$        & 0.32    	\\
$\vert\delta_T\vert/T^{\mathrm{fac}}$           & 0.15  &  $\vert \widetilde{A}_8^{15}\vert$           & 0.22    	\\
$\mathrm{arg}(\delta_A/T^{\mathrm{fac}})$       &  0.17 &  $\vert \widetilde{A}_8^{\bar{6}}\vert$      & 1.00    	\\
$\mathrm{arg}(\delta_T/T^{\mathrm{fac}})$       & $-3.06$  &  $\vert \widetilde{B}_1^3\vert$              & 0.67           \\
$\vert F_0^{K\pi}(m^2_{D_{(s)}})\vert$          &  3.54 &  $\vert \widetilde{B}_8^3\vert$              & 0.22    	\\
$F_0^{D K}(0)$                                  &  0.74 &  $\vert \widetilde{B}_8^{\bar{6}_1}\vert$    & 0.36    	\\
$F_0^{D\pi}(0)$                                 &  0.64 &  $\vert \widetilde{B}_8^{15_1}\vert$         & 0.39    	\\
$F_0^{D_s K}(0)/F_0^{D \pi}(0)$                 &  0.95 &  $\vert \widetilde{B}_8^{15_2}\vert$         & 0.29    	\\
$\mathrm{arg}(F_0^{K\pi}(m^2_{D_{(s)}}))$       &  $-1.74$ &  $\vert \widetilde{B}_{27}^{15_1}\vert$      & 0.18    	\\
$\vert\widetilde{C}/T^{\mathrm{fac}}\vert$                     &  1.10 &  $\vert \widetilde{B}_{27}^{15_2}\vert$      & 0.07    	\\
$\vert\widetilde{E}/T^{\mathrm{fac}}\vert$                     &  0.46 &  $\vert \widetilde{B}_{27}^{24_1}\vert$      & 0.13    	\\
$\vert P_{\mathrm{break}}/T^{\mathrm{fac}}\vert$               & 0.05  &  $\mathrm{arg}(A_{27}^{15})$                 & 1.44      	\\
$\mathrm{arg}(\widetilde{C})$                   & 2.47  &  $\mathrm{arg}(A_8^{15})$                    & $-2.53$   	\\
$\mathrm{arg}(C_1^{(1)})$                       & $-1.50$  &  $\mathrm{arg}(A_8^{\bar{6}})$               & 0.20      	\\
$\mathrm{arg}(C_2^{(1)})$                       & $-1.40$  &  $\mathrm{arg}(B_1^3)$                       & $-0.53$   	\\
$\mathrm{arg}(C_3^{(1)})$                       &  0.00  &  $\mathrm{arg}(B_8^3)$                       & $-0.96$   	\\
$\mathrm{arg}(\widetilde{E})$                   &  1.49  &  $\mathrm{arg}(B_8^{\bar{6}_1})$             & $-2.11$   	\\
$\mathrm{arg}(E_1^{(1)})$                       &  $-0.65$ &  $\mathrm{arg}(B_8^{15_1})$                  & $-1.35$   	\\
$\mathrm{arg}(E_2^{(1)})$                       &  $-0.93$ &  $\mathrm{arg}(B_8^{15_2})$                  & $-2.30$   	\\
$\mathrm{arg}(E_3^{(1)})$                       &  $-1.16$ &  $\mathrm{arg}(B_{27}^{15_1})$               & 2.56      	\\
$\mathrm{arg}(P_{\mathrm{break}})$              &  0.18 &  $\mathrm{arg}(B_{27}^{15_2})$               & 3.10      	\\
$\vert C_1^{(1)}/\widetilde{C}\vert $           &  0.07 &  $\mathrm{arg}(B_{27}^{24_1})$               & 0.00             \\
$\vert C_2^{(1)}/\widetilde{C}\vert $           &  0.16 &   &    \\
$\vert C_3^{(1)}/\widetilde{C}\vert $           &  0.19  &   & \\ 
$\vert E_1^{(1)}/\widetilde{E}\vert $           &  0.50  &   & \\ 
$\vert E_2^{(1)}/\widetilde{E}\vert $           &  0.50 &   & \\ 
$\vert E_3^{(1)}/\widetilde{E}\vert $           &  0.05 &   & \\
$\delta^{\prime,\mathrm{topo}}_X $              &  0.50  &   & \\
$\delta^{\prime,C}_X $                          &  0.50   &   & \\
$\delta^{\prime,E}_X $                          &  0.31   &   & \\
$\delta^{\prime,P_{\mathrm{break}}}_X $         &  0.07  &   & \\
$\chi^2$                                        &  0.27 &   & \\\hline\hline 
\end{tabular}
\caption{Fit example point I and corresponding {point in the
    SU(3)$_F$ decomposition with linear  SU(3)$_F$ breaking.} 
{The values quoted for $\widetilde{A}^i_j$ and 
$\widetilde{B}^i_j$ in the last column are normalized to the largest
SU(3)$_F$ limit matrix element. The fit is only sensitive to $\delta_T$ and $\delta_A$ in the combination
$(\delta_T-\delta_A)/T^{\mathrm{fac}} =  0.29 e^{-3.02\,i}$}.
\label{tab:matchingtoposu3-part-1}}
\end{table}

\begin{table}
\begin{tabular}{c|c||c|c}
\hline\hline
Topological parameter & Value & SU(3)$_F$ matrix element & Value \\\hline
$\vert\delta_A\vert/T^{\mathrm{fac}}$           &  0.15         &  $\vert \widetilde{A}_{27}^{15}\vert$  &  0.35        \\
$\vert\delta_T\vert/T^{\mathrm{fac}}$           &  0.15 &  $\vert \widetilde{A}_8^{15}\vert$             &  1.00        \\
$\mathrm{arg}(\delta_A/T^{\mathrm{fac}})$       & 1.27  &  $\vert \widetilde{A}_8^{\bar{6}}\vert$        &  0.19        \\
$\mathrm{arg}(\delta_T/T^{\mathrm{fac}})$       & $-1.88$  &  $\vert \widetilde{B}_1^3\vert$                &  0.36        \\
$\vert F_0^{K\pi}(m^2_{D_{(s)}})\vert$          &  4.50 &  $\vert \widetilde{B}_8^3\vert$                &  0.10        \\
$F_0^{D K}(0)$                                  & 0.74  &  $\vert \widetilde{B}_8^{\bar{6}_1}\vert$      &  0.15        \\
$F_0^{D\pi}(0)$                                 & 0.64  &  $\vert \widetilde{B}_8^{15_1}\vert$           &  0.49        \\
$F_0^{D_s K}(0)/F_0^{D \pi}(0)$                 &  0.95 &  $\vert \widetilde{B}_8^{15_2}\vert$           &  0.06        \\
$\mathrm{arg}(F_0^{K\pi}(m^2_{D_{(s)}}))$       & $-1.50$  &  $\vert \widetilde{B}_{27}^{15_1}\vert$        &  0.17        \\
$\vert\widetilde{C}/T^{\mathrm{fac}}\vert$                     & 1.17  &  $\vert \widetilde{B}_{27}^{15_2}\vert$        &  0.21        \\
$\vert\widetilde{E}/T^{\mathrm{fac}}\vert$                     & 2.05  &  $\vert \widetilde{B}_{27}^{24_1}\vert$        &  0.06        \\
$\vert P_{\mathrm{break}}/T^{\mathrm{fac}}\vert$               & 0.39  &  $\mathrm{arg}(A_{27}^{15})$                   & 0.94         \\
$\mathrm{arg}(\widetilde{C})$                   & 2.23  &  $\mathrm{arg}(A_8^{15})$                      & $-0.60$      \\
$\mathrm{arg}(C_1^{(1)})$                       & $-1.06$ &  $\mathrm{arg}(A_8^{\bar{6}})$                 & 1.47         \\
$\mathrm{arg}(C_2^{(1)})$                       & 1.48  &  $\mathrm{arg}(B_1^3)$                         & 2.15         \\
$\mathrm{arg}(C_3^{(1)})$                       & 0.22  &  $\mathrm{arg}(B_8^3)$                         & 0.55         \\
$\mathrm{arg}(\widetilde{E})$                   & 2.57  &  $\mathrm{arg}(B_8^{\bar{6}_1})$               & $-2.04$      \\
$\mathrm{arg}(E_1^{(1)})$                       & 0.23  &  $\mathrm{arg}(B_8^{15_1})$                    & $-2.00$      \\
$\mathrm{arg}(E_2^{(1)})$                       & 1.72  &  $\mathrm{arg}(B_8^{15_2})$                    & $-1.25$      \\
$\mathrm{arg}(E_3^{(1)})$                       & 0.75  &  $\mathrm{arg}(B_{27}^{15_1})$                 & $-2.84$      \\
$\mathrm{arg}(P_{\mathrm{break}})$              & 1.97  &  $\mathrm{arg}(B_{27}^{15_2})$                 & $-1.53$      \\
$\vert C_1^{(1)}/\widetilde{C}\vert $           & 0.22  &  $\mathrm{arg}(B_{27}^{24_1})$                 & 0.23         \\
$\vert C_2^{(1)}/\widetilde{C}\vert $           & 0.38  &      &        \\
$\vert C_3^{(1)}/\widetilde{C}\vert $           & 0.12  &   & \\ 
$\vert E_1^{(1)}/\widetilde{E}\vert $           & 0.06  &   & \\ 
$\vert E_2^{(1)}/\widetilde{E}\vert $           & 0.09  &   & \\ 
$\vert E_3^{(1)}/\widetilde{E}\vert $           & 0.31  &   & \\
$\delta^{\prime,\mathrm{topo}}_X $              &  0.50         &   & \\
$\delta^{\prime,C}_X $                          & 0.50          &   & \\
$\delta^{\prime,E}_X $                          &  0.16         &   & \\
$\delta^{\prime,P_{\mathrm{break}}}_X $         &  0.50         &   & \\
$\chi^2$                                        & 0.12  &   & \\\hline\hline 
\end{tabular}
\caption{Fit example point II and corresponding 
{point in the
    SU(3)$_F$ decomposition with linear  SU(3)$_F$ breaking.}
{Cf.\ } Tab.~\ref{tab:matchingtoposu3-part-1} for the notation.
{The fit is only sensitive to $\delta_T$ and $\delta_A$ in the combination
$(\delta_T-\delta_A)/T^{\mathrm{fac}} =  0.30 e^{-1.88\,i}$}.
\label{tab:matchingtoposu3-part-2}}
\end{table}

\begin{turnpage}
\begin{table*}[t]
\begin{center}
\begin{tabular}{c|c|c|c|c|c|c|c|c|c|c|c}
\hline \hline
 ME / $\mathcal{A}$  & $D^0\rightarrow K^+ K^-$  & $D^0\rightarrow \pi^+ \pi^-$ & $D^0\rightarrow \bar{K}^0 K^0$ & $D^+ \rightarrow \pi^0 \pi^+$  &  $D^+ \rightarrow \bar{K}^0 K^+$ & $D_s^+ \rightarrow  K^0 \pi^+$ & $D_s^+ \rightarrow  K^+ \pi^0$ & $D^0\rightarrow \bar{K}^0 \pi^0$ &  $D^+ \rightarrow \bar{K}^0 \pi^+$ & $D_s^+ \rightarrow \bar{K}^0 K^+$ & $D_s^+ \rightarrow K^0 K^+ $ \\\hline\hline
$A_{27}^{15}$  		&           0 & 0 & 0 & 0 & 0 & 0 & 0 & 0 & $\frac{2 \sqrt{2}}{3}$ & 0 & $\frac{\sqrt{2}}{3}$ \\\hline
$A_8^{15}$ 		&           $-\frac{5}{6 \sqrt{2}}$ & $\frac{5}{6 \sqrt{2}}$ & 0 & 0 & $\frac{5}{6 \sqrt{2}}$ & $-\frac{5}{6 \sqrt{2}}$ & 0 & $\frac{5}{6}$ & $-\frac{1}{3 \sqrt{2}}$ & $-\frac{5}{6 \sqrt{2}}$ & $-\frac{1}{6 \sqrt{2}}$ \\\hline
$A_8^{\bar 6}$ 		&           $\frac{\sqrt{5}}{2}$ & $-\frac{\sqrt{5}}{2}$ & 0 & 0 & $\frac{\sqrt{5}}{2}$ & $-\frac{\sqrt{5}}{2}$ & 0 & $\sqrt{\frac{5}{2}}$ & $-\sqrt{5}$ & $\frac{\sqrt{5}}{2}$ & $-\frac{\sqrt{5}}{2}$ \\\hline 
$B_1^{3}$ 		&           $\frac{16 \sqrt{\frac{35}{421}}}{3}$ & $\frac{16 \sqrt{\frac{35}{421}}}{3}$ & $-\frac{16 \sqrt{\frac{35}{421}}}{3}$ & $-\frac{10 \sqrt{\frac{70}{421}}}{3}$ & 0 & $\frac{2 \sqrt{\frac{35}{421}}}{3}$ & $\frac{2 \sqrt{\frac{70}{421}}}{3}$ & 0 & $-\frac{4 \sqrt{\frac{35}{421}}}{3}$ & 0 & $-\frac{4 \sqrt{\frac{35}{421}}}{3}$ \\\hline
$B_8^{3}$  		&           $\frac{20 \sqrt{\frac{7}{3937}}}{3}$ & $\frac{20 \sqrt{\frac{7}{3937}}}{3}$ & $\frac{40 \sqrt{\frac{7}{3937}}}{3}$ & $\frac{10 \sqrt{\frac{14}{3937}}}{3}$ & $20 \sqrt{\frac{7}{3937}}$ & $\frac{10 \sqrt{\frac{7}{3937}}}{3}$ & $-\frac{50 \sqrt{\frac{14}{3937}}}{3}$ & 0 & $-\frac{20 \sqrt{\frac{7}{3937}}}{3}$ & 0 & $-\frac{20 \sqrt{\frac{7}{3937}}}{3}$ \\\hline
$B_8^{\bar{6}_{1}}$  	&           $-20 \sqrt{\frac{7}{2869}}$ & $20 \sqrt{\frac{7}{2869}}$ & 0 & $2 \sqrt{\frac{14}{2869}}$ & $-20 \sqrt{\frac{7}{2869}}$ & $18 \sqrt{\frac{7}{2869}}$ & $-2 \sqrt{\frac{14}{2869}}$ & $-40 \sqrt{\frac{14}{2869}}$ & $60 \sqrt{\frac{7}{2869}}$ & $-40 \sqrt{\frac{7}{2869}}$ & $20 \sqrt{\frac{7}{2869}}$ \\\hline
$B_8^{15_{1}}$ 		&           $-460 \sqrt{\frac{7}{1330969}}$ & $-20 \sqrt{\frac{133}{70051}}$ & $-840 \sqrt{\frac{7}{1330969}}$ & $-78 \sqrt{\frac{14}{1330969}}$ & $460 \sqrt{\frac{7}{1330969}}$ & $626 \sqrt{\frac{7}{1330969}}$ & $246 \sqrt{\frac{14}{1330969}}$ & $-80 \sqrt{\frac{14}{1330969}}$ & $92 \sqrt{\frac{7}{1330969}}$ & $80 \sqrt{\frac{7}{1330969}}$ & $4 \sqrt{\frac{133}{70051}}$ \\\hline 
$B_8^{15_{2}}$  	&          $20 \sqrt{\frac{6}{871}}$ & $-10 \sqrt{\frac{6}{871}}$ & $10 \sqrt{\frac{6}{871}}$ & $28 \sqrt{\frac{3}{871}}$ & $-20 \sqrt{\frac{6}{871}}$ & $-6 \sqrt{\frac{6}{871}}$ & $-32 \sqrt{\frac{3}{871}}$ & $60 \sqrt{\frac{3}{871}}$ & $-4 \sqrt{\frac{6}{871}}$ & $-30 \sqrt{\frac{6}{871}}$ & $2 \sqrt{\frac{6}{871}}$ \\\hline
$B_{27}^{15_{1}}$	&           0 & 0 & 0 & $-34 \sqrt{\frac{14}{5281}}$ & 0 & $-22 \sqrt{\frac{7}{5281}}$ & $-22 \sqrt{\frac{14}{5281}}$ & 0 & $-\frac{92 \sqrt{\frac{7}{5281}}}{3}$ & 0 & $-\frac{76 \sqrt{\frac{7}{5281}}}{3}$ \\\hline
$B_{27}^{15_{2}}$	&           0 & 0 & 0 & $8 \sqrt{\frac{14}{453}}$ & 0 & $-2 \sqrt{\frac{7}{453}}$ & $-2 \sqrt{\frac{14}{453}}$ & 0 & $4 \sqrt{\frac{21}{151}}$ & 0 & $-2 \sqrt{\frac{21}{151}}$ \\\hline
$B_{27}^{24_{1}}$	&           0 & 0 & 0 & $\sqrt{\frac{14}{3}}$ & 0 & $-\sqrt{\frac{7}{3}}$ & $-\sqrt{\frac{14}{3}}$ & 0 & 0 & 0 & 0 \\\hline\hline 
\end{tabular}
\caption{The inverse of the SU(3)$_F$ coefficient matrix given in Tabs.~I and V of~\cite{Hiller:2012xm} which is used in order to map the fit example points of the topological approach onto the SU(3)$_F$ parameterization in Tabs.~\ref{tab:matchingtoposu3-part-1} and \ref{tab:matchingtoposu3-part-2}.  \label{tab:matchingtoposu3-part-3}}
\end{center}
\end{table*}
\end{turnpage}

\section{Approximate factorization formulas}

Below, we give the $1/N_c$-leading expressions for the tree and
annihilation diagrams.  Corrections of higher order in the
$1/N_c$-expansion are parameterized by $\delta_T$ and $\delta_A$
introduced in Sec.~\ref{sec:countingtopologies}.

\subsection{Factorization of tree amplitudes}
\label{sec:fac-tree-amps}
We use the following expressions for the $1/N_c$-leading contributions
to the tree diagrams.  {SU(3)$_F$ breaking in the $1/N_c^2$
  corrections} is of higher order in our power counting and
neglected,~i.e.,~ we use a flavor-universal correction parameter
$\delta_T$.  
{In our fit we vary}
\begin{align}
0 \leq \vert \delta_T\vert \leq 0.15\, T^{\mathrm{fac}}\,, \\
-\pi \leq \mathrm{arg}(\delta_T) \leq \pi\,,
\end{align}
with {$ T^{\mathrm{fac}}$ defined in \eq{eq:calctreefac}}
{and $ \delta_T= T - T^{\mathrm{fac}}$, see \eq{eq:fact}.}
The $1/N_c$-leading, factorizable contributions to the SCS
tree amplitudes are altogether given as: {\allowdisplaybreaks
\begin{align}
T^{\mathrm{fac}}_{D^0\rightarrow K^+ K^-}       &= \frac{G_F}{\sqrt{2}} a_1 f_K (m_D^2 - m_K^2 ) F_0^{DK}(m_K^2)\,, \\ 
T^{\mathrm{fac}}_{D^0\rightarrow \pi^+\pi^-}    &= -\frac{G_F}{\sqrt{2}} a_1 f_{\pi} (m_D^2 - m_{\pi}^2 ) F_0^{D\pi}(m_\pi^2)\,, \label{eq:facpp}\\ 
T^{\mathrm{fac}}_{D^+\rightarrow \pi^+ \pi^0}   &= -\frac{G_F}{\sqrt{2}} \frac{1}{\sqrt{2}} a_1 f_{\pi} (m_D^2 - m_{\pi}^2 ) F_0^{D\pi}(m_\pi^2) \,, \\ 
T^{\mathrm{fac}}_{D^+\rightarrow K^+ \bar{K}^0} &= \frac{G_F}{\sqrt{2}} a_1 f_K ( m_D^2 - m_K^2 ) F_0^{DK}(m_K^2)\,, \\ 
T^{\mathrm{fac}}_{D_s^+ \rightarrow \pi^+ K^0 } &= -\frac{G_F}{\sqrt{2}} a_1 f_{\pi} (m_{D_s}^2 - m_K^2 ) F_0^{D_sK}(m_\pi^2). 
\end{align}}

The $1/N_c$-leading, factorizable contributions to the CF tree amplitudes are given as:
{\allowdisplaybreaks
\begin{align}
T^{\mathrm{fac}}_{D^0\rightarrow K^-\pi^+} 	  &=  \frac{G_F}{\sqrt{2}} a_1  f_{\pi} ( m_D^2 - m_K^2 ) F_0^{DK}(m_\pi^2) \,, \\ 
T^{\mathrm{fac}}_{D^+\rightarrow \bar{K}^0 \pi^+} &=  \frac{G_F}{\sqrt{2}} a_1 f_{\pi} (m_D^2 - m_K^2 ) F_0^{DK}(m_\pi^2) .
\end{align}}

The $1/N_c$-leading, factorizable contributions to the DCS tree amplitudes are given as:
{\allowdisplaybreaks
\begin{align}
T^{\mathrm{fac}}_{D^0\rightarrow K^+\pi^- }  &=  \frac{G_F}{\sqrt{2}} a_1 f_K (m_D^2 - m_{\pi}^2 ) F_0^{D\pi}(m_K^2)  \,, \\ 
T^{\mathrm{fac}}_{D^+\rightarrow K^+ \pi^0 } &=  \frac{G_F}{\sqrt{2}} \frac{1}{\sqrt{2}} a_1 f_K (m_D^2 - m_{\pi}^2 ) F_0^{D\pi}(m_K^2) \,, \\ 
T^{\mathrm{fac}}_{D_s^+\rightarrow K^0 K^+}  &=  \frac{G_F}{\sqrt{2}} a_1 f_K (m_{D_s}^2 - m_K^2) F_0^{D_sK}(m_K^2)  .
\end{align}} The matrix element of the vector current can be
parameterized by the vector and scalar form factors as
\cite{Koponen:2013tua}
\begin{align}
\bra{P} V^\mu \ket{D} &= F_+^{D\rightarrow P}(q^2) \left[
	p_D^\mu + p_P^\mu - \frac{m_D^2 - m_K^2}{q^2} q^\mu
	\right]\nn\\
	&+ F_0^{D\rightarrow K}(q^2) \frac{m_D^2 - m_K^2}{q^2} q^\mu\,, 
\end{align}
with the vector form factor $F_+^{D\rightarrow P}$ and the scalar form
factor $F_0^{D\rightarrow K}$ {obeying} \cite{Koponen:2013tua}
\begin{align}
\bra{P}S\ket{D} &= F_0^{D\rightarrow P}(q^2) \frac{m_D^2 - m_P^2}{m_c - m_{p}}.  \label{eq:scalarFormFactor}
\end{align}
{Here the same renormalization scheme and scale must be used for $S$
and $m_c - m_{p}$.}

We calculate the form factors that appear in the tree amplitudes using the 
overall scaling factor appearing in the $z$-parameterization,~i.e.,~a pole factor \cite{Barger:1979fu, Koponen:2013tua} 
\begin{align}
F_0^{DK}(m_P^2)    &= \frac{F_0^{DK}(0)}{1 - m_P^2/m^2_{D_{s0}^*}(2317)^\pm}\,, \\
F_0^{D_s K}(m_P^2) &= \frac{F_0^{D_s K}(0)}{1 - m_P^2/m^2_{D_0^*}(2400)^\pm}\,, \\
F_0^{D \pi}(m_P^2) &= \frac{F_0^{D\pi}(0)}{1 - m_P^2/m^2_{D_0^*}(2400)^\pm}\,,
\end{align}
with the scalar resonances \cite{Beringer:1900zz} 
\begin{align}
m_{D_{s0}^*}(2317)^\pm &= (2317.8 \pm0.6) \, \mathrm{MeV}\,, \\
m_{D_0^*}(2400)^\pm    &= (2403\pm 40)    \, \mathrm{MeV}.
\end{align}
The used input values for $F_0^{DK}(0)$, $F_0^{D_s K}(0)$ and
$F_0^{D\pi}(0)$ are given in Tab.~\ref{tab:inputformfactor}.

As we assume isospin {symmetry} in the topological-amplitude
decomposition we {ignore} the smallish isospin breaking between
charged and neutral masses of kaons and pions for consistency. We use
the neutral masses {in all amplitudes}. However, in the phase space
factors of the branching ratios we take the {isospin} mass splittings
into account.

\subsection{Factorization of annihilation amplitudes}
\label{sec:fac-anni}

We use the following expressions for the $1/N_c$-leading contributions
to the annihilation diagrams.  
{As in the case of tree amplitudes we vary $\delta_A$
  of \eq{eq:defda} as}
\begin{align}
0 &\leq \vert \delta_A\vert \leq 0.15\, T^{\mathrm{fac}}\,,\\
-\pi &\leq \mathrm{arg}({\delta_A}) \leq \pi\,.
\end{align}
The $1/N_c$-leading, factorizable contributions \cite{Buras:1985xv} to
the SCS annihilation amplitudes are given as:
\begin{align}
A^{\mathrm{fac}}_{D^+\rightarrow \bar{K}^0 K^+}   &= 0\,,\\
A^{\mathrm{fac}}_{D_s^+\rightarrow K^0\pi^+}	  &=  \frac{G_F}{\sqrt{2}} a_1 f_{D_s} F_0^{K \pi}(m_{D_s}^2) \left(m^2_K - m^2_{\pi} \right)\,, \\ 
A^{\mathrm{fac}}_{D_s^+\rightarrow K^+\pi^0}	  &= - \frac{G_F}{\sqrt{2}} \frac{1}{\sqrt{2}}  a_1 f_{D_s} F_0^{K \pi}(m_{D_s}^2) \left(m^2_K - m^2_{\pi} \right)\,. 
\end{align}
The $1/N_c$-leading, factorizable contribution to the CF annihilation amplitude is given as:
\begin{align}
A^{\mathrm{fac}}_{D_s^+\rightarrow \bar{K}^0 K^+}  &= 0\,.
\end{align}
The $1/N_c$-leading, factorizable contributions to the DCS annihilation amplitudes are given as: 
\begin{align}
A^{\mathrm{fac}}_{D^+\rightarrow K^0 \pi^+}       &=  \frac{G_F}{\sqrt{2}} a_1 f_D F_0^{K \pi}(m_D^2) \left(m_K^2 - m_{\pi}^2\right)\,,\\
A^{\mathrm{fac}}_{D^+\rightarrow K^+ \pi^0}       &= -  \frac{G_F}{\sqrt{2}} \frac{1}{\sqrt{2}} a_1 f_D F_0^{K \pi}(m_D^2) \left(m_K^2 - m_{\pi}^2\right).
\end{align}
Note that the $1/N_c$-leading SCS annihilation amplitude $A^{\mathrm{fac}}_{D^+\rightarrow \bar{K}^0 K^+}$ and the 
$1/N_c$-leading CF annihilation amplitude $A^{\mathrm{fac}}_{D_s^+\rightarrow \bar{K}^0 K^+}$
can be neglected due to isospin symmetry \cite{Ali:1998eb}. However, the corresponding $1/N_c^2$ corrections are of 
course taken into account (as for the others) and specified 
in Tab.~\ref{tab:topoparametrization1Nc}.

Constraints on $\vert F_0^{K\pi}(m_{D_{(s)}}^2)\vert$ can be taken from
$\tau$ decays.  In order to accommodate the measurements of
$\tau\rightarrow K_S\pi^- \nu_\tau$ from Belle \cite{Epifanov:2007rf} we
vary the form factor {in the interval}
\begin{align}
1 &\lesssim \vert F_0^{K\pi}(m_{D_{(s)}}^2)\vert \lesssim 4.5\,, \label{eq:Kpiabs} \\
-\pi &\lesssim \mathrm{arg}\left(F_0^{K\pi}(m_{D_{(s)}}^2)\right) \lesssim \pi\,, \label{eq:Kpiarg}
\end{align}
setting
\begin{align}
F_0^{K\pi}(m_{D_{s}}^2) = F_0^{K\pi}(m_D^2)\,.
\end{align}

\section{Diagrammatic representation of sum rules}
\label{sec:sumrules}
 
In Tabs.~\ref{tab:sumrule-1}--\ref{tab:sumrule-6} we give the diagrammatic representation of the 
six Grossman-Robinson SU(3)$_F$ sum rules {which hold to linear
  order in} SU(3)$_F$ breaking \cite{Grossman:2012ry}. 

\begin{table*}
\begin{center}
\begin{tabular}{c|c|c|c|c}
\hline \hline
{Decay amplitude} & $T$ & $C$  & $E$  & $P_{\mathrm{break}}$  \\\hline\hline
 $+\frac{1}{\sqrt{2}} \mathcal{A}(D^0\rightarrow \pi^+ \pi^-) $ 	&	
 $-\frac{1}{\sqrt{2}}\times$\vcenteredhbox{\includegraphics[width=0.08\linewidth]{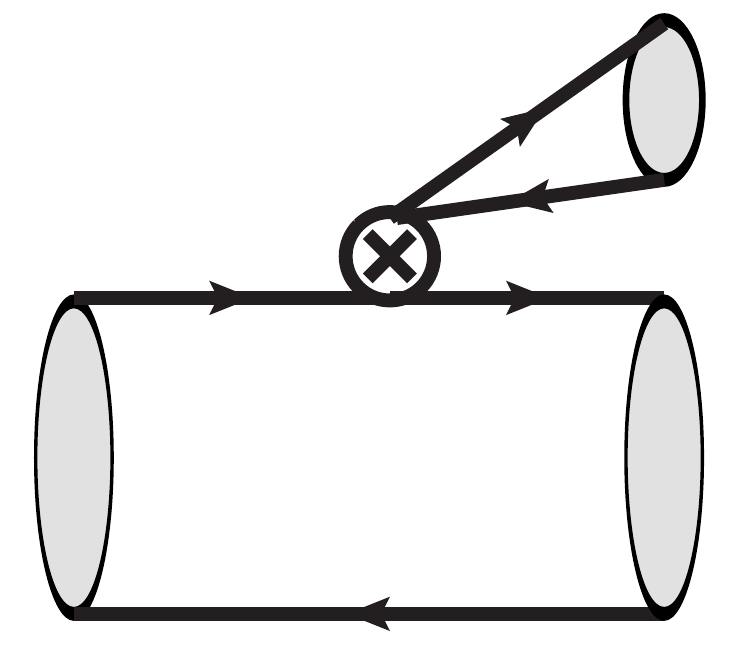}} & 
 0  & 
 $-\frac{1}{\sqrt{2}}\times$ \vcenteredhbox{\includegraphics[width=0.08\linewidth]{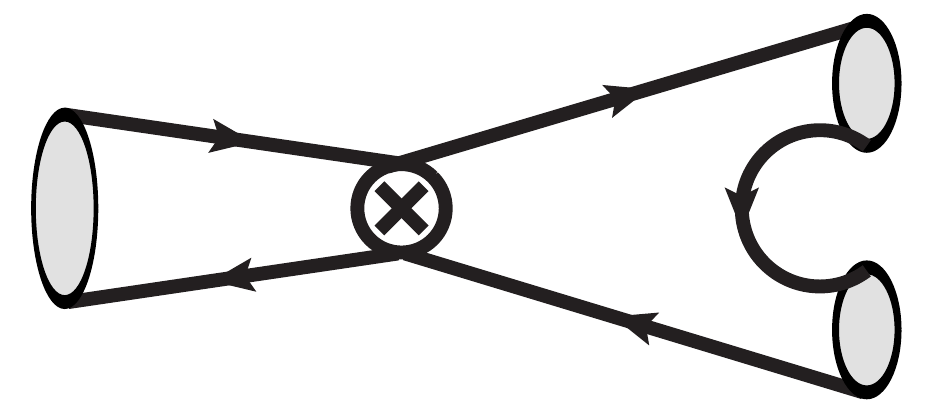}} & 
 $+\frac{1}{\sqrt{2}}\times $ \vcenteredhbox{\includegraphics[width=0.08\linewidth]{su3break-penguin-1-x.pdf}} \\\hline
 $+\mathcal{A}(D^0 \rightarrow \pi^0 \pi^0)$ & 
 0 &  
 $-\frac{1}{\sqrt{2}}\times$ \vcenteredhbox{\includegraphics[width=0.08\linewidth]{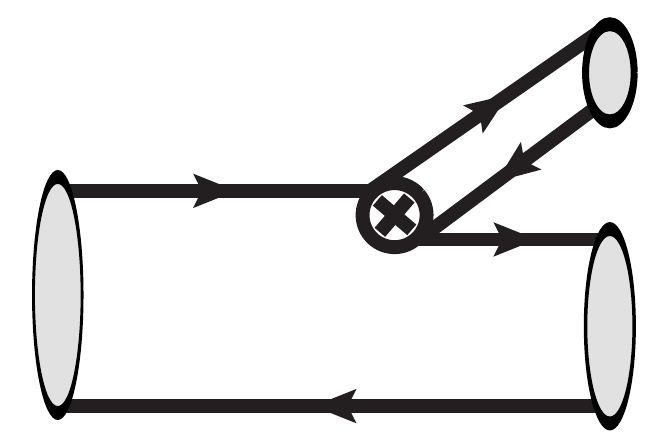}} & 
 $+\frac{1}{\sqrt{2}}\times $ \vcenteredhbox{\includegraphics[width=0.08\linewidth]{sumrule-su3limit-emission.pdf}} &  
 $-\frac{1}{\sqrt{2}}\times$ \vcenteredhbox{\includegraphics[width=.08\linewidth]{su3break-penguin-1-x.pdf}}\\\hline
 $-\mathcal{A}(D^+ \rightarrow \pi^0 \pi^+)$ & 
 $+\frac{1}{\sqrt{2}}\times$ \vcenteredhbox{\includegraphics[width=0.08\linewidth]{sumrule-su3limit-tree.pdf}} & 
 $+\frac{1}{\sqrt{2}}\times$ \vcenteredhbox{\includegraphics[width=0.08\linewidth]{sumrule-su3limit-color.pdf}} & 
 0 & 
 0 \\\hline\hline
\end{tabular}
\caption{Diagrammatic representation of sum rule I, $ \frac{1}{\sqrt{2}} \mathcal{A}(D^0\rightarrow \pi^+ \pi^-) + \mathcal{A}(D^0\rightarrow \pi^0 \pi^0) - \mathcal{A}(D^+\rightarrow \pi^0 \pi^+ ) = 0$.\label{tab:sumrule-1}  }
\end{center}
\end{table*}

\begin{table*}
\begin{center}
\begin{tabular}{c|c|c|c|c|c|c}
\hline \hline
{Decay amplitude} 
& $T$ & $T_1^{(1)}$ & $C$ & $C_1^{(1)}$ & $E$  & $E_1^{(1)}$ \\\hline\hline
$\frac{1}{\sqrt{2}} \mathcal{A}(D^0\rightarrow K^- \pi^+)$   	&	
 $\frac{1}{\sqrt{2}}\times$\vcenteredhbox{\includegraphics[width=0.08\linewidth]{sumrule-su3limit-tree.pdf}} & 
 $\frac{1}{\sqrt{2}}\times$\vcenteredhbox{\includegraphics[width=0.08\linewidth]{su3break-tree-1-x.pdf}} &  
 $0$                    & 
 $0$                    & 
 $\frac{1}{\sqrt{2}}\times$\vcenteredhbox{\includegraphics[width=0.08\linewidth]{sumrule-su3limit-emission.pdf}}              & 
 $\frac{1}{\sqrt{2}}\times$\vcenteredhbox{\includegraphics[width=0.08\linewidth]{su3break-emission-1-x.pdf}} \\\hline
$+\mathcal{A}(D^0\rightarrow \bar{K}^0 \pi^0)$ & 	
 $0$ & 
 $0$ &  
 $\frac{1}{\sqrt{2}}\times$ \vcenteredhbox{\includegraphics[width=0.08\linewidth]{sumrule-su3limit-color.pdf}}  & 
 $\frac{1}{\sqrt{2}}\times$ \vcenteredhbox{\includegraphics[width=0.08\linewidth]{su3break-color-1-x.pdf}} & 
 $\frac{-1}{\sqrt{2}}\times$ \vcenteredhbox{\includegraphics[width=0.08\linewidth]{sumrule-su3limit-emission.pdf}}  & 
 $\frac{-1}{\sqrt{2}}\times$ \vcenteredhbox{\includegraphics[width=0.08\linewidth]{su3break-emission-1-x.pdf}} \\\hline
$-\frac{1}{\sqrt{2}} \mathcal{A}(D^+ \rightarrow \bar{K}^0 \pi^+)$ &	
 $-\frac{1}{\sqrt{2}} \times$\vcenteredhbox{\includegraphics[width=0.08\linewidth]{sumrule-su3limit-tree.pdf}} & 
 $-\frac{1}{\sqrt{2}} \times$\vcenteredhbox{\includegraphics[width=0.08\linewidth]{su3break-tree-1-x.pdf}} &  
 $-\frac{1}{\sqrt{2}} \times$\vcenteredhbox{\includegraphics[width=0.08\linewidth]{sumrule-su3limit-color.pdf}}   & 
 $-\frac{1}{\sqrt{2}} \times$\vcenteredhbox{\includegraphics[width=0.08\linewidth]{su3break-color-1-x.pdf}} & 
 $0$ & 
 $0$  \\\hline
\end{tabular}
\caption{Diagrammatic representation of sum rule II, $\frac{1}{\sqrt{2}} \mathcal{A}(D^0\rightarrow K^-\pi^+ ) + \mathcal{A}(D^0\rightarrow \bar{K}^0 \pi^0 )  - \frac{1}{\sqrt{2}} \mathcal{A}(D^+\rightarrow \bar{K}^0 \pi^+) = 0$. \label{tab:sumrule-2} }
\end{center}
\end{table*}

\begin{table*}
\begin{center}
\begin{tabular}{c|c|c|c|c|c|c|c|c|c|c|c|c|c|c|c}
\hline \hline
{Decay amplitude} 
& $T$ & $T_2^{(1)}$ &  $A$ & $A_2^{(1)}$ & $C$ &  $C_2^{(1)}$ &  $E$  & $E_2^{(1)}$  \\\hline\hline
$\mathcal{A}(D^0 \rightarrow K^+ \pi^-)$ 	& 
 $1\times$\vcenteredhbox{\includegraphics[width=0.05\linewidth]{sumrule-su3limit-tree.pdf}} & 
 $1\times$\vcenteredhbox{\includegraphics[width=0.05\linewidth]{su3break-tree-2-x.pdf}} & 
 0 & 
 0 & 
 0 & 
 0 & 
 $1\times$ \vcenteredhbox{\includegraphics[width=0.05\linewidth]{sumrule-su3limit-emission.pdf}}& 
 $1\times$ \vcenteredhbox{\includegraphics[width=0.05\linewidth]{su3break-emission-2-x.pdf}}\\\hline
$\sqrt{2} \mathcal{A}(D^0 \rightarrow K^0 \pi^0)$ & 
 0 & 
 0 & 
 0 & 
 0 & 
 $1\times$  \vcenteredhbox{\includegraphics[width=0.05\linewidth]{sumrule-su3limit-color.pdf}} & 
 $1\times$  \vcenteredhbox{\includegraphics[width=0.05\linewidth]{su3break-color-2-x.pdf}}     & 
 $-1\times$ \vcenteredhbox{\includegraphics[width=0.05\linewidth]{sumrule-su3limit-emission.pdf}}& 
 $-1\times$ \vcenteredhbox{\includegraphics[width=0.05\linewidth]{su3break-emission-2-x.pdf}}\\\hline
$-\mathcal{A}(D^+ \rightarrow K^0\pi^+)$ 	& 
 0 &  
 0 &  
 $-1\times$ \vcenteredhbox{\includegraphics[width=0.05\linewidth]{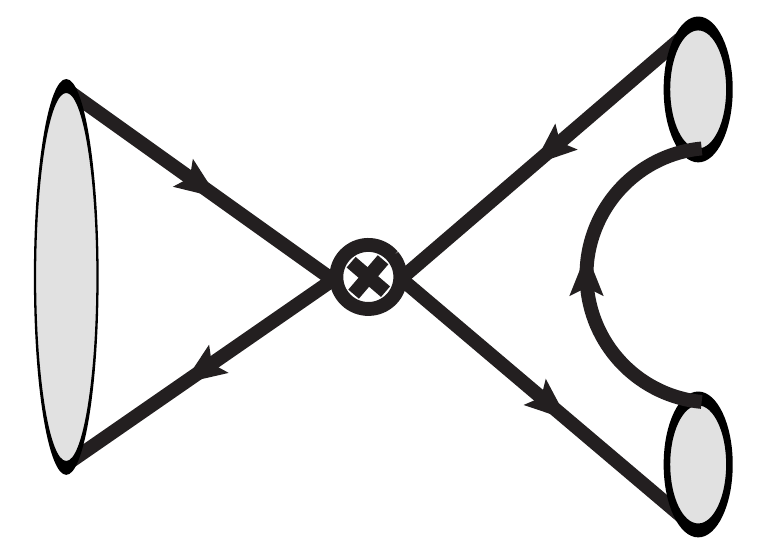}}  & 
 $-1\times$ \vcenteredhbox{\includegraphics[width=0.05\linewidth]{su3break-anni-2-x.pdf}}      & 
 $-1\times$ \vcenteredhbox{\includegraphics[width=0.05\linewidth]{sumrule-su3limit-color.pdf}}& 
 $-1\times$ \vcenteredhbox{\includegraphics[width=0.05\linewidth]{su3break-color-2-x.pdf}}    & 
 0 & 
 0 \\\hline
$-\sqrt{2} \mathcal{A}(D^+ \rightarrow K^+ \pi^0)$ & 
 $-1\times$\vcenteredhbox{\includegraphics[width=0.05\linewidth]{sumrule-su3limit-tree.pdf}} & 
 $-1\times$\vcenteredhbox{\includegraphics[width=0.05\linewidth]{su3break-tree-2-x.pdf}} & 
 $1\times$ \vcenteredhbox{\includegraphics[width=0.05\linewidth]{sumrule-su3limit-anni.pdf}} & 
 $1\times$ \vcenteredhbox{\includegraphics[width=0.05\linewidth]{su3break-anni-2-x.pdf}}     & 
 0 & 
 0 & 
 0 & 
 0 \\\hline\hline
\end{tabular}
\caption{Diagrammatic representation of sum rule III, 
$\mathcal{A}(D^0\rightarrow K^+ \pi^- ) +\sqrt{2} \mathcal{A}(D^0\rightarrow K^0 \pi^0 ) - \mathcal{A}(D^+\rightarrow K^0 \pi^+ ) - \sqrt{2} \mathcal{A}(D^+\rightarrow K^+ \pi^0 ) = 0$. \label{tab:sumrule-3} }
\end{center}
\end{table*}

\begin{table*}
\begin{center}
\begin{tabular}{c|c|c|c|c|c|c|c|c}
\hline \hline
{Decay amplitude} 
& $T$ & $T_1^{(1)}$ & $T_2^{(1)}$ & $E$  & $E_1^{(1)}$ & $E_2^{(1)}$ & $P_{\mathrm{break}}$  \\\hline\hline
$\mathcal{A}(D^0\rightarrow K^+ K^-)$  	        & 
 $1\times$\vcenteredhbox{\includegraphics[width=0.08\linewidth]{sumrule-su3limit-tree.pdf}}         &    
 $1\times$\vcenteredhbox{\includegraphics[width=0.08\linewidth]{su3break-tree-1-x.pdf}}     & 
 $1\times$\vcenteredhbox{\includegraphics[width=0.08\linewidth]{su3break-tree-2-x.pdf}}     & 
 $1\times$\vcenteredhbox{\includegraphics[width=0.08\linewidth]{sumrule-su3limit-emission.pdf}}     & 
 $1\times$\vcenteredhbox{\includegraphics[width=0.08\linewidth]{su3break-emission-1-x.pdf}} & 
 $1\times$\vcenteredhbox{\includegraphics[width=0.08\linewidth]{su3break-emission-2-x.pdf}} & 
 $1\times$\vcenteredhbox{\includegraphics[width=0.08\linewidth]{su3break-penguin-1-x.pdf}}  \\\hline
$-\mathcal{A}(D^0\rightarrow \pi^+ \pi^-) $ & 
 $1\times$\vcenteredhbox{\includegraphics[width=0.08\linewidth]{sumrule-su3limit-tree.pdf}} & 
 0 & 
 0 & 
 $1\times$\vcenteredhbox{\includegraphics[width=0.08\linewidth]{sumrule-su3limit-emission.pdf}} & 
 0 & 
 0 & 
 $-1\times$\vcenteredhbox{\includegraphics[width=0.08\linewidth]{su3break-penguin-1-x.pdf}} \\\hline
$-\mathcal{A}(D^0\rightarrow K^- \pi^+)$ & 
 $-1\times$\vcenteredhbox{\includegraphics[width=0.08\linewidth]{sumrule-su3limit-tree.pdf}}      &
 $-1\times$\vcenteredhbox{\includegraphics[width=0.08\linewidth]{su3break-tree-1-x.pdf}}  & 
 0 & 
 $-1\times$\vcenteredhbox{\includegraphics[width=0.08\linewidth]{sumrule-su3limit-emission.pdf}}     & 
 $-1\times$\vcenteredhbox{\includegraphics[width=0.08\linewidth]{su3break-emission-1-x.pdf}} & 
 0 & 
 0 \\\hline
$-\mathcal{A}(D^0 \rightarrow K^+ \pi^-)$ & 
 $-1\times$\vcenteredhbox{\includegraphics[width=0.08\linewidth]{sumrule-su3limit-tree.pdf}}  &
 0 & 
 $-1\times$\vcenteredhbox{\includegraphics[width=0.08\linewidth]{su3break-tree-2-x.pdf}} & 
 $-1\times$\vcenteredhbox{\includegraphics[width=0.08\linewidth]{sumrule-su3limit-emission.pdf}} & 
 0 & 
 $-1\times$\vcenteredhbox{\includegraphics[width=0.08\linewidth]{su3break-emission-2-x.pdf}} & 
 0 \\\hline
\end{tabular}
\caption{Diagrammatic representation of sum rule IV, 
$\mathcal{A}(D^0\rightarrow K^- K^+ ) - \mathcal{A}(D^0\rightarrow \pi^+ \pi^-) - \mathcal{A}(D^0\rightarrow K^-\pi^+ ) - \mathcal{A}(D^0\rightarrow K^+ \pi^- )   = 0$. \label{tab:sumrule-4} }
\end{center}
\end{table*}

\begin{turnpage}
\begin{table}
\begin{center}
\begin{tabular}{c|c|c|c|c|c|c|c|c|c|c|c|c|c|c|c|c|c}
\hline \hline
{Decay amplitude}
& $T$ & $T_1^{(1)}$ & $T_2^{(1)}$ & $T_3^{(1)}$  & $A$ & $A_1^{(1)}$ & $A_2^{(1)}$ & $A_3^{(1)}$ & $C$ & $C_1^{(1)}$ & $C_2^{(1)}$ & $C_3^{(1)}$ & $P_{\mathrm{break}}$  \\\hline\hline
$-\mathcal{A}(D^+ \rightarrow \bar{K}^0 K^+)$ &	
 $-1\times$\vcenteredhbox{\includegraphics[width=0.03\linewidth]{sumrule-su3limit-tree.pdf}} & 
 $-1\times$\vcenteredhbox{\includegraphics[width=0.03\linewidth]{su3break-tree-1-x.pdf}} & 
 $-1\times$\vcenteredhbox{\includegraphics[width=0.03\linewidth]{su3break-tree-2-x.pdf}} & 
 0 & 
 $1\times$\vcenteredhbox{\includegraphics[width=0.03\linewidth]{sumrule-su3limit-anni.pdf}} & 
 0 & 
 0 & 
 $1\times$\vcenteredhbox{\includegraphics[width=0.03\linewidth]{su3break-anni-3-x.pdf}} & 
 0 & 
 0 & 
 0 & 
 0 &  
 $-1\times$\vcenteredhbox{\includegraphics[width=0.03\linewidth]{su3break-penguin-1-x.pdf}} \\\hline
$\mathcal{A}(D_s^+ \rightarrow  K^0 \pi^+)$ 	&	
 $-1\times$\vcenteredhbox{\includegraphics[width=0.03\linewidth]{sumrule-su3limit-tree.pdf}} & 
 0 & 
 0 & 
 $-1\times$\vcenteredhbox{\includegraphics[width=0.03\linewidth]{su3break-tree-3-x.pdf}} & 
 $1\times$\vcenteredhbox{\includegraphics[width=0.03\linewidth]{sumrule-su3limit-anni.pdf}} & 
 $1\times$\vcenteredhbox{\includegraphics[width=0.03\linewidth]{su3break-anni-1-x.pdf}} & 
 $1\times$\vcenteredhbox{\includegraphics[width=0.03\linewidth]{su3break-anni-2-x.pdf}} & 
 0 & 
 0 & 
 0 & 
 0 & 
 0 &  
 $1\times$\vcenteredhbox{\includegraphics[width=0.03\linewidth]{su3break-penguin-1-x.pdf}} \\\hline
$\mathcal{A}(D^+ \rightarrow \bar{K}^0 \pi^+)$ &	
 $1\times$\vcenteredhbox{\includegraphics[width=0.03\linewidth]{sumrule-su3limit-tree.pdf}} & 
 $1\times$\vcenteredhbox{\includegraphics[width=0.03\linewidth]{su3break-tree-1-x.pdf}} & 
 0 & 
 0 & 
 0 & 
 0 & 
 0 & 
 0 & 
 $1\times$\vcenteredhbox{\includegraphics[width=0.03\linewidth]{sumrule-su3limit-color.pdf}} & 
 $1\times$\vcenteredhbox{\includegraphics[width=0.03\linewidth]{su3break-color-1-x.pdf}} & 
 0 & 
 0 &  
 0 \\\hline
$-\mathcal{A}(D_s^+ \rightarrow \bar{K}^0 K^+)$ &
 0 & 
 0 & 
 0 & 
 0 & 
 $-1\times$\vcenteredhbox{\includegraphics[width=0.03\linewidth]{sumrule-su3limit-anni.pdf}} & 
 $-1\times$\vcenteredhbox{\includegraphics[width=0.03\linewidth]{su3break-anni-1-x.pdf}} &  
 0 & 
 $-1\times$\vcenteredhbox{\includegraphics[width=0.03\linewidth]{su3break-anni-3-x.pdf}} & 
 $-1\times$\vcenteredhbox{\includegraphics[width=0.03\linewidth]{sumrule-su3limit-color.pdf}} & 
 $-1\times$\vcenteredhbox{\includegraphics[width=0.03\linewidth]{su3break-color-1-x.pdf}} & 
 0 & 
 $-1\times$\vcenteredhbox{\includegraphics[width=0.03\linewidth]{su3break-color-3-x.pdf}} &  
 0 \\\hline
$-\mathcal{A}(D^+ \rightarrow K^0\pi^+) $ 	&	
 0 & 
 0 & 
 0 & 
 0 & 
 $-1\times$\vcenteredhbox{\includegraphics[width=0.03\linewidth]{sumrule-su3limit-anni.pdf}} & 
 0 & 
 $-1\times$\vcenteredhbox{\includegraphics[width=0.03\linewidth]{su3break-anni-2-x.pdf}} & 
 0 &  
 $-1\times$\vcenteredhbox{\includegraphics[width=0.03\linewidth]{sumrule-su3limit-color.pdf}} & 
 0 & 
 $-1\times$\vcenteredhbox{\includegraphics[width=0.03\linewidth]{su3break-color-2-x.pdf}} & 
 0 &  
 0 \\\hline
$\mathcal{A}(D_s^+ \rightarrow K^0 K^+) $ 	&	
 $1\times$\vcenteredhbox{\includegraphics[width=0.03\linewidth]{sumrule-su3limit-tree.pdf}} & 
 0 & 
 $1\times$\vcenteredhbox{\includegraphics[width=0.03\linewidth]{su3break-tree-2-x.pdf}} & 
 $1\times$\vcenteredhbox{\includegraphics[width=0.03\linewidth]{su3break-tree-3-x.pdf}} & 
 0 & 
 0 & 
 0 & 
 0 & 
 $1\times$\vcenteredhbox{\includegraphics[width=0.03\linewidth]{sumrule-su3limit-color.pdf}} & 
 0 & 
 $1\times$\vcenteredhbox{\includegraphics[width=0.03\linewidth]{su3break-color-2-x.pdf}} & 
 $1\times$\vcenteredhbox{\includegraphics[width=0.03\linewidth]{su3break-color-3-x.pdf}} &  
 0 \\\hline\hline
\end{tabular}
\caption{Diagrammatic representation of sum rule V,  
$  - \mathcal{A}(D^+ \rightarrow \bar{K}^0 K^+ )
   + \mathcal{A}(D_s^+ \rightarrow K^0 \pi^+ ) 
   + \mathcal{A}(D^+ \rightarrow \bar{K}^0 \pi^+ ) 
   - \mathcal{A}(D_s^+ \rightarrow \bar{K}^0 K^+ ) 
   - \mathcal{A}(D^+ \rightarrow  K^0 \pi^+  ) 
   + \mathcal{A}( D_s^+\rightarrow K^0 K^+ ) = 0$. \label{tab:sumrule-5} }
\end{center}
\end{table}

\begin{table*}
\begin{center}
\begin{tabular}{c|c|c|c|c|c|c|c|c|c|c|c|c}
\hline \hline
{Decay amplitude} 
& $T$ & $T_1^{(1)}$ & $T_2^{(1)}$ & $A$ & $A_1^{(1)}$ & $A_2^{(1)}$ & $A_3^{(1)}$ & $C$ & $C_1^{(1)}$ & $C_2^{(1)}$ & $C_3^{(1)}$ & $P_{\mathrm{break}}$  \\\hline\hline
$\sqrt{2} \mathcal{A}(D^+ \rightarrow \pi^0 \pi^+)$  	&	
 $-1\times$\vcenteredhbox{\includegraphics[width=0.03\linewidth]{sumrule-su3limit-tree.pdf}}  & 
 0 & 
 0 & 
 0 &
 0 & 
 0 & 
 0 & 
 $-1\times$\vcenteredhbox{\includegraphics[width=0.03\linewidth]{sumrule-su3limit-color.pdf}} & 
 0 & 
 0 & 
 0 & 
 0 \\\hline
$-\mathcal{A}(D^+ \rightarrow \bar{K}^0 K^+)$ &	
 $-1\times$\vcenteredhbox{\includegraphics[width=0.03\linewidth]{sumrule-su3limit-tree.pdf}} & 
 $-1\times$\vcenteredhbox{\includegraphics[width=0.03\linewidth]{su3break-tree-1-x.pdf}} & 
 $-1\times$\vcenteredhbox{\includegraphics[width=0.03\linewidth]{su3break-tree-2-x.pdf}} & 
 $1\times$\vcenteredhbox{\includegraphics[width=0.03\linewidth]{sumrule-su3limit-anni.pdf}}  & 
 0 & 
 0 & 
 $1\times$\vcenteredhbox{\includegraphics[width=0.03\linewidth]{su3break-anni-3-x.pdf}} & 
 0 & 
 0 & 
 0 & 
 0 &  
 $-1\times$\vcenteredhbox{\includegraphics[width=0.03\linewidth]{su3break-penguin-1-x.pdf}} \\\hline
$-\sqrt{2} \mathcal{A}(D_s^+ \rightarrow  K^+ \pi^0)$ 	&	
 0 & 
 0 & 
 0 & 
 $1\times$\vcenteredhbox{\includegraphics[width=0.03\linewidth]{sumrule-su3limit-anni.pdf}} & 
 $1\times$\vcenteredhbox{\includegraphics[width=0.03\linewidth]{su3break-anni-1-x.pdf}} & 
 $1\times$\vcenteredhbox{\includegraphics[width=0.03\linewidth]{su3break-anni-2-x.pdf}} & 
 0 & 
 $1\times$\vcenteredhbox{\includegraphics[width=0.03\linewidth]{sumrule-su3limit-color.pdf}} & 
 0 & 
 0 & 
 $1\times$\vcenteredhbox{\includegraphics[width=0.03\linewidth]{su3break-color-3-x.pdf}} &  
 $1\times$\vcenteredhbox{\includegraphics[width=0.03\linewidth]{su3break-penguin-1-x.pdf}} \\\hline
$\mathcal{A}(D^+ \rightarrow \bar{K}^0 \pi^+)$ &	
 $1\times$\vcenteredhbox{\includegraphics[width=0.03\linewidth]{sumrule-su3limit-tree.pdf}} & 
 $1\times$\vcenteredhbox{\includegraphics[width=0.03\linewidth]{su3break-tree-1-x.pdf}} & 
 0 & 
 0 & 
 0 &  
 0 & 
 0 & 
 $1\times$\vcenteredhbox{\includegraphics[width=0.03\linewidth]{sumrule-su3limit-color.pdf}} & 
 $1\times$\vcenteredhbox{\includegraphics[width=0.03\linewidth]{su3break-color-1-x.pdf}} & 
 0 & 
 0 &  
 0 \\\hline
$-\mathcal{A}(D_s^+ \rightarrow \bar{K}^0 K^+)$ &	
 0 & 
 0 & 
 0 & 
 $-1\times$\vcenteredhbox{\includegraphics[width=0.03\linewidth]{sumrule-su3limit-anni.pdf}} & 
 $-1\times$\vcenteredhbox{\includegraphics[width=0.03\linewidth]{su3break-anni-1-x.pdf}} & 
 0 & 
 $-1\times$\vcenteredhbox{\includegraphics[width=0.03\linewidth]{su3break-anni-3-x.pdf}} & 
 $-1\times$\vcenteredhbox{\includegraphics[width=0.03\linewidth]{sumrule-su3limit-color.pdf}} & 
 $-1\times$\vcenteredhbox{\includegraphics[width=0.03\linewidth]{su3break-color-1-x.pdf}} & 
 0 & 
 $-1\times$\vcenteredhbox{\includegraphics[width=0.03\linewidth]{su3break-color-3-x.pdf}} &  
 0 \\\hline
$+\sqrt{2} \mathcal{A}(D^+ \rightarrow K^+ \pi^0)$ 	&	
 $1\times$\vcenteredhbox{\includegraphics[width=0.03\linewidth]{sumrule-su3limit-tree.pdf}} & 
 0 & 
 $1\times$\vcenteredhbox{\includegraphics[width=0.03\linewidth]{su3break-tree-2-x.pdf}} & 
 $-1\times$\vcenteredhbox{\includegraphics[width=0.03\linewidth]{sumrule-su3limit-anni.pdf}} & 
 0 & 
 $-1\times$\vcenteredhbox{\includegraphics[width=0.03\linewidth]{su3break-anni-2-x.pdf}} & 
 0 & 
 0 & 
 0 & 
 0 & 
 0 &  
 0 \\\hline
\end{tabular}
\caption{Diagrammatic representation of sum rule VI, 
$     + \sqrt{2} \mathcal{A}(D^+\rightarrow \pi^0\pi^+) 
      - \mathcal{A}(D^+\rightarrow \bar{K}^0 K^+ ) 
      - \sqrt{2} \mathcal{A}(D_s^+\rightarrow K^+ \pi^0 ) 
      + \mathcal{A}(D^+\rightarrow \bar{K}^0 \pi^+ ) 
      - \mathcal{A}( D_s^+\rightarrow \bar{K}^0 K^+ )
      +\sqrt{2} \mathcal{A}(D^+\rightarrow K^+ \pi^0 ) = 0$. \label{tab:sumrule-6} }
\end{center}
\end{table*}
\end{turnpage}

\clearpage


\bibliography{mns-14.bib}

\end{document}